\documentclass[english,notitlepage,nofootinbib]{revtex4-1}
\usepackage[T1]{fontenc}
\usepackage[latin9]{inputenc}
\usepackage{babel}
\usepackage{amsmath}
\usepackage{amssymb}
\usepackage{graphicx}
\usepackage[unicode=true,
 bookmarks=false,
 breaklinks=false,pdfborder={0 0 1},backref=false,colorlinks=false]
 {hyperref}

\makeatletter

\usepackage{babel}

\makeatother

\begin{document}
\title{Exclusive photoproduction of heavy quarkonia pairs}
\author{Sebastián Andradé, Marat Siddikov, Iván Schmidt}
\affiliation{Departamento de Física, Universidad Técnica Federico Santa María,~~~~\\
 y Centro Científico - Tecnológico de Valparaíso, Casilla 110-V, Valparaíso,
Chile}
\begin{abstract}
In this paper we study the high energy exclusive photoproduction of heavy quarkonia
pairs in the leading order of the strong coupling constant $\alpha_{s}$.
In the suggested mechanism the quarkonia pairs are produced with opposite charge parities,
and predominantly have oppositely directed transverse momenta. Using
the Color Glass Condensate approach, we estimated numerically the
production cross-sections in the kinematics of the forthcoming electron-proton
colliders, as well as proton-ion colliders in ultraperipheral
collisions. We found that the cross-sections are within the reach
of planned experiments and can be measured with reasonable precision.
The suggested mechanism has significantly larger cross-section than
that of the same $C$-parity quarkonia pair production.
\end{abstract}
\maketitle

\section{Introduction}

The production of heavy quarkonia is frequently considered as a clean
probe for the study of gluon dynamics in high-energy interactions,
since in the limit of heavy quark mass $m_{Q}$ the running coupling
becomes small, and it is possible to apply perturbative methods for the
description of quark-gluon interactions. In many scattering problems
the small size of the color singlet heavy quarkonium provides additional
twist suppression~\cite{Korner:1991kf,Neubert:1993mb}, thus facilitating the
applicability of perturbative treatments. The modern NRQCD framework
allows to use quarkonia production as a powerful probe of strong interactions,
systematically taking into account various perturbative corrections~\cite{Bodwin:1994jh,Maltoni:1997pt,Brambilla:2008zg,Feng:2015cba,Brambilla:2010cs,Cho:1995ce,Cho:1995vh,Baranov:2002cf,Baranov:2007dw,Baranov:2011ib,Baranov:2016clx,Baranov:2015laa}.

For precision studies of hadronic interactions, \emph{exclusive}
production presents a special interest in view of its simpler structure.
However, up to now most of the experimental data on exclusive heavy
quarkonia production were limited to channels with single quarkonia
in the final state. This limitation was largely motivated by probable smaller cross-sections
of events with more than one quarkonia in the final state. Nevertheless,
processes with two mesons in the final state present a lot of
interest and have been the subject of studies since early days of QCD~\cite{Brodsky:1986ds,Lepage:1980fj,Berger:1986ii,Baek:1994kj}.
A recent discovery of all-heavy tetraquarks, which might be consider molecular
states of two quarkonia, has significantly reinvigorated interest
in the study of this channel~\cite{Bai:2016int,Heupel:2012ua,Lloyd:2003yc,Vijande:2006vu,Vijande:2012jw,Chen:2019vrj,Esposito:2018cwh,Cardinale:2018zus,Aaij:2018zrb,Capriotti:2019huu,LHCb:2020bwg}. 

In LHC kinematics most of the previous studies of exclusive double quarkonia
production~\cite{Goncalves:2015sfy,Goncalves:2019txs,Goncalves:2006hu,Baranov:2012vu,Yang:2020xkl}
focused on the so-called two-photon mechanism, $\gamma\gamma\to M_{1}M_{2},$
 which gives the dominant contribution for the production of quarkonia
pairs with the same $C$-parity in ultraperipheral collisions.
Studies beyond the double photon fusion show that, in a TMD factorization
approach, the exclusive double quarkonia production could allow to
measure the currently unknown generalized transverse momentum distributions~
(GTMDs) of gluons~\cite{Bhattacharya:2018lgm}. However, in LHC kinematics
the cross-section of this process can get sizable contributions from the so-called
multiparton scattering diagrams. Such contributions depend on the
poorly known multigluon distributions, leading to potential ambiguities
in the theoretical interpretation of the data. 

Electron-proton collisions have a significant advantage for studies
of heavy quarkonia pair production, due to a smaller number of production mechanisms
compared to hadron-hadron collisions. Moreover, precision studies of double quarkonia
production in $ep$ collisions could become possible after the launch
of new high luminosity facilities, such as the forthcoming Electron Ion
Collider (EIC)~\cite{Accardi:2012qut,DOEPR,BNLPR,AbdulKhalek:2021gbh},
the future Large Hadron electron Collider (LHeC)~\cite{AbelleiraFernandez:2012cc},
the Future Circular Collider (FCC-he)~\cite{Mangano:2017tke,Agostini:2020fmq,Abada:2019lih}
and the CEPC collider~\cite{CEPCStudyGroup:2018rmc,CEPCStudyGroup:2018ghi}.
The main objective of this manuscript is the study of exclusive production
of heavy quarkonia pairs, $\gamma p\to M_{1}M_{2}p$, in the kinematics of the above-mentioned
electron-proton colliders. Potentially such production might be
also probed in ultraperipheral heavy ion and proton-ion collisions.
However, in these cases the analysis becomes more complicated in view of
possible contributions of other mechanisms~\cite{Goncalves:2015sfy,Goncalves:2019txs,Goncalves:2006hu,Baranov:2012vu,Yang:2020xkl}. The large
mass $m_{q}$ of the heavy flavors justifies the perturbative treatment
in a wide kinematic range, without additional restrictions on the
virtuality of the incoming photon $Q^{2}$ or the invariant mass of
the produced quarkonia pair. In absence of imposed kinematic constraints,
the dominant contribution to the cross-section will come from events
induced by quasi-real photons with small $Q^{2}\approx0$ and relatively
small values of $x_{B}\ll1$. In this kinematics it is appropriate
to use the language of color dipole amplitudes and apply the color
dipole (also known as Color Glass Condensate or CGC) framework~\cite{GLR,McLerran:1993ni,McLerran:1993ka,McLerran:1994vd,MUQI,MV,gbw01:1,Kopeliovich:2002yv,Kopeliovich:2001ee}.
At high energies the color dipoles are eigenstates of interaction,
and thus can be used as universal elementary building blocks, automatically
accumulating both the hard and soft fluctuations~\cite{Nikolaev:1994kk}.
The light-cone color dipole framework has been developed and successfully
applied to the phenomenological description of both hadron-hadron and
lepton-hadron collisions~\cite{Kovchegov:1999yj,Kovchegov:2006vj,Balitsky:2008zza,Kovchegov:2012mbw,Balitsky:2001re,Cougoulic:2019aja,Aidala:2020mzt,Ma:2014mri},
and for this reason we will use it for our estimates. 

The paper is structured as follow. Below, in Section~\ref{sec:Formalism},
we evaluate theoretically the cross-section of exclusive photoproduction
of heavy quarkonia pairs in the CGC approach.  In Section~\ref{sec:Numer}
we present our numerical estimates, in the kinematics of the future
$ep$ colliders (EIC, LHeC and FCC-he) and ultraperipheral $pA$ collisions
at LHC. Finally, in Section~\ref{sec:Conclusions} we draw conclusions.

\section{Exclusive meson pair photoproduction}

\label{sec:Formalism} 

\subsection{Kinematics of the process}

We would like to start our discussion of the theoretical framework with
a short description of the kinematics of the process. Our choice of
the light cone decomposition of particles momenta is similar to that
of earlier studies of pion pair~\cite{LehmannDronke:1999vvq,LehmannDronke:2000hlo,Clerbaux:2000hb,Diehl:1999cg,ZEUS:1998xpo}
and single-meson production~\cite{Ji:1998xh,Collins:1998be,Mueller:1998fv,Ji:1996nm,Ji:1998pc,Radyushkin:1996nd,Radyushkin:1997ki,Radyushkin:2000uy,Collins:1996fb,Brodsky:1994kf,Goeke:2001tz,Diehl:2000xz,Belitsky:2001ns,Diehl:2003ny,Belitsky:2005qn,Kubarovsky:2011zz,Dupre:2017hfs}.
However, we should take into account that the mass of the quarkonium,
in contrast to that of pion, is quite large, and thus cannot be disregarded
as a kinematic higher twist correction. Besides, for photoproduction
this mass can appear as one of the hard scales in the problem.

In what follows we will use the notations: $q$ for the photon momentum,
$P$ and $P'$ for the momentum of the proton before and after the
collision, and $p_{1},\,p_{2}$ for the 4-momenta of produced heavy
quarkonia. For sake of generality we will assume temporarily that
the photon can have a nonzero virtuality $-q^{2}=Q^{2}$, taking
later that for photoproduction $Q^{2}=0$. We also will
use the notation $\Delta$ for the momentum transfer to the proton,
$\Delta=P'-P$, and the notation $t$ for its square, $t\equiv\Delta^{2}$.
The light-cone expansion of the above-mentioned momenta in the lab
frame is given by~\footnote{In earlier theoretical studies~~\cite{Goeke:2001tz,Gui98,LehmannDronke:1999vvq,LehmannDronke:2000hlo,Clerbaux:2000hb,Diehl:1999cg}
the evaluations were done in the so-called symmetric frame, in which the
axis $z$ is chosen in such a way that the vectors $q$ and $\bar{P}\equiv P+P'$
do not have transverse components. Besides, all evaluations were done
in the Bjorken limit, assuming infinitely large $Q^{2}$ and negligibly
small masses of the produced mesons (pions). In our studies we consider
quasi-real photons, with $Q^{2}\approx0$, and moreover the heavy mass
of the quarkonia does not allow to drop certain ``higher twist''
terms. For this reason the kinematic expressions in the symmetric
frame become quite complicated, and there is no advantage in its
use for photoproduction. } 
\begin{align}
q & =\,\left(q^{+},\,\frac{Q^{2}}{2q^{+}},\,\,\boldsymbol{0}_{\perp}\right),\quad q^{+}=E_{\gamma}+\sqrt{E_{\gamma}^{2}+Q^{2}}\approx2E_{\gamma}\label{eq:qPhoton}\\
P & =\left(\frac{m_{N}^{2}}{2P^{-}},\,P^{-},\,\,\boldsymbol{0}_{\perp}\right),\quad P^{-}=E_{p}+\sqrt{E_{p}^{2}-m_{N}^{2}}\approx2E_{p}\\
p_{a} & =\left(M_{a}^{\perp}\,e^{y_{a}}\,,\,\frac{M_{a}^{\perp}e^{-y_{a}}}{2},\,\,\boldsymbol{p}_{a}^{\perp}\right),\quad a=1,2,\label{eq:MesonLC}\\
 & M_{a}^{\perp}\equiv\sqrt{M_{a}^{2}+\left(\boldsymbol{p}_{a}^{\perp}\right)^{2}},
\end{align}
where $\left(y_{a},\,\boldsymbol{p}_{a}^{\perp}\right)$ are the rapidity
and transverse momentum of the quarkonium $a$, and $M_{a}$ is
its mass. Using conservation of 4-momentum, we may obtain
for the momentum transfer to the proton 
\begin{align}
\Delta & =P'-P=q-p_{1}-p_{2}=\\
 & =\left(q^{+}-M_{1}^{\perp}\,e^{y_{1}}-M_{2}^{\perp}\,e^{y_{2}}\,,\,\frac{Q^{2}}{2q^{+}}-\frac{M_{1}^{\perp}e^{-y_{1}}+M_{2}^{\perp}e^{-y_{2}}}{2},-\boldsymbol{p}_{1}^{\perp}-\boldsymbol{p}_{2}^{\perp}\right),\nonumber 
\end{align}
and for the variable $t\equiv\Delta^{2}$ 
\begin{align}
t & =\Delta^{2}=\left(q^{+}-M_{1}^{\perp}\,e^{y_{1}}-M_{2}^{\perp}\,e^{y_{2}}\right)\left(\frac{Q^{2}}{q^{+}}-M_{1}^{\perp}e^{-y_{1}}-M_{2}^{\perp}e^{-y_{2}}\right)-\left(\boldsymbol{p}_{1}^{\perp}+\boldsymbol{p}_{2}^{\perp}\right)^{2}.
\end{align}
After the interaction the 4-momentum of the proton  is given by 
\begin{equation}
P'=P+\Delta=\left(q^{+}+\frac{m_{N}^{2}}{2P^{-}}-M_{1}^{\perp}\,e^{y_{1}}-M_{2}^{\perp}\,e^{y_{2}}\,,\,P^{-}+\frac{Q^{2}}{2q^{+}}-\frac{M_{1}^{\perp}e^{-y_{1}}+M_{2}^{\perp}e^{-y_{2}}}{2},-\boldsymbol{p}_{1}^{\perp}-\boldsymbol{p}_{2}^{\perp}\right),
\end{equation}
and the onshellness condition $\left(P+\Delta\right)^{2}=m_{N}^{2}$
allows to get an additional constraint 
\begin{align}
q\cdot P & \equiv q^{+}P^{-}=P^{-}\left(M_{1}^{\perp}\,e^{y_{1}}+M_{2}^{\perp}\,e^{y_{2}}\right)-\frac{m_{N}^{2}+t}{2}+\frac{m_{N}^{2}}{4P^{-}}\left(M_{1}^{\perp}e^{-y_{1}}+M_{2}^{\perp}e^{-y_{2}}-\frac{Q^{2}}{q^{+}}\right).\label{qPlus}
\end{align}
Solving Equation~(\ref{qPlus}) with respect to $q\cdot P$,
we get 
\begin{align}
q\cdot P & =\frac{P^{-}\left(M_{1}^{\perp}\,e^{y_{1}}+M_{2}^{\perp}\,e^{y_{2}}\right)-\frac{m_{N}^{2}+t}{2}+\frac{m_{N}^{2}}{4P^{-}}\left(M_{1}^{\perp}e^{-y_{1}}+M_{2}^{\perp}e^{-y_{2}}\right)}{2}\pm\label{qPlus-1}\\
 & \pm\frac{1}{2}\sqrt{\left(P^{-}\left(M_{1}^{\perp}\,e^{y_{1}}+M_{2}^{\perp}\,e^{y_{2}}\right)-\frac{m_{N}^{2}+t}{2}+\frac{m_{N}^{2}}{4P^{-}}\left(M_{1}^{\perp}e^{-y_{1}}+M_{2}^{\perp}e^{-y_{2}}\right)\right)^{2}+Q^{2}m_{N}^{2}}\,,\nonumber 
\end{align}
which allows to express the energy of the photon $E_{\gamma}\approx q^{+}/2$
in terms of the kinematic variables $\left(y_{a},\,\boldsymbol{p}_{a}^{\perp}\right)$
of the produced quarkonia. In the kinematics of all experiments which
we consider below, the typical values $q^{+},P^{-}\gg\{Q,\,M_{a},\,m_{N},\,t\}$,
and for this reason we may approximate~(\ref{qPlus-1}) as 
\begin{align}
q\cdot P & \equiv q^{+}P^{-}\approx P^{-}\left(M_{1}^{\perp}\,e^{y_{1}}+M_{2}^{\perp}\,e^{y_{2}}\right),\qquad{\rm or}\label{eq:qPlusApprox}\\
 & q^{+}\approx M_{1}^{\perp}\,e^{y_{1}}+M_{2}^{\perp}\,e^{y_{2}}\label{eq:qPlus}
\end{align}
From comparison of~(\ref{eq:MesonLC}) and~(\ref{qPlus}) we may
see that at high energies the light-cone plus-component of the photon
momentum $q^{+}$ is shared between the momenta of the produced quarkonia,
whereas the momentum transfer to the proton (vector $\Delta$) has a negligibly
small plus-component, in agreement with the eikonal picture expectations.
The expressions~(\ref{qPlus-1},~\ref{eq:qPlusApprox}) allow to
express the Bjorken variable $x_{B}$, which appears in the analysis of
this process in Bjorken kinematics, using its conventional definition
$x_{B}=Q^{2}/2\left(p\cdot q\right)\approx Q^2/(Q^2+W^2)$. As was discussed in~\cite{Tung:2001mv,Kowalski:2006hc,Kowalski:2003hm,Rezaeian:2012ji},
in phenomenological studies usually it is assumed that for heavy quarks all the gluon densities and forward dipole amplitudes should depend on the so-called ``rescaling variable''
\begin{equation}
x=x_{B}\left(1+\frac{(4m_{Q})^{2}}{Q^{2}}\right)=\frac{Q^{2}+(4m_{Q})^{2}}{2\left(p\cdot q\right)},
\end{equation}
which was introduced in~\cite{Tung:2001mv} in order to improve description of the near-threshold heavy quarkonia production. While the color dipole framework is usually applied far from the near-threshold kinematics, the use of the variable $x$ instead of $x_B$ for heavy quarks improves agreement of dipole approach predictions with experimental data. In Bjorken limit the variable $x$ coincides with $x_B$. For small $Q^2\approx 0$ (photoproduction regime) the variable $x_B$ vanishes, whereas $x$ remains finite and is given by the approximate expression
\begin{align}
x & =\left.\frac{Q^{2}+(4m_{Q})^{2}}{2\left(p\cdot q\right)}\right|_{Q\approx 0}\approx\frac{8m_{Q}^{2}}{P^{-}\left(M_{1}^{\perp}\,e^{y_{1}}+M_{2}^{\perp}\,e^{y_{2}}\right)}+\mathcal{O}\left(\frac{Q^{2}}{m_{Q}^{2}}\right)\approx\frac{4m_{Q}^{2}}{E_{p}\left(M_{1}^{\perp}\,e^{y_{1}}+M_{2}^{\perp}\,e^{y_{2}}\right)}.
\end{align}
In this study we are interested in the production of both quarkonia at
central rapidities (in lab frame) by high-energy photon-proton collision.
In this kinematics the variable $x$ is very small, which suggests
that the amplitude of this process should be analyzed in frameworks
with built-in saturation, such as color glass condensate (CGC). In contrast,
in the Bjorken limit ($Q^{2}\to\infty,\,Q^{2}/2p\cdot q={\rm const}$)
we observe that the variable $x$ can be quite large, so it
is more appropriate to analyze this kinematics using collinear or
$k_{T}$-factorization. The latter case requires a separate study
and will be presented elsewhere.

In the photoproduction
approximation the invariant energy of the $\gamma p$ collision can be written as 
\begin{equation}
W^{2}\equiv s_{\gamma p}=\left(q+P\right)^{2}=-Q^{2}+m_{N}^{2}+2q\cdot P\approx-m_{N}^{2}+P^{-}\left(M_{1}^{\perp}\,e^{y_{1}}+M_{2}^{\perp}\,e^{y_{2}}\right),\label{eq:W2}
\end{equation}
whereas the invariant mass of the produced heavy quarkonia pair is
given by
\begin{equation}
M_{12}^{2}=\left(p_{1}+p_{2}\right)^{2}=M_{1}^{2}+M_{2}^{2}+2\left(M_{1}^{\perp}M_{2}^{\perp}\cosh\left(y_1-y_2\right)-\boldsymbol{p}_{1}^{\perp}\cdot\boldsymbol{p}_{2}^{\perp}\right).\label{eq:M12}
\end{equation}

In electron-proton collisions the cross-section of heavy meson pairs 
is dominated by a single-photon exchange between its leptonic and hadronic
parts, and for this reason can be represented as 
\begin{equation}
\frac{d\sigma_{ep\to eM_{1}M_{2}p}}{dQ^{2}\,dy_{1}d^{2}\boldsymbol{p}_{1}^\perp dy_{2}d^{2}\boldsymbol{p}_{2}^\perp}=\frac{\alpha_{{\rm em}}}{\pi\,Q^{2}}\,\left[\left(1-y\right)\frac{d\sigma_{L}}{dy_{1}d^{2}\boldsymbol{p}_{1}^\perp dy_{2}d^{2}\boldsymbol{p}_{2}^\perp}+\left(1-y+\frac{y^{2}}{2}\right)\frac{d\sigma_{T}}{dy_{1}d^{2}\boldsymbol{p}_{1}^\perp dy_{2}d^{2}\boldsymbol{p}_{2}^\perp}\right],\label{eq:LTSep}
\end{equation}
where we use the standard DIS notation $y$ for the elasticity (fraction
of electron energy which passes to the photon, not to be confused
with the rapidities $y_{a}$ of the produced quarkonia). The subscript
letters $L,T$ in the right-hand side of~(\ref{eq:LTSep}) stand
for the contributions of longitudinally and transversely polarized
photons respectively. The structure of~(\ref{eq:LTSep}) suggests
that the dominant contribution to the cross-section comes from the
region of small $Q^{2}$. In this kinematics the contribution of $d\,\sigma_{L}$
is suppressed compared to the term $d\,\sigma_{T}$. This expectation
is partially corroborated by the experimental data from ZEUS~\cite{ZEUS:2004yeh}
and H1~\cite{H1:2005dtp}, which found that for \emph{single }quarkonia
production in the region $Q^{2}\lesssim1\,{\rm GeV}^{2}$ the
longitudinal cross-section $d\sigma_{L}$ constitutes less than 10\%
of the transverse cross-section $d\sigma_{T}$. For this reason in
this paper we'll disregard the cross-section $d\sigma_{L}$ altogether,
while the  relevant cross-section $d\sigma_{T}$ is  
\begin{align}
\frac{d\sigma_{T}}{dy_{1}\,d\left|p_{1}^\perp\right|^{2}dy_{2}\,d\left|p_{2}^\perp\right|^{2}d\phi} & \approx\frac{1}{256\pi^{4}}\left|\mathcal{A}_{\gamma_{T}p\to M_{1}M_{2}p}\right|^{2}\delta\left(\frac{M_{1}^{\perp}\,e^{y_{1}}+M_{2}^{\perp}\,e^{y_{2}}}{q^{+}}-1\right)\label{eq:AmpXSection}
\end{align}
where $\mathcal{A}_{\gamma_{T}p\to M_{1}M_{2}p}$ is the amplitude
of the exclusive process, induced by a transversely polarized photon,
and $\phi$ is the angle between the vectors $\boldsymbol{p}_{1}$
and $\boldsymbol{p}_{2}$ in the transverse plane. The $\delta$-function
in~(\ref{eq:AmpXSection}) reflects conservation of plus-component
of momentum, discussed earlier in~(\ref{qPlus}). 

Similarly, for exclusive \emph{hadro}production $pA\to pAM_{1}M_{2}$
in ultraperipheral kinematics we may obtain the cross-section using
the equivalent photon (Weizsäcker-Williams) approximation, 
\begin{equation}
\frac{d\sigma\left(p+A\to p+A+M_{1}+M_{2}\right)}{dy_{1}d^{2}\boldsymbol{p}_{1}^{\perp}dy_{2}d^{2}\boldsymbol{p}_{2}^{\perp}}=\int dn_{\gamma}\left(\omega\equiv E_{\gamma},\,\boldsymbol{q}_{\perp}\right)\,\frac{d\sigma_{T}\left(\gamma+p\to\gamma+p+M_{1}+M_{2}\right)}{dy_{1}d^{2}\boldsymbol{p}_{1}^{*}dy_{2}d^{2}\boldsymbol{p}_{2}^{*}}\label{eq:EPA_q}
\end{equation}
where $dn_{\gamma}\left(\omega\equiv E_{\gamma},\,\boldsymbol{q}_{\perp}\right)$
is the spectral density of the flux of photons created by the nucleus,
$\boldsymbol{q}_{\perp}$ is the transverse momentum of the photon
with respect to the nucleus, and the energy $E_{\gamma}$ of the photon
can be related to the kinematics of produced quarkonia using Eq.~(\ref{qPlus-1},~\ref{eq:qPlusApprox}).
The explicit expression for $dn_{\gamma}\left(\omega\equiv E_{\gamma},\,\boldsymbol{q}_{\perp}\right)$
can be found in~\cite{Budnev:1975poe}. The momenta $\boldsymbol{p}_{i}^{*}=\boldsymbol{p}_{i}^{\perp}-\boldsymbol{q}_{\perp}$are
the transverse parts of the quarkonia momenta with respect to the
produced photon. Due to nuclear form factors the typical values of
momenta $\boldsymbol{q}_{\perp}$ are controlled by the nuclear radius
$R_{A}$ and are quite small, $\left\langle \boldsymbol{q}_{\perp}^{2}\right\rangle \sim\left\langle Q^{2}\right\rangle \sim\left\langle R_{A}^{2}\right\rangle ^{-1}\lesssim\left(0.2\,{\rm GeV}/A^{1/3}\right)^{2}$.
For this reason, for very heavy ions ($A\gg1$) we may expect that
the $p_{T}$-dependence of the cross-sections in the left-hand side
of~(\ref{eq:EPA_q}) largely repeats the $p_{T}$-dependence of the
cross-section in the integrand in the right-hand side. For the special
and experimentally important case of $\boldsymbol{p}^{\perp}$-integrated
cross-section, the expression~(\ref{eq:EPA_q}) simplifies and can
be rewritten as 
\begin{align}
\frac{d\sigma\left(p+A\to p+A+M_{1}+M_{2}\right)}{dy_{1}dy_{2}} & =\int dE_{\gamma}\frac{dN_{\gamma}\left(\omega\equiv E_{\gamma}\right)}{dE_{\gamma}}\,\frac{d\sigma_{T}\left(\gamma+p\to\gamma+p+M_{1}+M_{2}\right)}{dy_{1}dy_{2}},\label{eq:EPA_q-1}
\end{align}
where 
\begin{equation}
N_{\gamma}\left(\omega\right)\equiv\int d^{2}\boldsymbol{q}_{\perp}\,\frac{dn_{\gamma}\left(\omega,\,\boldsymbol{q}_{\perp}\right)}{d\omega\,d^{2}\boldsymbol{q}_{\perp}}.
\end{equation}
In the following subsection~\ref{subsec:Evaluation} we evaluate
the amplitude $\mathcal{A}_{\gamma_{T}p\to M_{1}M_{2}p}$ which 
determines the cross-sections of photoproduction processes.

\subsection{Amplitude of the process in the color dipole picture}

\label{subsec:Evaluation}Since the formation time of rapidly moving
heavy quarkonia significantly exceeds the size of the proton, the
quarkonia formation occurs far outside the interaction region. For
this reason the amplitudes of the quarkonia production processes can
be represented as a convolution of the quarkonia wave functions with
hard amplitudes, which characterize the production of the small pairs of
nearly onshell heavy quarks in the gluonic field of the target. In
what follows we will refer to these nearly onshell quarks as ``produced''
or ``final state'' quarks. For exclusive production the cross-section
falls rapidly as function of transverse momenta $p_{T}$ of the produced
quarkonia, and for this reason we expect that the quarkonia will be
produced predominantly with small momenta. In this kinematical region it is
possible to disregard completely the color octet contributions~\cite{Cho:1995ce,Cho:1995vh}.
As was shown in~\cite{Kowalski:2003hm,Kowalski:2006hc,Rezaeian:2012ji},
this assumption gives very good description of the exclusive production
of \emph{single} quarkonia.

The general rules for the evaluation of different hard amplitudes in terms
of the color singlet forward dipole amplitude were introduced in~\cite{GLR,McLerran:1993ka,McLerran:1994vd,MUQI,MV,gbw01:1,Kopeliovich:2002yv,Kopeliovich:2001ee}
and are briefly summarized in Appendix~\ref{subsec:Derivation}.
This approach is based on the high energy eikonal picture, and therefore
the partons transverse coordinates and helicities remain essentially
frozen during propagation in the gluonic field of the target. The
hard scale, which controls the interaction of a heavy quark with the strong
gluonic field, is its mass $m_{Q}$, so in the heavy mass limit we
may treat this interaction perturbatively. However, the interaction
of gluons with each other, as well as with light quarks, remains strongly 
nonperturbative in the deeply saturated regime. 

\begin{figure}
\includegraphics[scale=0.65]{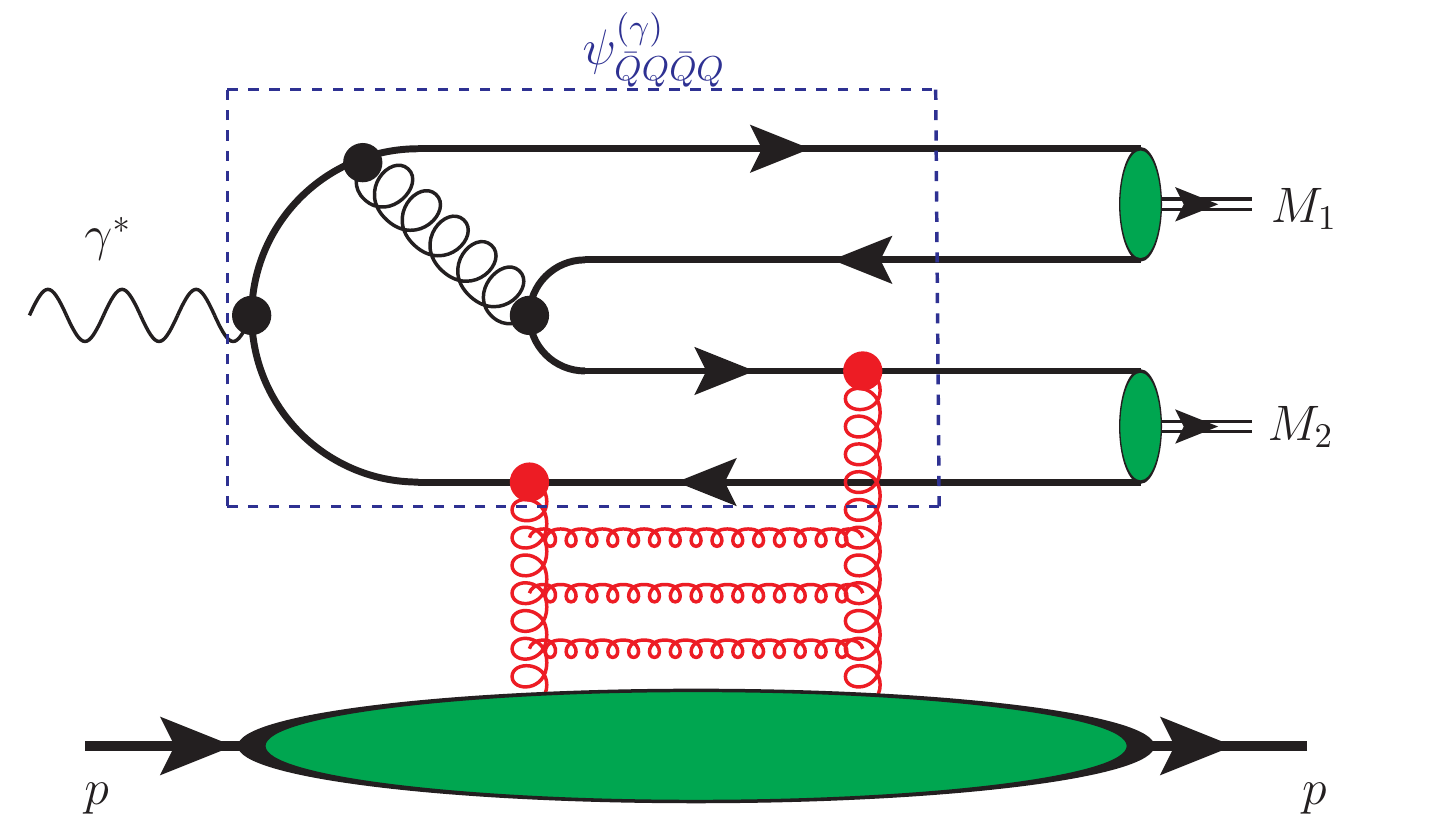}\includegraphics[scale=0.65]{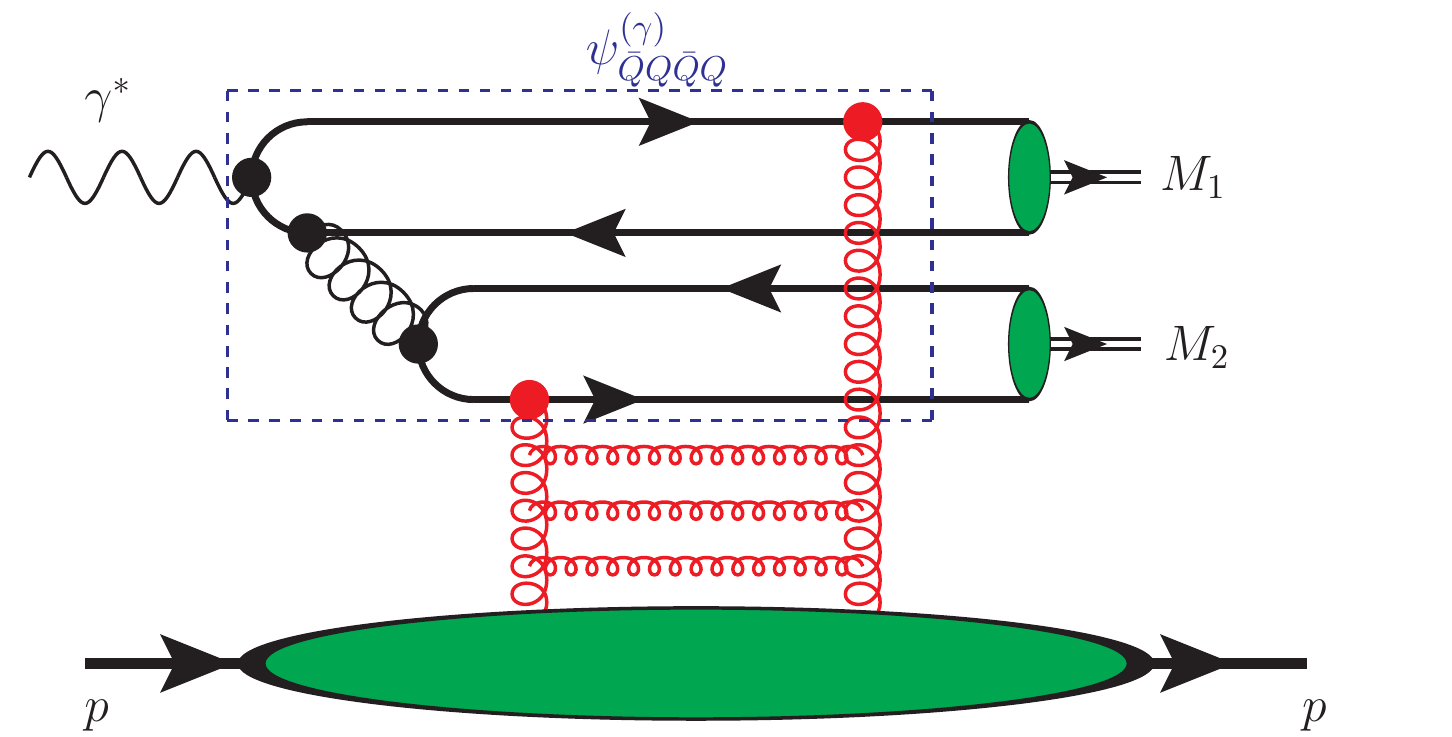}\caption{\label{fig:Photoproduction}Main classes of diagrams which contribute
in the leading order over $\alpha_{s}\left(m_{Q}\right)$ to exclusive
photoproduction of quarkonia pairs (type-$A$ and type-$B$ diagrams).
The eikonal interactions are shown schematically as exchanges of $t$-channel
gluons, indicated by the red wavy lines. In both plots it is implied:
(a) summation over all possible attachments of $t$-channel gluons
to partons in the upper part of diagram (b) inclusion of diagrams with
inverted direction of heavy quark lines (\textquotedblleft charge
conjugation\textquotedblright ). In the right diagram the $t$-channel
gluons must be connected to different quark loops in order to guarantee a
\emph{color singlet} $\bar{Q}Q$ in the final state. The blue dashed rectangle
schematically shows part of the diagrams which (in absence of eikonal
interactions) contribute to the $\bar{Q}Q\bar{Q}Q$-component of the
photon wave function $\psi_{\bar{Q}Q\bar{Q}Q}^{(\gamma)}$.}
\end{figure}

In the leading order over the strong coupling $\alpha_{s}\left(m_{Q}\right)$,
there are a few dozen Feynman diagrams which contribute to the
exclusive photoproduction of meson pairs. In what follows, it is convenient
to represent them as one of the two main classes shown schematically
in Figure~\ref{fig:Photoproduction}. For the sake of definiteness
we'll call ``type-$A$'' all the diagrams in which quarkonia are
formed from different heavy quark lines, as shown in the left panel
of Figure~\ref{fig:Photoproduction}. The opposite case, when
quarkonia are formed from the same quark lines, as shown in the right
panel of the Figure~\ref{fig:Photoproduction}, will be referred
to as ``type-$B$'' diagrams. This classification is convenient
for a discussion of symmetries, as well as for analysis of quarkonia production
with mixed flavors. For example, production of $B_{c}^{+}B_{c}^{-}$
pairs clearly gets contributions only from type-$A$ diagrams,
whereas production of mixed flavor hidden-charm and hidden-bottom
quarkonia (\emph{e.g}. $J/\psi+\eta_{b}$) gets contributions only
from type-$B$ diagrams.

In configuration space the eikonal interactions with the target do not
affect the impact parameters of the partons, so the interaction basically
reduces to a mere multiplication of target-dependent factors,
as discussed in Appendix~\ref{subsec:Derivation}. This allows to
express the amplitude of the whole process as a convolution of the
4-quark Fock component wave function $\psi_{\bar{Q}Q\bar{Q}Q}^{(\gamma)}$
of the photon with dipole amplitudes and wave functions of the produced
quarkonia.  The amplitude of the process $\gamma^{*}p\to M_{1}M_{2}p$
can be represented as a sum

\begin{align}
\mathcal{A}\left(y_{1},\boldsymbol{p}_{1}^\perp,y_{2},\boldsymbol{p}_{2}^\perp\right) & =\mathcal{A}_{1}\left(y_{1},\boldsymbol{p}_{1}^\perp,y_{2},\boldsymbol{p}_{2}^\perp\right)+\mathcal{A}_{2}\left(y_{1},\boldsymbol{p}_{1}^\perp,y_{2},\boldsymbol{p}_{2}^\perp\right),\label{eq:Amp}
\end{align}
where $\mathcal{A}_{1}$ and $\mathcal{A}_{2}$ stand for contributions
of all type-$A$ and type-$B$ diagrams. Explicitly, these amplitudes
are given by

\begin{align}
\mathcal{A}_{1}\left(y_{1},\boldsymbol{p}_{1}^\perp,y_{2},\boldsymbol{p}_{2}^\perp\right) & =\prod_{i=1}^{4}\left(\int d\alpha_{i}d^{2}\boldsymbol{x}_{i}\right)\delta\left(\sum_{k}\alpha_{k}-1\right)\tilde{\sum}_{\ell n}\sigma_{\ell}\sigma_{n}\,c_{\ell n}\gamma\left(\boldsymbol{b}_{\ell}\right)\gamma\left(\boldsymbol{b}_{n}\right)\times\label{eq:Amp-1}\\
 & \times\left[\Psi_{M_{1}}^{\dagger}\left(\alpha_{14},\,\boldsymbol{r}_{14}\right)\Psi_{M_{2}}^{\dagger}\left(\alpha_{23},\,\boldsymbol{r}_{23}\right)e^{i\left(\boldsymbol{p}_{1}^{\perp}\cdot\boldsymbol{b}_{14}+\boldsymbol{p}_{2}^{\perp}\cdot\boldsymbol{b}_{23}\right)}\delta\left(y_{1}-\mathcal{Y}_{14}\right)\delta\left(y_{2}-\mathcal{Y}_{23}\right)\right.\nonumber \\
 & +\left.\Psi_{M_{1}}^{\dagger}\left(\alpha_{23},\,\boldsymbol{r}_{23}\right)\Psi_{M_{2}}^{\dagger}\left(\alpha_{14},\,\boldsymbol{r}_{14}\right)e^{i\left(\boldsymbol{p}_{1}^{\perp}\cdot\boldsymbol{b}_{23}+\boldsymbol{p}_{2}^{\perp}\cdot\boldsymbol{b}_{14}\right)}\delta\left(y_{1}-\mathcal{Y}_{23}\right)\delta\left(y_{2}-\mathcal{Y}_{14}\right)\right]\nonumber \\
 & \times\psi_{\bar{Q}Q\bar{Q}Q}^{(\gamma)}\left(\alpha_{1},\boldsymbol{x}_{1};\,\alpha_{2},\,\boldsymbol{x}_{2};\,\alpha_{3},\,\boldsymbol{x}_{3};\,\alpha_{4},\,\boldsymbol{x}_{4};\,q\right).\nonumber 
\end{align}
\begin{align}
\mathcal{A}_{2}\left(y_{1},\boldsymbol{p}_{1}^\perp,y_{2},\boldsymbol{p}_{2}^\perp\right) & =\prod_{i=1}^{4}\left(\int d\alpha_{i}d^{2}\boldsymbol{x}_{i}\right)\delta\left(\sum_{k}\alpha_{k}-1\right)\tilde{\sum}_{\ell n}\sigma_{\ell}\sigma_{n}\,c_{\ell n}\gamma\left(\boldsymbol{b}_{\ell}\right)\gamma\left(\boldsymbol{b}_{n}\right)\times\label{eq:Amp-2}\\
 & \times\left[\Psi_{M_{1}}^{\dagger}\left(\alpha_{12},\,\boldsymbol{r}_{12}\right)\Psi_{M_{2}}^{\dagger}\left(\alpha_{34},\,\boldsymbol{r}_{34}\right)e^{i\left(\boldsymbol{p}_{1}^{\perp}\cdot\boldsymbol{b}_{12}+\boldsymbol{p}_{2}^{\perp}\cdot\boldsymbol{b}_{34}\right)}\delta\left(y_{1}-\mathcal{Y}_{12}\right)\delta\left(y_{2}-\mathcal{Y}_{34}\right)\right.\nonumber \\
 & +\left.\Psi_{M_{1}}^{\dagger}\left(\alpha_{34},\,\boldsymbol{r}_{34}\right)\Psi_{M_{2}}^{\dagger}\left(\alpha_{12},\,\boldsymbol{r}_{12}\right)e^{i\left(\boldsymbol{p}_{1}^{\perp}\cdot\boldsymbol{b}_{34}+\boldsymbol{p}_{2}^{\perp}\cdot\boldsymbol{b}_{12}\right)}\delta\left(y_{1}-\mathcal{Y}_{34}\right)\delta\left(y_{2}-\mathcal{Y}_{12}\right)\right]\nonumber \\
 & \times\psi_{\gamma^{*}\to\bar{Q}Q\bar{Q}Q}\left(\alpha_{1},\boldsymbol{x}_{1};\,\alpha_{2},\,\boldsymbol{x}_{2};\,\alpha_{3},\,\boldsymbol{x}_{3};\,\alpha_{4},\,\boldsymbol{x}_{4};\,q\right),\nonumber 
\end{align}
where in the expressions~(\ref{eq:Amp-1},~\ref{eq:Amp-2}) we introduced
a few shorthand notations, which characterize the pair of heavy
partons $i$ and $j$: the relative distance between them $\boldsymbol{r}_{ij}=\boldsymbol{x}_{i}-\boldsymbol{x}_{j}$,
the light-cone fraction $\alpha_{ij}=\alpha_{i}/\left(\alpha_{i}+\alpha_{j}\right)$
carried by the quark in the pair ($ij$), and the transverse coordinate
of its center of mass $\boldsymbol{b}_{ij}=\left(\alpha_{i}\boldsymbol{x}_{i}+\alpha_{j}\boldsymbol{x}_{j}\right)/\left(\alpha_{i}+\alpha_{j}\right)$.
The notation $\tilde{\sum}_{\ell n}$ in the first line of~(\ref{eq:Amp-1},~\ref{eq:Amp-2})
implies summation over all possible attachments of $t$-channel gluons
to the partons in the upper part of the diagram. For type-$A$ diagrams
the variables $\ell,\,n$ may take independently six different values,
which correspond to connections to final quarks, virtual quark or virtual
gluon. For type-$B$ diagrams both produced quark pairs must
be in a color singlet state, which translates into the additional constraint
that $\ell,\,n$ should be connected to \emph{different} quark loops
(either upper or lower quark-antiquark pairs). The factors
$\sigma_{\ell},\,\sigma_{n}$ in the first line of~(\ref{eq:Amp-1},~\ref{eq:Amp-2})
have the value $+1$ if the corresponding $t$-channel gluon is connected
to a quark line or gluon, and -1 otherwise. On the other hand, the color
factors $c_{\ell n}$ depend on the topology of the diagram under
consideration; more precisely, \emph{how} the $t$-channel gluons are connected
to the quark lines. For type-$A$ diagrams, the color factor $c_{\ell n}=\mathcal{C}_{1}\equiv\frac{1}{N_{c}^{2}-1}{\rm tr}_{c}(t_{a}t_{a}t_{b}t_{b})=\left(N_{c}^{2}-1\right)/4N_{c}$,
if both $t$-channel gluons are connected to the same quark line,
or quark and antiquark lines of opposite color (\emph{e.g}. quark-antiquark
lines originating from colorless photon or leading to formation of
colorless quarkonium). If the vertices of the $t$-channel gluons
are separated by color changing vertex of a virtual gluon, then the
color factor is given by $c_{\ell n}=\mathcal{C}_{2}\equiv\frac{1}{N_{c}^{2}-1}{\rm tr}_{c}(t_{a}t_{b}t_{a}t_{b})=-1/4N_{c}$.
For the diagrams with one 3-gluon vertex, when one of the $t$-channel
gluons is attached to a virtual gluon, the corresponding color factor is
$c_{\ell n}=\pm\mathcal{C}_{3}=\pm N_{c}/4$, where the sign is positive
for the diagram with attachment of the other $t$-channel gluon to
the upper quark-antiquark pair (i.e. partons 1,2), and negative otherwise.
Finally, for the diagram when both $t$-channel gluons are attached
to a virtual (intermediate) gluon, the corresponding factor is $c_{\ell n}=\mathcal{C}_{4}\equiv N_{c}/2$.
For type-$B$ diagrams, the corresponding color factor is $c_{\ell n}=\frac{1}{N_{c}^{2}-1}\left[{\rm tr}_{c}\left(t_{a}t_{c}\right)\right]^{2}=\frac{1}{4}$
for all possible connections of $t$-channel gluons. The functions
$\gamma(...)$ characterize the interaction of the parton with the target,
and can be related to the dipole amplitude as explained in Appendix~(\ref{subsec:Derivation}).
The variables $\boldsymbol{b}_{\ell},\,\boldsymbol{b}_{n}$ in the
arguments of $\gamma(...)$-functions stand for the transverse coordinate
of the parton which interacts with a $t$-channel gluon. For the final
quarks this variable corresponds to the transverse coordinates of
these partons (the integration variables $\boldsymbol{x}_{i}$). For
intermediate partons this variable is the position of
the center of mass of all final quarks which are produced at later
stages, 
\begin{equation}
\boldsymbol{b}_{j_{1}...j_{n}}=\frac{\sum_{j=j_{1}...j_{n}}\alpha_{j}\boldsymbol{x}_{j}}{\sum_{j=j_{1}...j_{n}}\alpha_{j}}\label{eq:ImpFac}
\end{equation}
where the summation is done over all final quarks $j_{1},\,...j_{n}$
which stem from a given parton. The notations $\Psi_{M_{1}},\,\Psi_{M_{2}}$
are used for the wave functions of the final state quarkonia $M_{1}$
and $M_{2}$ (for a moment we disregard completely their spin indices),
and $\psi_{\bar{Q}Q\bar{Q}Q}^{(\gamma)}\left(\left\{ \alpha_{i},\boldsymbol{x}_{i}\right\} ;q\right)$
is the 4-quark light-cone wave function of the virtual photon $\gamma^{*}$,
which is evaluated in Appendix~\ref{sec:WF}. The product $\tilde{\sum}_{\ell n}\sigma_{\ell}\sigma_{n}\,c_{\ell n}\gamma\left(\boldsymbol{b}_{\ell}\right)\gamma\left(\boldsymbol{b}_{n}\right)$
can be expressed as a linear superposition of the color singlet
dipole amplitudes $N\left(x,\boldsymbol{r}_{ij},\,\boldsymbol{b}_{ij}\right)$
(see derivation in Appendix~\ref{sec:DipAmp}). For the type-$A$
contribution, the final result is 
\begin{align}
\tilde{\sum}_{\ell n}\sigma_{\ell}\sigma_{n}\,c_{\ell n}\gamma\left(\boldsymbol{b}_{\ell}\right)\gamma\left(\boldsymbol{b}_{n}\right) & =\left\{ \frac{2-N_{c}^{2}}{4N_{c}}N\left(x,\,\boldsymbol{r}_{14},\,\boldsymbol{b}_{14}\right)-\frac{1}{2N_{c}}N\left(x,\,\boldsymbol{r}_{34},\,\boldsymbol{b}_{34}\right)-\frac{3+5N_{c}^{2}}{4N_{c}}N\left(x,\,\boldsymbol{r}_{12},\,\boldsymbol{b}_{12}\right)+\right.\label{eq:N_dip-1}\\
 & +\frac{1}{4N_{c}}\left[N\left(x,\,\boldsymbol{r}_{23},\,\boldsymbol{b}_{23}\right)-N\left(x,\,\frac{\alpha_{1}\boldsymbol{r}_{14}+\alpha_{3}\boldsymbol{r}_{34}}{1-\alpha_{2}},\,\boldsymbol{b}_{1344}\right)\right]\nonumber \\
 & +\frac{N_{c}^{2}+2}{4N_{c}}N\left(x,\,\boldsymbol{r}_{13},\,\boldsymbol{b}_{13}\right)+\frac{3N_{c}^{2}-2}{4N_{c}}\,N\left(x,\frac{\alpha_{1}\boldsymbol{r}_{21}+\alpha_{3}\boldsymbol{r}_{23}+\alpha_{4}\boldsymbol{r}_{24}}{1-\alpha_{2}},\boldsymbol{b}_{1234}\right)\nonumber \\
 & +\frac{3N_{c}}{2}N\left(x,\frac{\alpha_{3}\boldsymbol{r}_{13}+\alpha_{4}\boldsymbol{r}_{14}}{\alpha_{3}+\alpha_{4}},\boldsymbol{b}_{134}\right)+2N_{c}\,N\left(x,\frac{\alpha_{3}\boldsymbol{r}_{23}+\alpha_{4}\boldsymbol{r}_{24}}{\alpha_{3}+\alpha_{4}},\boldsymbol{b}_{234}\right)\nonumber \\
 & +\frac{N_{c}^{2}+1}{4N_{c}}\left[N\left(x,\frac{\alpha_{3}\boldsymbol{r}_{13}+\alpha_{4}\boldsymbol{r}_{14}}{1-\alpha_{2}},\boldsymbol{b}_{1134}\right)+N\left(x,\,\boldsymbol{r}_{24},\,\boldsymbol{b}_{24}\right)\right]\nonumber \\
 & -\frac{N_{c}}{2}\left[N\left(x,\frac{\alpha_{4}\boldsymbol{r}_{34}}{\alpha_{3}+\alpha_{4}},\boldsymbol{b}_{334}\right)+N\left(x,-\frac{\alpha_{3}\boldsymbol{r}_{34}}{\alpha_{3}+\alpha_{4}},\boldsymbol{b}_{344}\right)\right]\nonumber \\
 & -\frac{N_{c}}{2}\,N\left(x,-\frac{\alpha_{1}\left(\alpha_{3}\boldsymbol{r}_{13}+\alpha_{4}\boldsymbol{r}_{14}\right)}{\left(\alpha_{3}+\alpha_{4}\right)\left(\alpha_{1}+\alpha_{3}+\alpha_{4}\right)},\boldsymbol{b}_{34,134}\right)\nonumber \\
 & -\left.\frac{N_{c}^{2}-1}{4N_{c}}N\left(x,\,\frac{\alpha_{1}\boldsymbol{r}_{31}+\alpha_{4}\boldsymbol{r}_{34}}{1-\alpha_{2}},\,\boldsymbol{b}_{1334}\right)\right\} ,\nonumber 
\end{align}
whereas for the type-$B$ contribution it is given by 
\begin{align}
\tilde{\sum}_{\ell n}\sigma_{\ell}\sigma_{n}\,c_{\ell n}\gamma\left(\boldsymbol{b}_{\ell}\right)\gamma\left(\boldsymbol{b}_{n}\right) & =\frac{1}{4}\left[N\left(x,\boldsymbol{r}_{23},\boldsymbol{b}_{23}\right)-N\left(x,\boldsymbol{r}_{24},\boldsymbol{b}_{24}\right)+N\left(x,\boldsymbol{r}_{3,234},\boldsymbol{b}_{2334}\right)-N\left(x,\boldsymbol{r}_{4,234},\boldsymbol{b}_{2344}\right)+\right.\label{eq:N_dip-2}\\
 & +\left.2N\left(x,\boldsymbol{r}_{14},\boldsymbol{b}_{24}\right)-2N\left(x,\boldsymbol{r}_{13},\boldsymbol{b}_{13}\right)\right]\nonumber 
\end{align}
The variables $\mathcal{Y}_{ij}$ in~(\ref{eq:Amp-1},~\ref{eq:Amp-2})
stand for the lab-frame rapidity of quark-antiquark pair made of partons
$i,\,j$. Explicitly it is given by 
\begin{equation}
\mathcal{Y}_{ij}=\ln\left(\frac{\left(\alpha_{i}+\alpha_{j}\right)q^{+}}{M_{\perp}}\right),\label{eq:Rapidity}
\end{equation}
where $\alpha_{i}$ and $\alpha_{j}$ are light-cone fractions of
the heavy quarks which form a given quarkonium.

\begin{figure}
\includegraphics[scale=0.65]{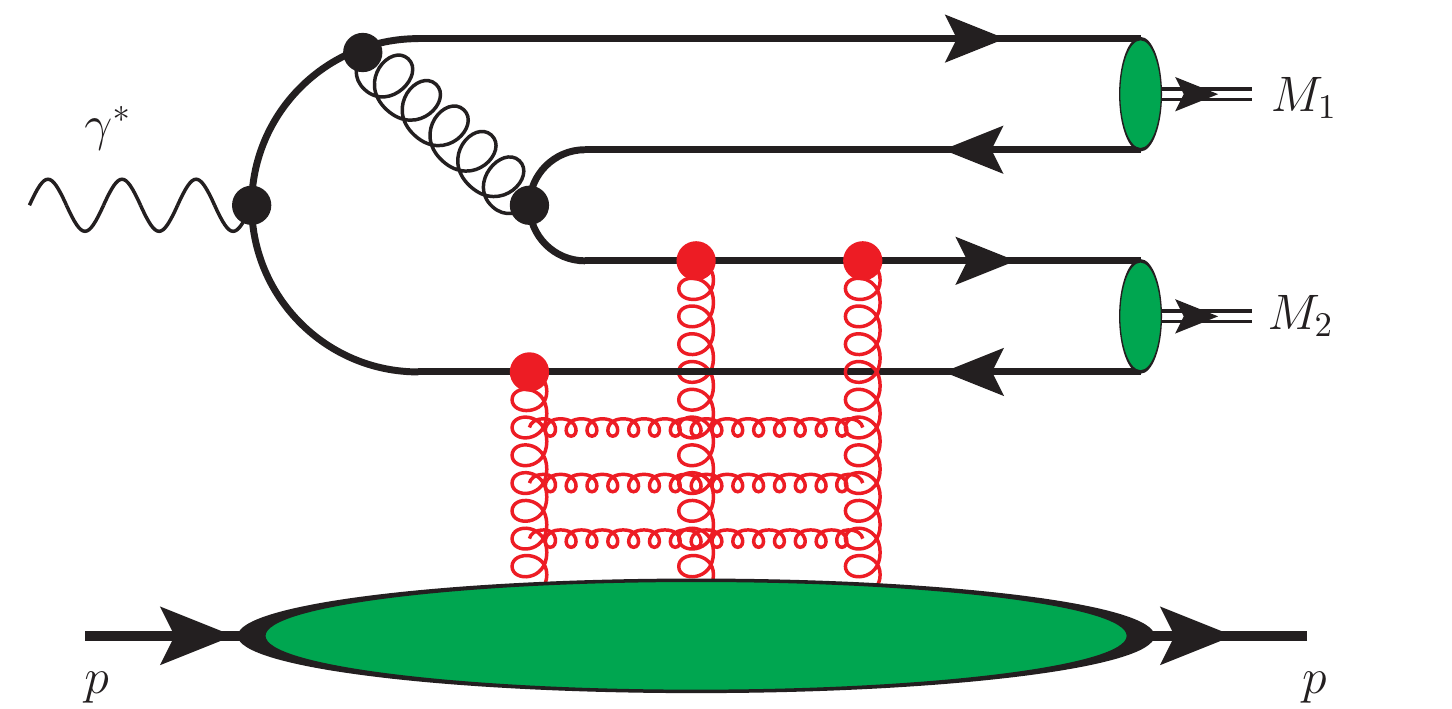}\includegraphics[scale=0.65]{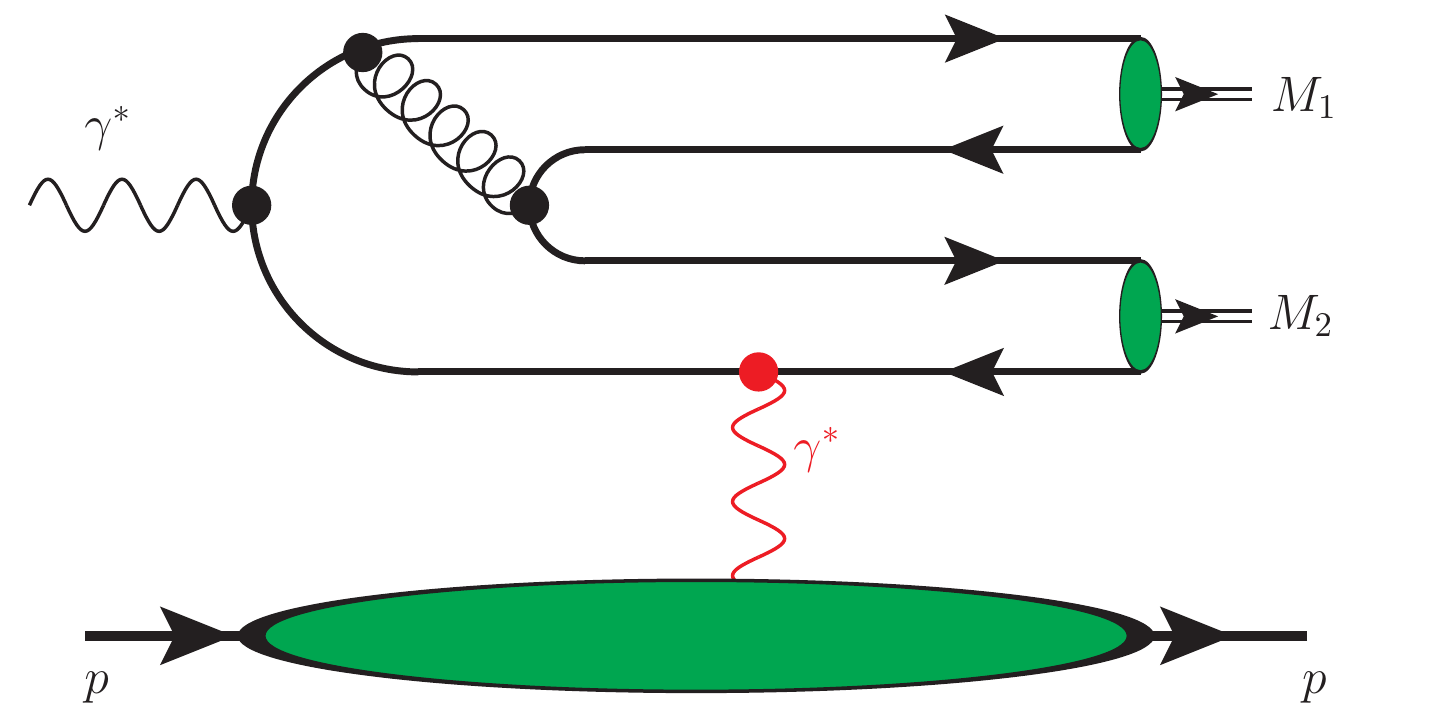}\caption{\label{fig:Photoproduction-HO}Examples of higher order contributions,
which become relevant for the exclusive production of quarkonia with
the same $C$-parity. The left diagram corresponds to exchange of
odderon (3-gluon ladder) in the $t$-channel, whereas the right diagram
corresponds to photon exchange in the $t$-channel. In both plots it is
implied summation over all possible attachments of $t$-channel gluons
and photon (red) to black-colored partonic lines. As explained in
the text, both types of contributions are suppressed compared to diagrams
from Figure~\ref{fig:Photoproduction} and will be disregarded in
what follows.}
\end{figure}

The dipole amplitude, which appears in~(\ref{eq:N_dip-1},\ref{eq:N_dip-2}),
effectively takes into account a sum of different pomeron ladders~\cite{Kovchegov:1999yj,Kovchegov:2012mbw},
and for this reason it corresponds to exchange of vacuum quantum numbers
in the $t$-channel. This fact imposes certain constraints on possible
quantum numbers of heavy quarkonia produced via the $\gamma+IP\to M_{1}M_{2}$
subprocess. Since the $C$-parity of a photon is negative, the neutral
quarkonia $M_{1},\,M_{2}$ must have opposite $C$-parities. This
explicitly excludes production of quarkonia with the same quantum
numbers ($M_{1}=M_{2}$). For the case when quarkonia are charged
(\emph{e.g}. $B_{c}^{+}B_{c}^{-}$), this implies that they necessarily
must be produced with odd value of the mutual angular momentum $L$ . Finally,
we need to mention that at higher orders the interaction with the
target should be supplemented by the exchange of $C$-odd three-gluon
ladders (so-called odderons) in the $t$-channel~\cite{Hatta:2005as}
potentially giving contributions of odderon exchange,
as shown in the right panel of Figure~\ref{fig:Photoproduction-HO}.
Such interactions are suppressed at high energies, because the odderon
has a smaller intercept than the pomeron. Besides, formally such contribution
is also suppressed by $\mathcal{O}\left(\alpha_{s}(m_{Q})\right)$. Another
possibility to produce $C$-even pair of quarkonia is via exchange
of a ($C$-odd) photon, as shown in the right panel
of Figure~\ref{fig:Photoproduction-HO}. Formally such contributions
are suppressed by $\sim\alpha_{{\rm em}}/\alpha_{s}^{2}\left(m_{Q}\right)$,
which is a small parameter for charm and bottom quarks, yet could
get enhanced in the infinitely heavy quark mass limit $m_{Q}\to\infty$
due to suppression of $\alpha_{s}\left(m_{Q}\right)$ in the denominator.
Besides, this contribution can be enhanced in the very forward kinematics
by the photon propagator $\sim1/t$, where $t\equiv\left(p_{f}-p_{i}\right)^{2}$
is very small~\footnote{Numerical estimates show that the invariant momentum transfer $t$
for photoproduction of a pair of quarkonia $M_{1},M_{2}$ is restricted
by 
\[
\left|t\right|\gtrsim\left|t_{{\rm min}}(W)\right|\approx\frac{m_{N}^{2}M_{12}^{2}}{W^{2}}+\mathcal{O}\left(\frac{m_{N}^{2}}{s},\,\frac{M_{12}^{2}}{s}\right),
\]
where $m_{N}$ is the mass of the nucleon, $W^{2}\equiv s_{\gamma p}=\left(q+P\right)^{2}$,
and $M_{12}^{2}=\left(p_{M_{1}}+p_{M_{2}}\right)^{2}$ is the invariant
mass of the quarkonia pair (clearly, $M_{12}\ge M_{1}+M_{2}$). Already
for EIC energies $W\sim$100 GeV, so we can see that it is possible
to achieve the kinematics of very small $t$ even for heavy quarkonia.}. According to phenomenological analyses~~\cite{Goncalves:2015sfy,Goncalves:2019txs,Goncalves:2006hu,Baranov:2012vu,Yang:2020xkl},
the cross-sections of this mechanism numerically is much smaller than
that of the mechanism suggested in this paper. For this reason in
what follows we will focus on the production of opposite-parity quarkonia,
and will disregard the contributions of $t$-channel odderons and
photons altogether.

\section{Numerical results}

\label{sec:Numer} The framework developed in the previous section
is valid for heavy quarkonia of both $c$ and $b$ flavors. In what
follows we will focus on the all-charm sector and present results
for $J/\psi+\eta_{c}$ production, for which the cross-section is
larger and thus is easier to study experimentally~\footnote{According to our estimates, for bottomonia the cross-sections are
at least an order of magnitude smaller due to the heavier quark mass}.

For the wave function of the $J/\psi$-mesons we will use a simple
ansatz suggested in~\cite{Dosch:1996ss,BoostedGaussian}, 
\begin{align}
 & \Psi_{J/\psi}\left(z,\boldsymbol{r},M=0\right)=\frac{\delta_{h,-\bar{h}}}{\sqrt{2}}z(1-z)\varphi\left(z,\,\boldsymbol{r}\right),\qquad\varphi(z,\,\boldsymbol{r})=\frac{\sqrt{2}\pi f_{V}}{\sqrt{N_{c}}\hat{e}_{V}}f(z)e^{-\omega^{2}r^{2}/2},\label{eq:WF_JPsi}\\
 & \Psi_{J/\psi}\left(z,\boldsymbol{r},M=\pm1\right)=\frac{1}{M_{V}}\left[iMe^{iM\theta}\left(\bar{z}\delta_{h,-M}\delta_{\bar{h},M}-z\delta_{h,M}\delta_{\bar{h},-M}\right)\partial_{r}+m_{Q}\delta_{h,M}\delta_{\bar{h},M}\right]\varphi\left(z,\,\boldsymbol{r}\right)\\
 & f(z)=\sqrt{z(1-z)}e^{-M_{V}^{2}\left(z-1/2\right)^{2}/2\omega^{2}}
\end{align}
where $M$ is the helicity of $J/\psi$, $\boldsymbol{r}$ is the
distance between the quark and antiquark, $h,\,\bar{h}$ are the helicities
of the quark and antiquark, and $f_{V},e_{V},\,\omega$ are some numerical
constants. This result can be trivially extended to the case of
$\eta_{c}$-meson, which differs from the $J/\psi$ meson only by the
orientation of the quark spins. Taking into account the structure
of the Clebsch-Gordan coefficients for the $1/2\times1/2$ product, we
may immediately write out the corresponding wave functions for $\eta_{c}$,
modifying the corresponding $M=0$ component of the $J/\psi$ wave
function, 
\begin{align}
\Psi{}_{\eta_{c}}\left(z,\boldsymbol{r}\right) & =\frac{\varepsilon_{h,\bar{h}}}{\sqrt{2}}z(1-z)\varphi\left(z,\,\boldsymbol{r}\right),\qquad\varepsilon_{ab}=-\varepsilon_{ba}=\delta_{a,-b}{\rm sign}(a).\label{eq:WF_Etac}
\end{align}
Alternatively, the wave functions of quarkonia can be constructed
using potential models or the well-known Brodsky-Huang-Lepage-Terentyev
(BHLT) prescription~\cite{BHT,Brodsky:2003pw,Terentev:1976jk} which
allows to convert the rest frame wave function $\psi_{{\rm RF}}$
into a light-cone wave function $\Psi_{{\rm LC}}$. It is known that
in the small-$r$ region, which is relevant for estimates, the wave
functions of the $S$-wave heavy quarkonia in different schemes are
quite close to each other~\cite{Stadler:2018hjv,Daniel:1990ah,Kawanai:2011xb,Kawanai:2013aca},
and for this reason in what follows we will use the ansatz of~(\ref{eq:WF_JPsi}-\ref{eq:WF_Etac}),
in view of its simplicity.

For our numerical evaluations we also need a parametrization of the
dipole amplitude. In what follows we will we use the impact parameter
($b$) dependent ``bCGC'' parametrization of the dipole cross-section~\cite{Kowalski:2006hc,RESH},
\begin{align}
N\left(x,\,\boldsymbol{r},\,\boldsymbol{b}\right) & =\left\{ \begin{array}{cc}
N_{0}\,\left(\frac{r\,Q_{s}(x)}{2}\right)^{2\gamma_{{\rm eff}}(r)}, & r\,\le\frac{2}{Q_{s}(x)}\\
1-\exp\left(-\mathcal{A}\,\ln\left(\mathcal{B}r\,Q_{s}\right)\right), & r\,>\frac{2}{Q_{s}(x)}
\end{array}\right.~,\label{eq:CGCDipoleParametrization}\\
 & \mathcal{A}=-\frac{N_{0}^{2}\gamma_{s}^{2}}{\left(1-N_{0}\right)^{2}\ln\left(1-N_{0}\right)},\quad\mathcal{B}=\frac{1}{2}\left(1-N_{0}\right)^{-\frac{1-N_{0}}{N_{0}\gamma_{s}}},\\
 & Q_{s}(x,\,\boldsymbol{b})=\left(\frac{x_{0}}{x}\right)^{\lambda/2}T_{G}(b),\,\,\gamma_{{\rm eff}}(r)=\gamma_{s}+\frac{1}{\kappa\lambda Y}\ln\left(\frac{2}{r\,Q_{s}(x)}\right),\label{eq:gamma_eff}\\
\gamma_{s} & =0.66,\quad\lambda=0.206,\quad x_{0}=1.05\times10^{-3},\quad T_{G}(b)=\exp\left(-\frac{b^{2}}{2\gamma_{s}B_{{\rm CGC}}}\right).
\end{align}

\begin{figure}
\includegraphics[height=6cm]{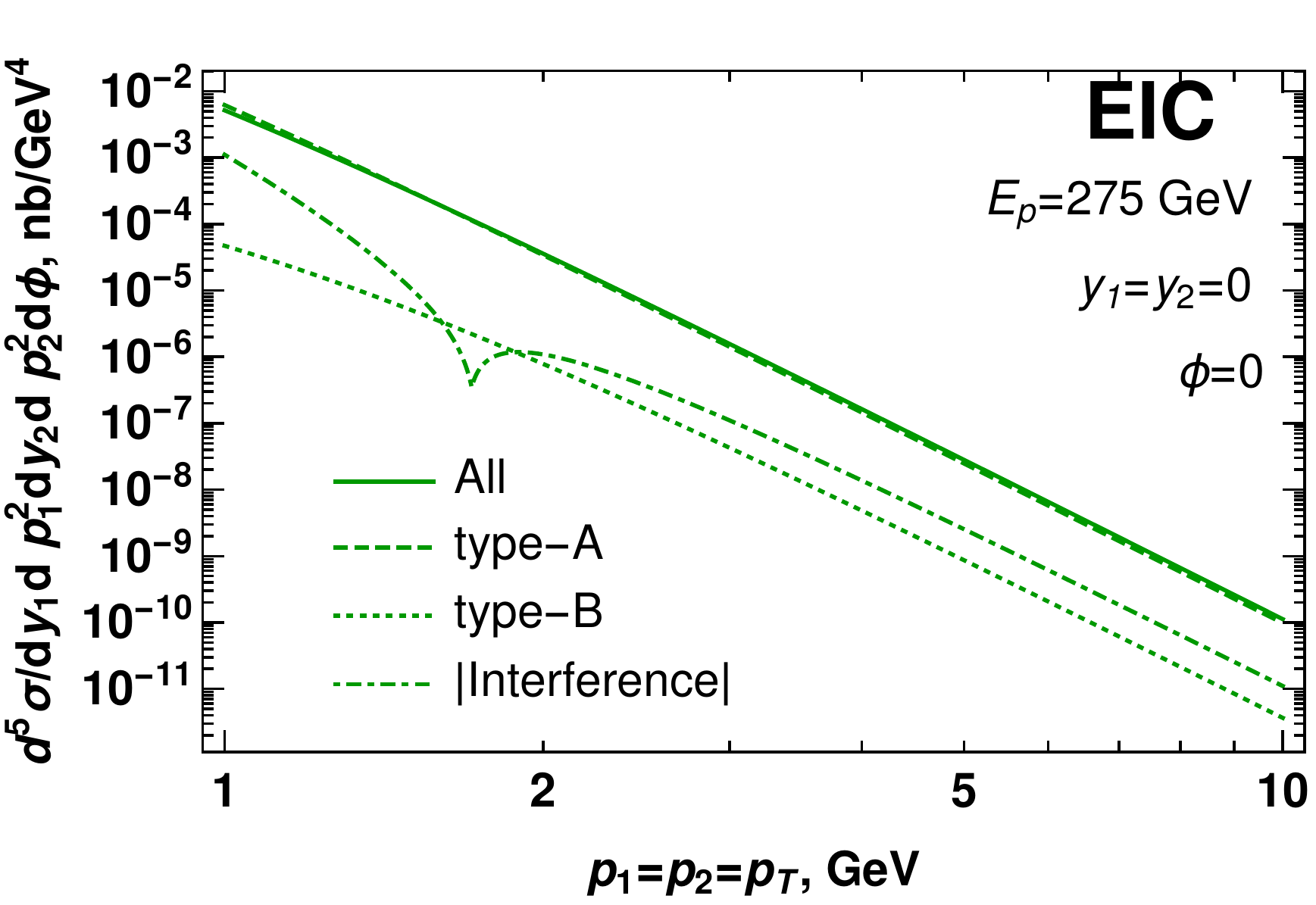}\includegraphics[height=6cm]{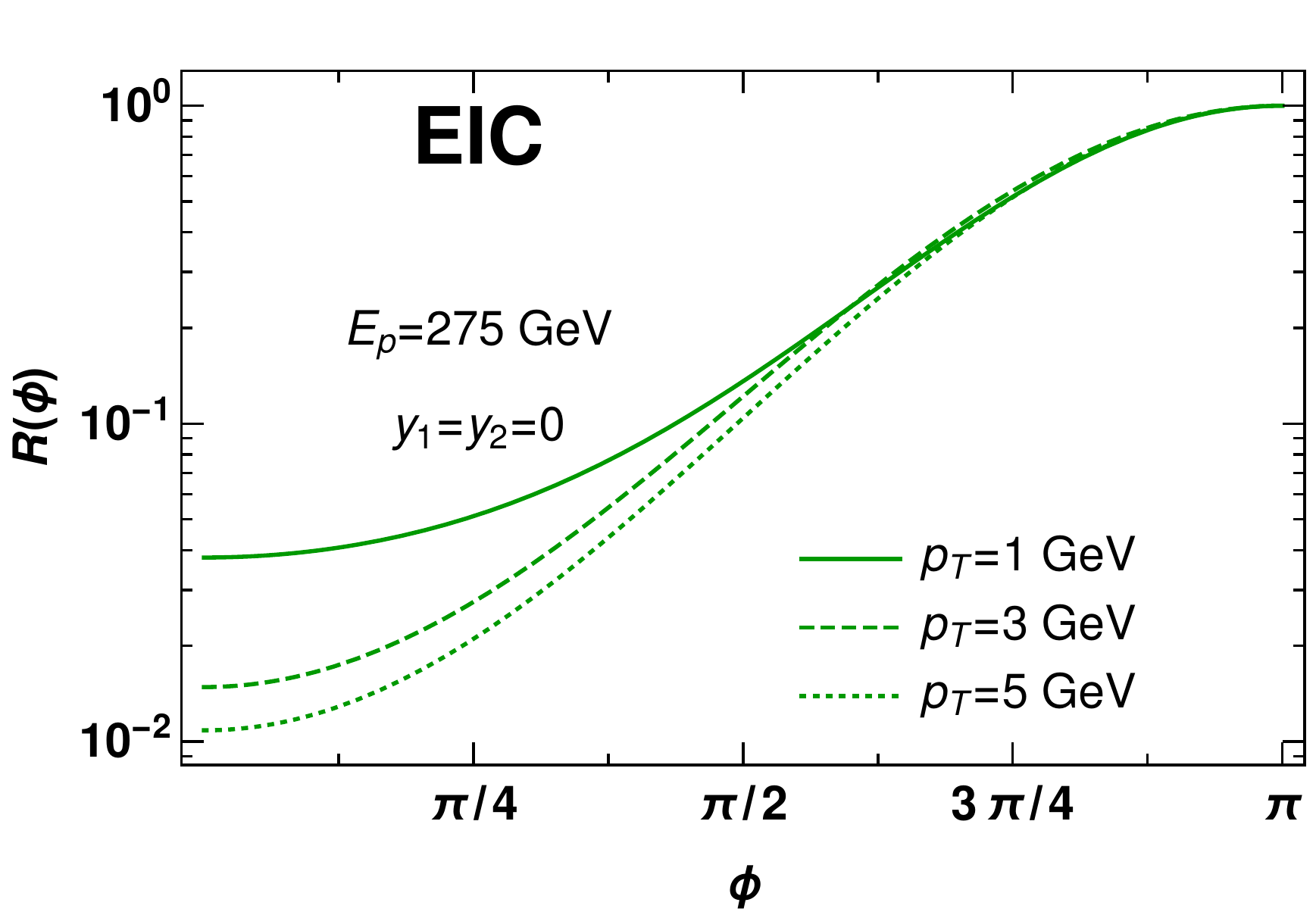}\caption{\label{fig:RelRolePhi}Left plot: Different contributions to charmonia
pair photoproduction in EIC kinematics: the type-$A$ and type-$B$ diagrams,
as well as their interference. Right plot: The dependence of the normalized
ratio $R(\phi)$, defined in~(\ref{eq:RatioR}), on the angle $\phi$
(difference between azimuthal angles of both quarkonia). The appearance
of a sharp peak in back-to-back kinematics is explained in the text.
For definiteness we considered the case when both quarkonia
are produced at central rapidities ($y_{1}=y_{2}=0$) in the lab frame;
for other rapidities the $\phi$-dependence has a similar shape.}
\end{figure}

We would like to start the presentation of numerical results from a discussion
of the relative contribution of type-$A$ and type-$B$ diagrams introduced
in the previous section. From the left panel of Figure~\ref{fig:RelRolePhi}
we can see that the dominant contribution comes from the type-$A$
diagrams. Partially this enhancement can be explained by larger
color factors in the large-$N_{c}$ limit. The interference of type-$A$
and type-$B$ contributions represents approximately a 10\% correction
and moreover, has a node, whose position depends on the produced quarkonia kinematics.
As expected, the cross-section is suppressed as a function of $p_{T}$
(we considered $\left|\boldsymbol{p}_{J/\psi}^{\perp}\right|=\left|\boldsymbol{p}_{\eta}^{\perp}\right|=p_{T}$
for the sake of definiteness). In the right panel of the same Figure~\ref{fig:Photoproduction}
we present the dependence of the yields on the azimuthal angle $\phi$
between the transverse momenta of the $J/\psi$ and $\eta_{c}$ mesons.
For definiteness, we assumed that the transverse momenta $\boldsymbol{p}_{J/\psi}^{\perp},\,\boldsymbol{p}_{\eta}^{\perp}$
of both quarkonia have equal absolute values. In order to make meaningful
comparison of the cross-sections, which differ by orders of magnitude,
we plotted the normalized ratio 
\begin{align}
R(\phi) & =\frac{d\sigma\left(...,\,\phi\right)/dy_{1}dp_{1}^{2}dy_{2}dp_{2}^{2}d\phi}{d\sigma\left(...,\,\phi=\pi\right)/ddy_{1}dp_{1}^{2}dy_{2}dp_{2}^{2}d\phi},\qquad R(\phi=\pi)\equiv1\label{eq:RatioR}
\end{align}
We can see that the ratio has a sharp peak in the back-to-back region
($\phi=\pi$), which happens because in this kinematics the momentum
transfer to the target $|t|=\left|\Delta^{2}\right|$ is minimal.
In contrast, for the angle $\phi\approx0$, which maximizes the variable
$|t|=\left|\Delta^{2}\right|$, the ratio has a pronounced dip. For
$p_{1}\not=p_{2}$ the dependence on $\phi$ is qualitatively similar,
although the maximum and minimum are less pronounced.

\begin{figure}
\includegraphics[width=9cm]{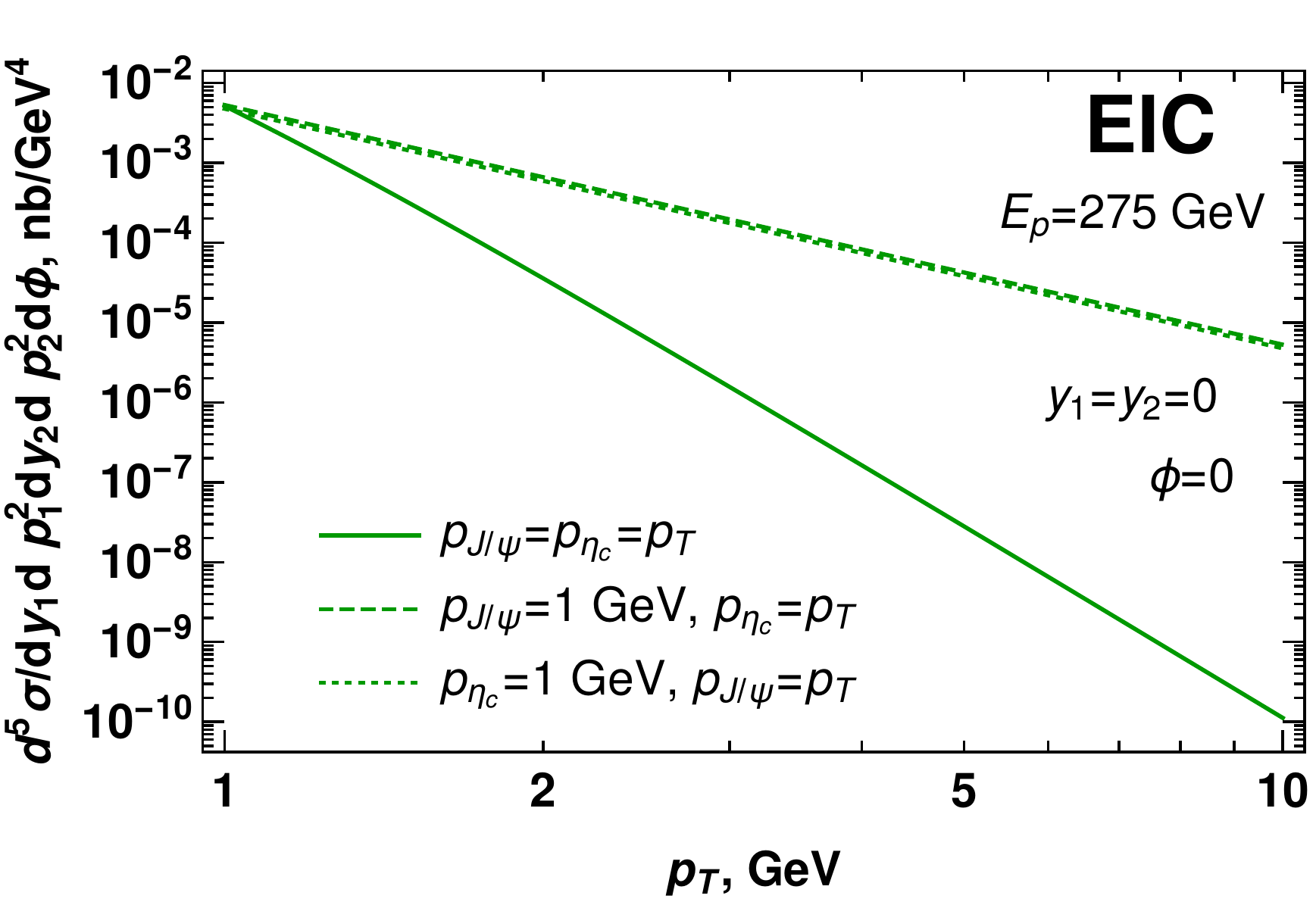}\includegraphics[width=9cm]{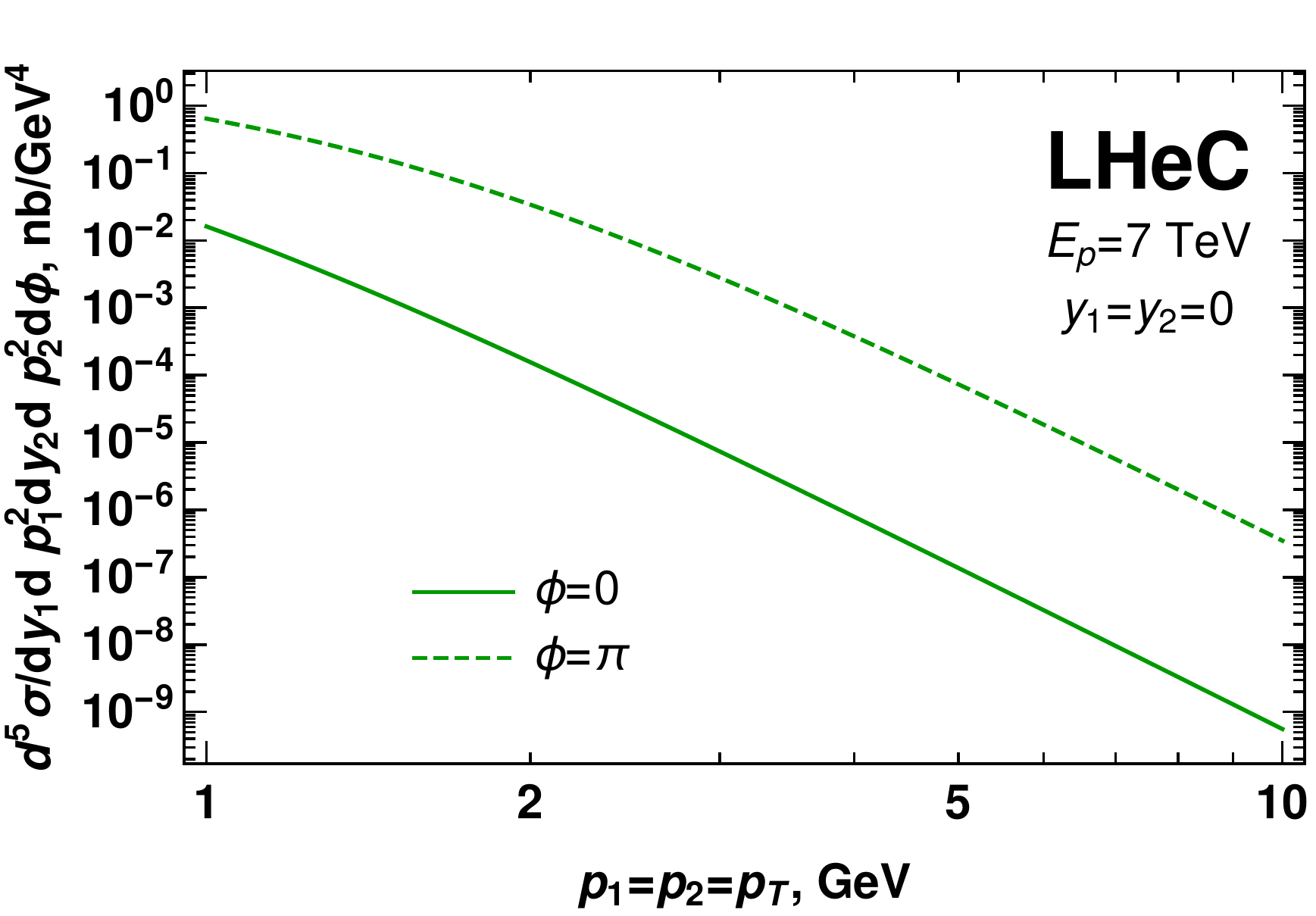}\caption{\label{fig:SoftHard}Left plot: The $p_{T}$-dependence of the charmonia
pair photoproduction cross-section. Comparison of the case when both quarkonia
have large transverse momentum (solid line), with the cases when one of the produced
quarkonia is small (dashed and dot-dashed lines). Within errors of numerical evaluation, there is
no difference if the soft transverse momentum $p_{T}\approx1\,{\rm GeV}$
is assigned to $J/\psi$ or $\eta_{c}$ mesons. Right plot: The $p_{T}$-dependence
of the cross-section in LHeC kinematics. For definiteness
we considered the case when both quarkonia are produced at central
rapidities ($y_{1}=y_{2}=0$) in the lab frame.}
\end{figure}

In the left panel of the Figure~\ref{fig:SoftHard} we analyze the
$p_{T}$-dependence, for the case when one of the quarkonia has a small
transverse momentum $p_{i}\sim1\,{\rm GeV}$. As expected, in this
case the cross-section has a significantly milder suppression compared
to the case when both quarkonia share the same transverse momentum.
This result indicates that the quarkonia pair are predominantly produced
with small transverse momenta $p_{1}^{\perp}\sim p_{2}^{\perp}\lesssim1\,{\rm GeV}$
and opposite directions in the transverse plane ( $\phi\equiv\phi_{1}-\phi_{2}\approx\pi$).
In the right panel of the same Figure~\ref{fig:SoftHard}, we 
show the $p_{T}$-dependence of the cross-section in LHeC kinematics.
While the absolute value increases in this case, we may observe that
qualitatively the dependence on $p_{T}$ and angle $\phi$ remains
the same.

\begin{figure}
\includegraphics[width=9cm]{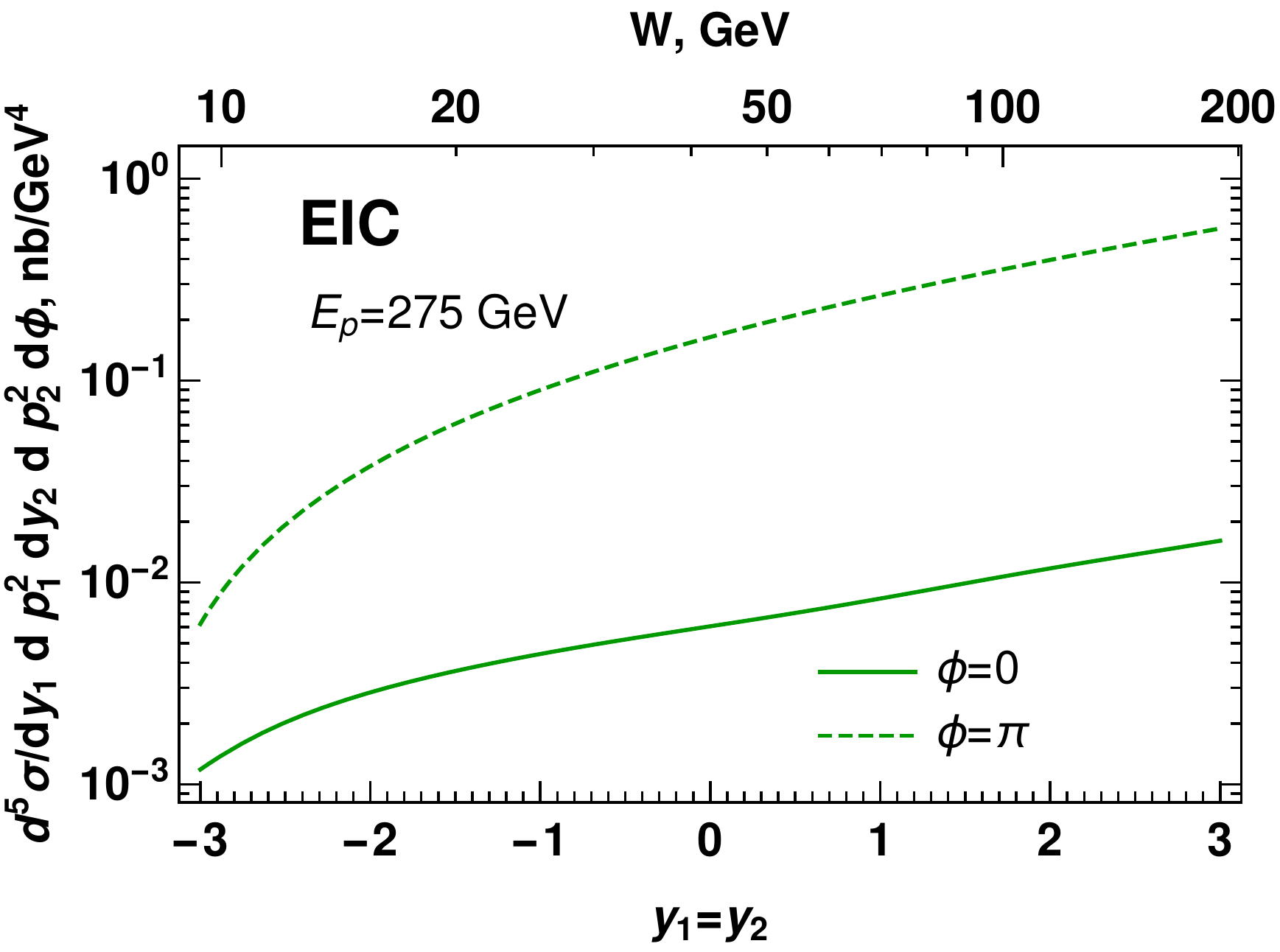}\includegraphics[width=9cm]{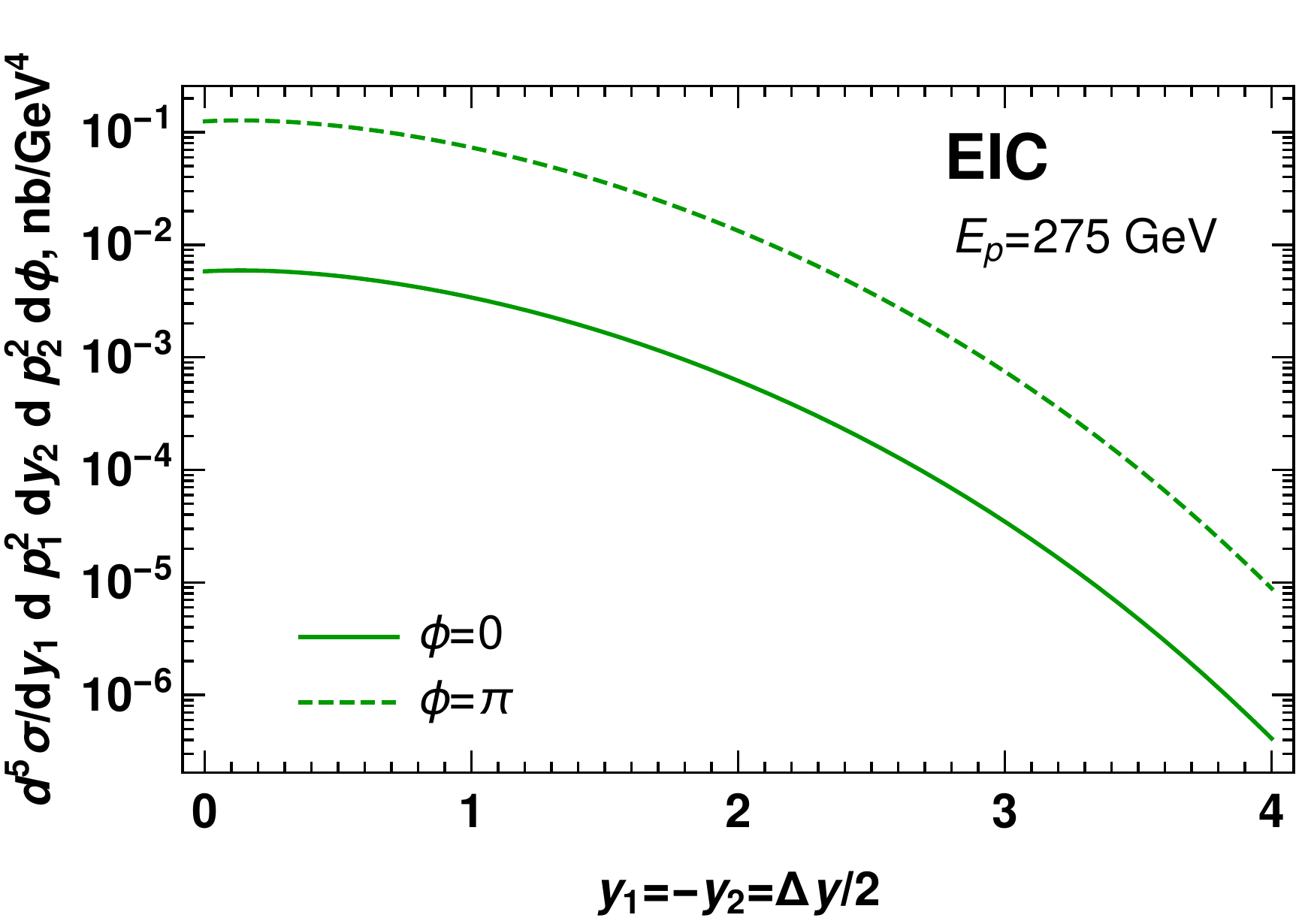}

\caption{\label{fig:Rapidities}Left plot: The rapidity dependence of the photoproduction cross-section
in EIC kinematics, assuming equal rapidities of the produced quarkonia,
$y_{1}=y_{2}$. The upper horizontal scale illustrates the corresponding
value of invariant energy $W\equiv\sqrt{s_{\gamma p}}$ defined in~(\ref{eq:W2}). Right plot:
The dependence on the rapidity difference between the produced quarkonia,
$y_{1}=-y_{2}=\Delta y/2$. For the sake of definiteness we assumed
that both quarkonia are produced at central rapidities ($y_{1}=y_{2}=0$) with transverse momenta $p_1=p_2=1$ GeV in the lab frame.}
\end{figure}

In Figure~\ref{fig:Rapidities} we analyze the dependence of
the cross-section on rapidities of the quarkonia. In the left panel
we consider the special case when both quarkonia are produced with
the same transverse momenta $p_{1}^{\perp}\sim p_{2}^{\perp}\sim1\,{\rm GeV}$
and the same rapidities $y_{1}=y_{2}$ in the lab frame. The variables
$y_{1,2}$ in this case can be unambiguously related to the invariant
photon-proton energy $W_{\gamma p}\sim\sqrt{s_{\gamma p}}$ (shown
in the upper horizontal axis), and as expected, the cross-section
grows as a function of energy. In the right panel of the same Figure~~\ref{fig:Rapidities}
we analyze the dependence of the cross-section on the rapidity difference
$\Delta y$ between two heavy mesons. For the sake of definiteness
we consider that both quarkonia have opposite rapidities in the lab
frame, $y_{1}=-y_{2}=\Delta y/2$. We observe that in this case
the cross-section becomes suppressed as a function of $\Delta y$,
which illustrates the fact that the quarkonia are predominantly produced
with the same rapidities. 

\begin{figure}
\includegraphics[width=9cm]{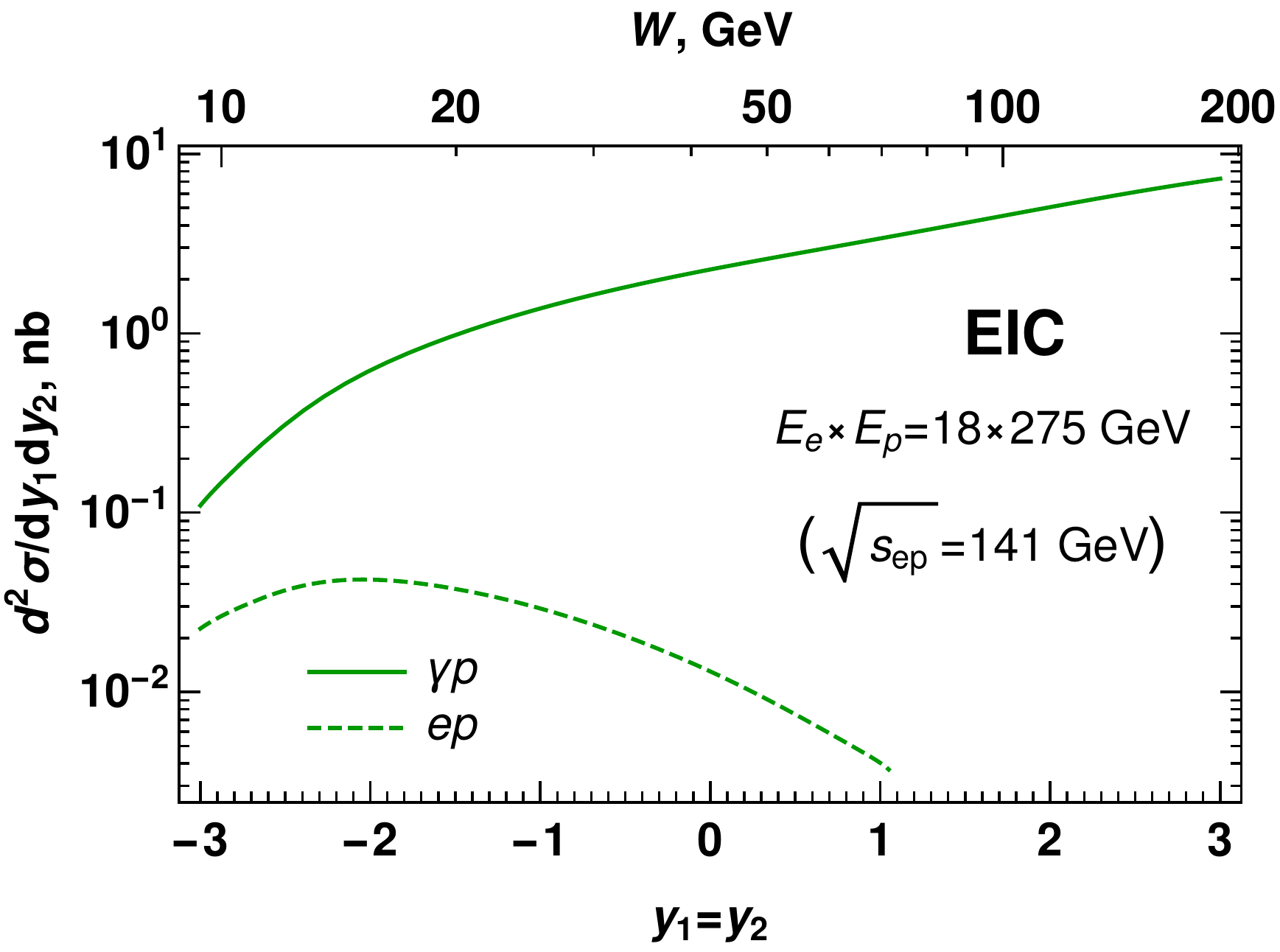}\includegraphics[width=9cm]{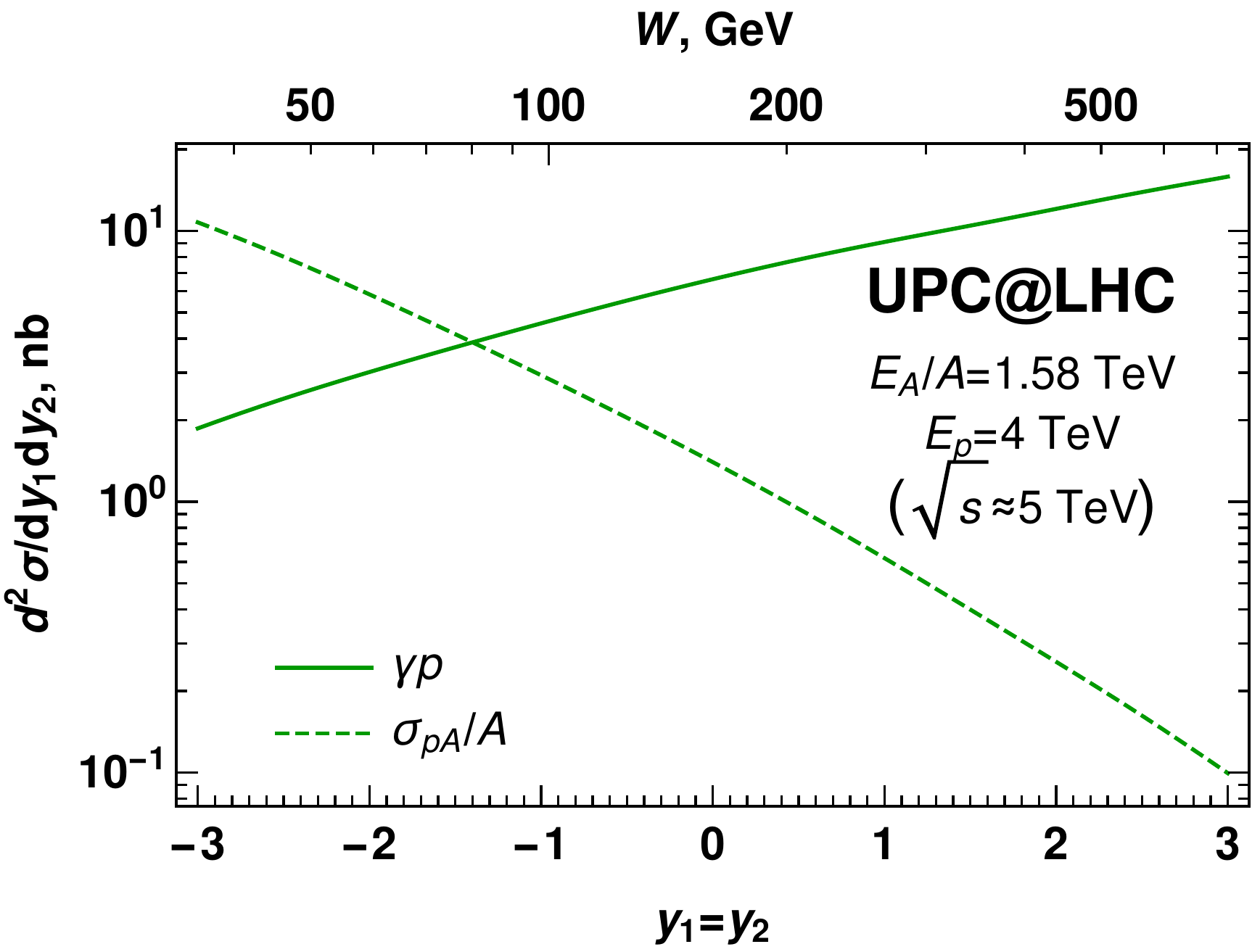}

\includegraphics[width=9cm]{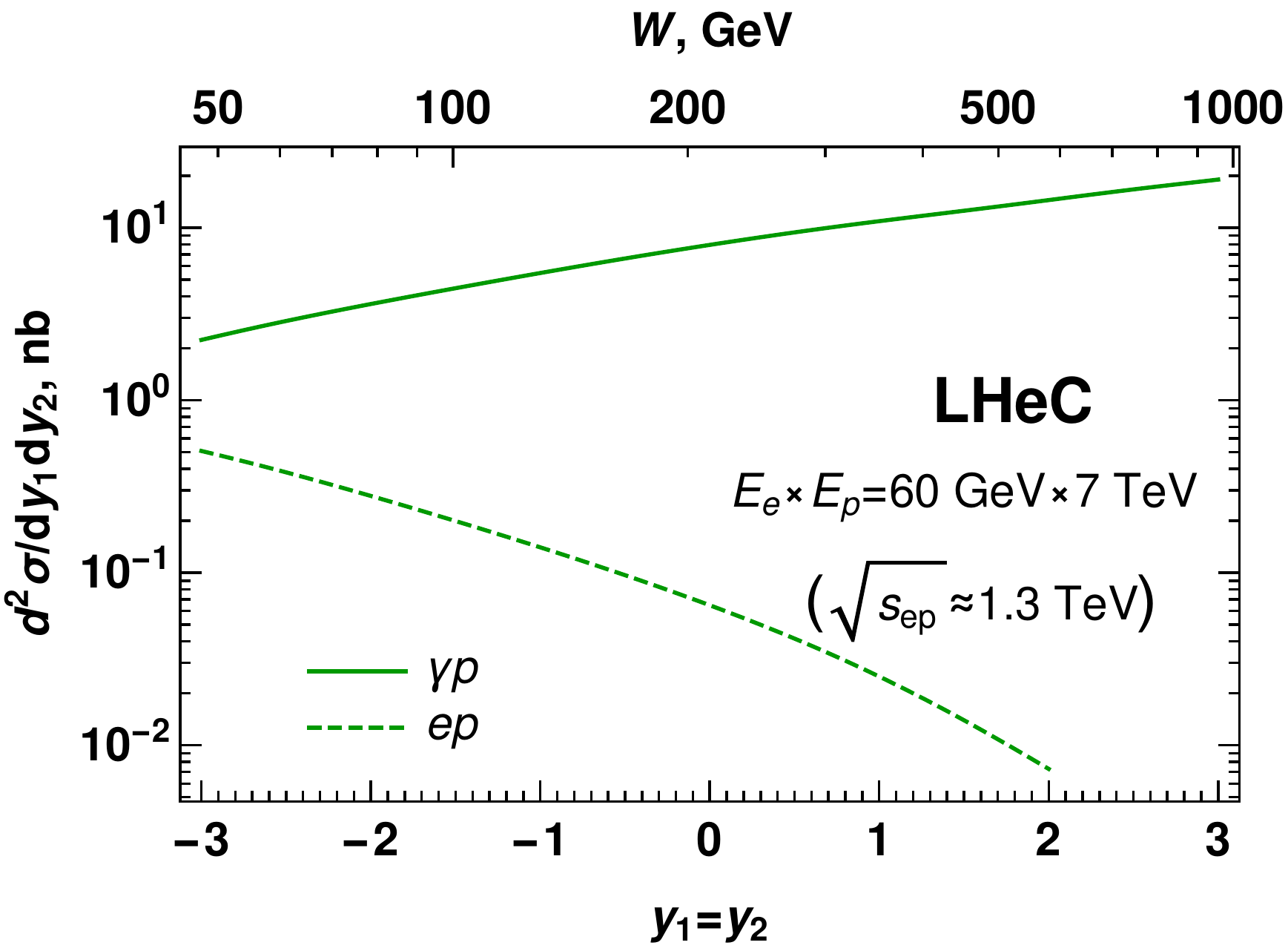}\includegraphics[width=9cm]{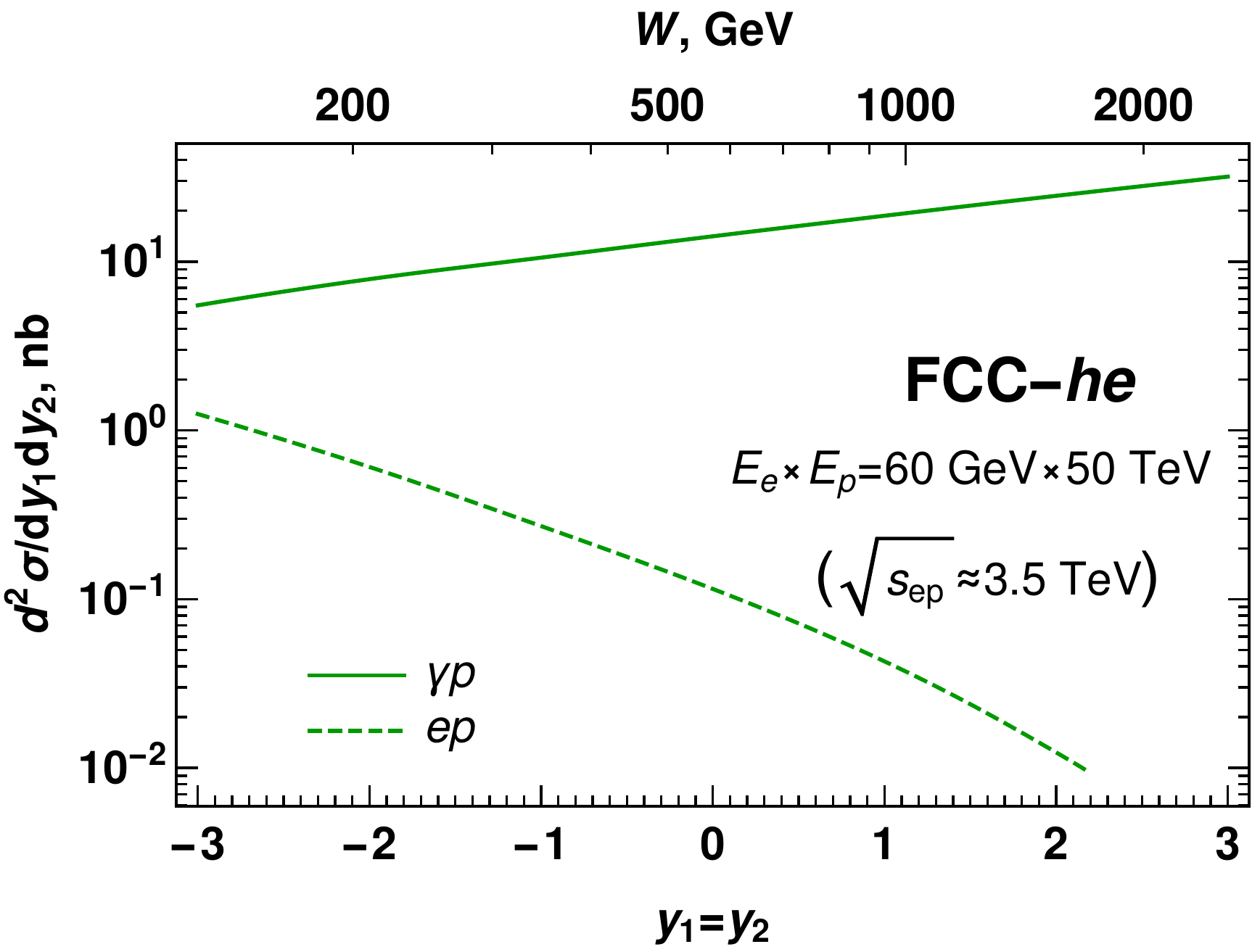}

\caption{\label{fig:Rapidities-Integrated}The rapidity dependence of the $p_{T}$-integrated
cross-section in the kinematics of ultraperipheral collisions at LHC and
in the kinematics of the future $ep$ colliders. A positive sign of rapidity
is chosen in the direction of electron or emitted quasi-real photon. For
UPC collisions the positive direction of rapidity is that of a heavy
lead ion, and the cross-sections are given per nucleon. The solid
curves correspond to cross-section of the $\gamma p\to M_{1}M_{2}p$ subprocess,
whereas dotted lines correspond to the cross-sections of the complete
physically observable $ep$ or $Ap$ processes. We assume for definiteness
that the rapidities of both quarkonia are equal to each other in the
lab frame, $y_{1}=y_{2}=y$. The upper horizontal scale illustrates
the corresponding value of invariant energy $W\equiv\sqrt{s_{\gamma p}}$,
as defined in~(\ref{eq:W2}). }
\end{figure}

\begin{figure}
\includegraphics[width=9cm]{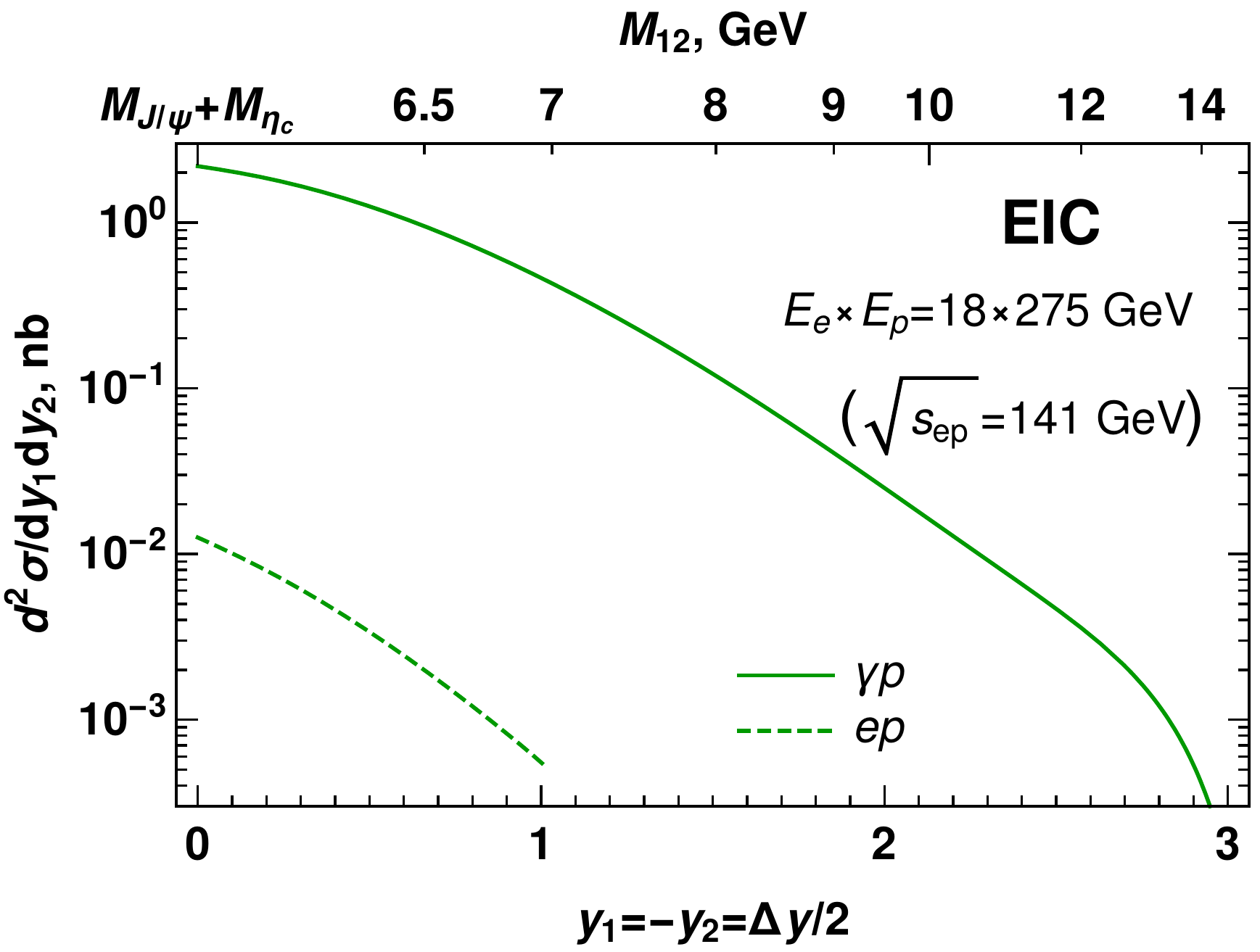}\includegraphics[width=9cm]{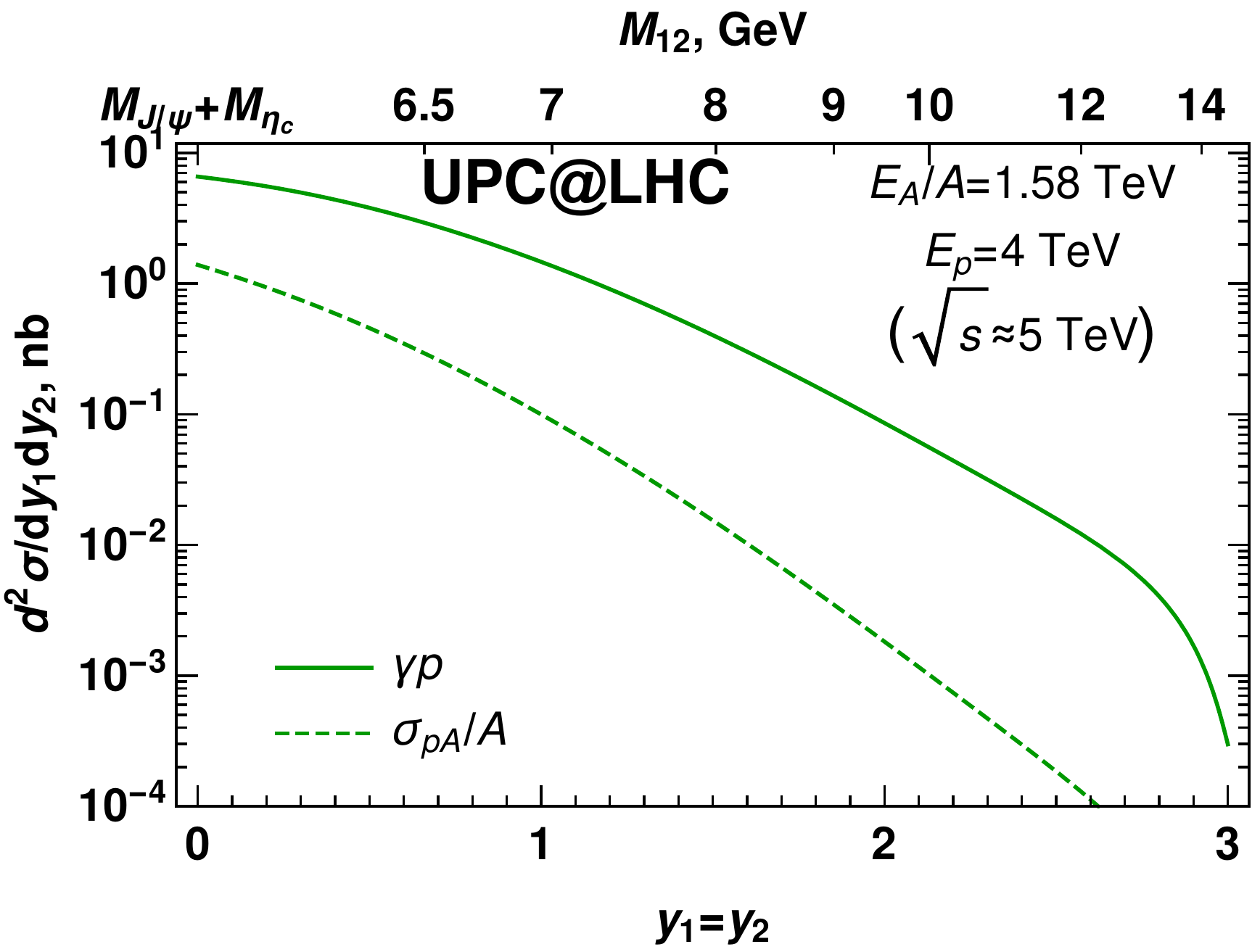}

\includegraphics[width=9cm]{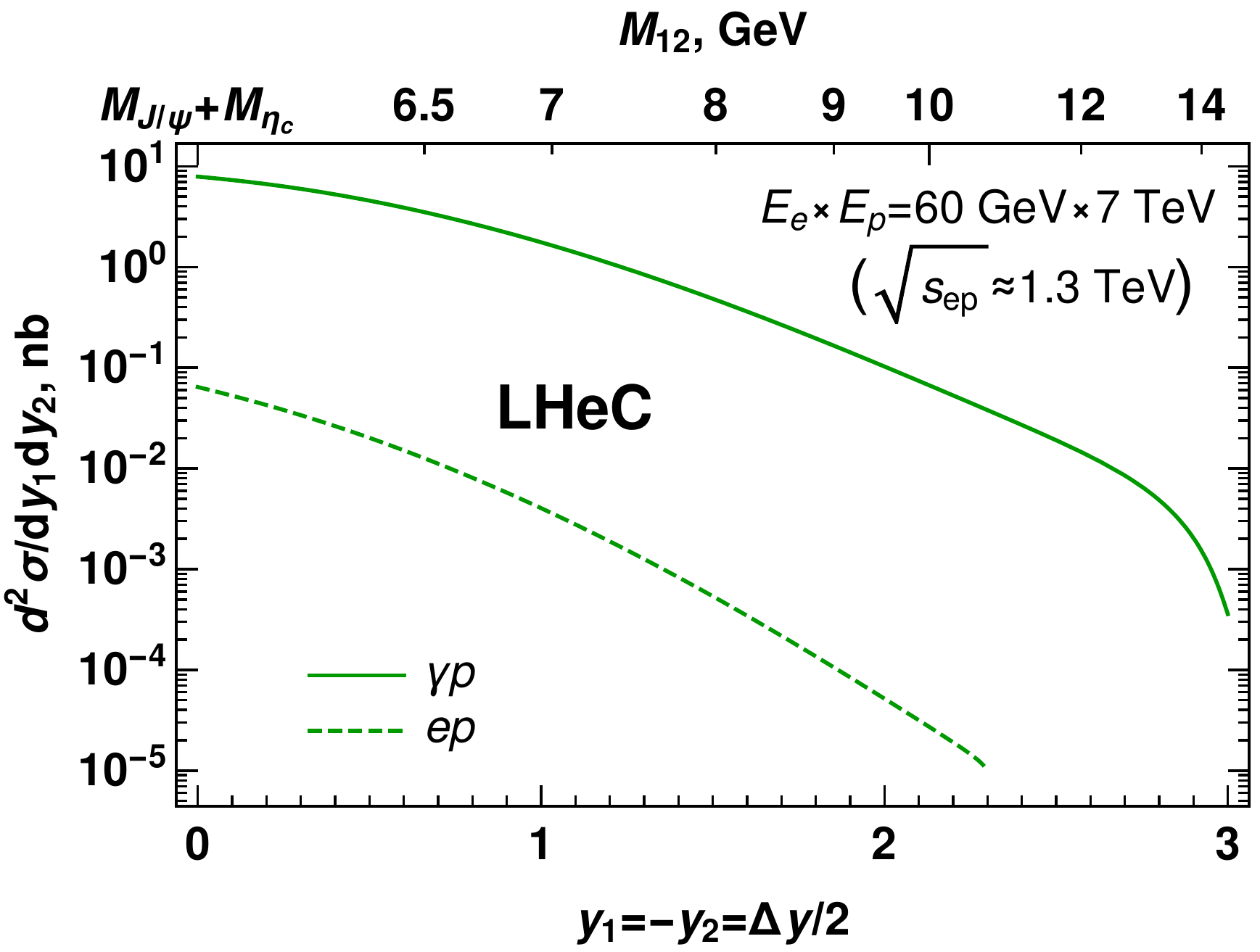}\includegraphics[width=9cm]{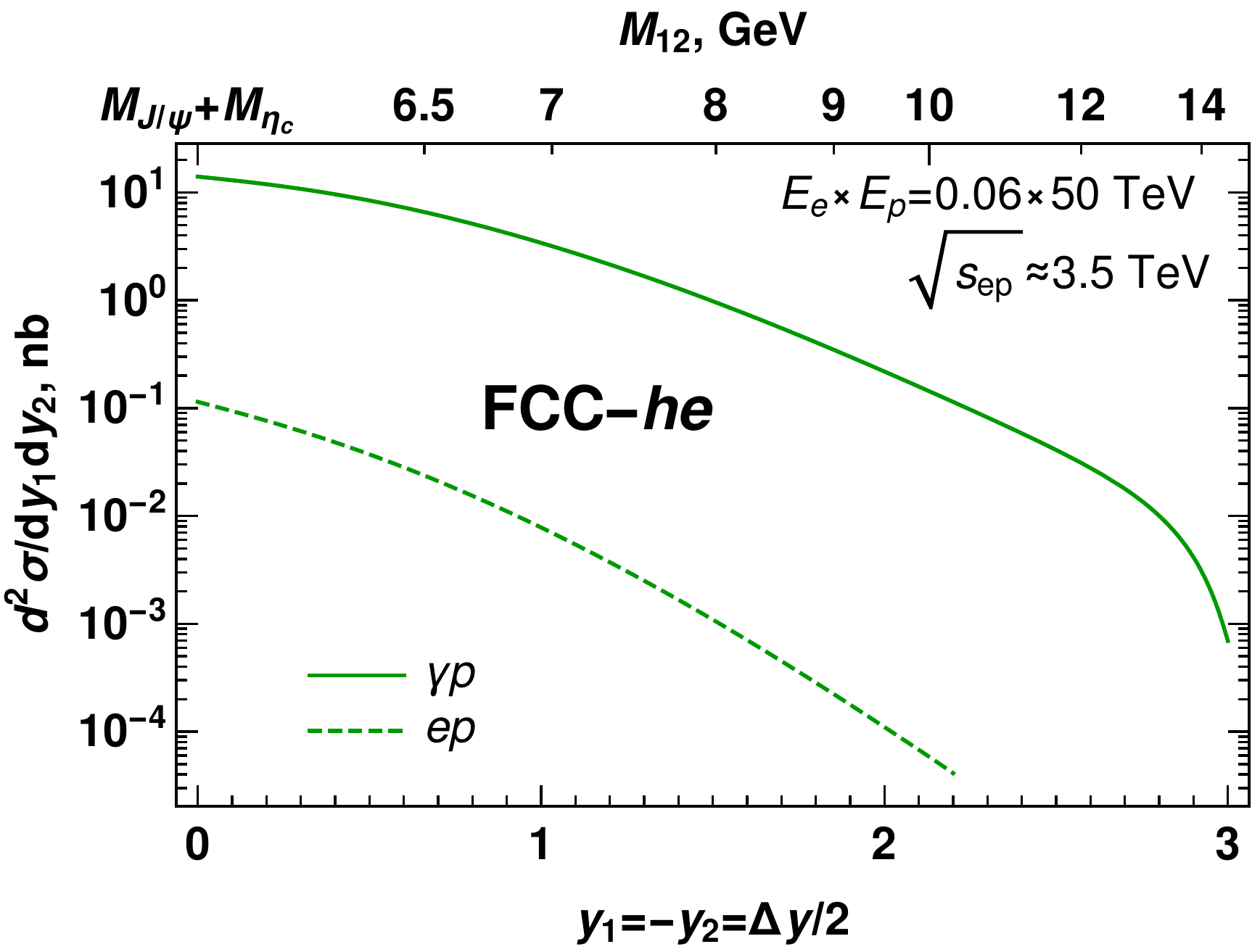}

\caption{\label{fig:dRapidities-Integrated} The dependence on rapidity difference
for the $p_{T}$-integrated cross-section, in the kinematics of ultraperipheral
collisions, at LHC and future electron-proton colliders. The positive
sign of rapidity is chosen in the direction of electron or emitted quasi-real
photon. For UPC collisions the positive direction of rapidity is that
of a heavy lead ion, and the cross-sections are given per nucleon.
For the sake of definiteness we assume that in the lab frame the quarkonia
have opposite rapidities, $y_{1}=-y_{2}=\Delta y/2$. The upper horizontal
scale illustrates the corresponding value of the invariant mass  $M_{12}\equiv\sqrt{\left(p_{J/\psi}+p_{\eta_c}\right)^2}$,
as defined in~(\ref{eq:M12}). Dotted curves correspond to the cross-sections
of the complete process (electron-proton or heavy ion-proton).}
\end{figure}

\begin{figure}
\includegraphics[width=18cm]{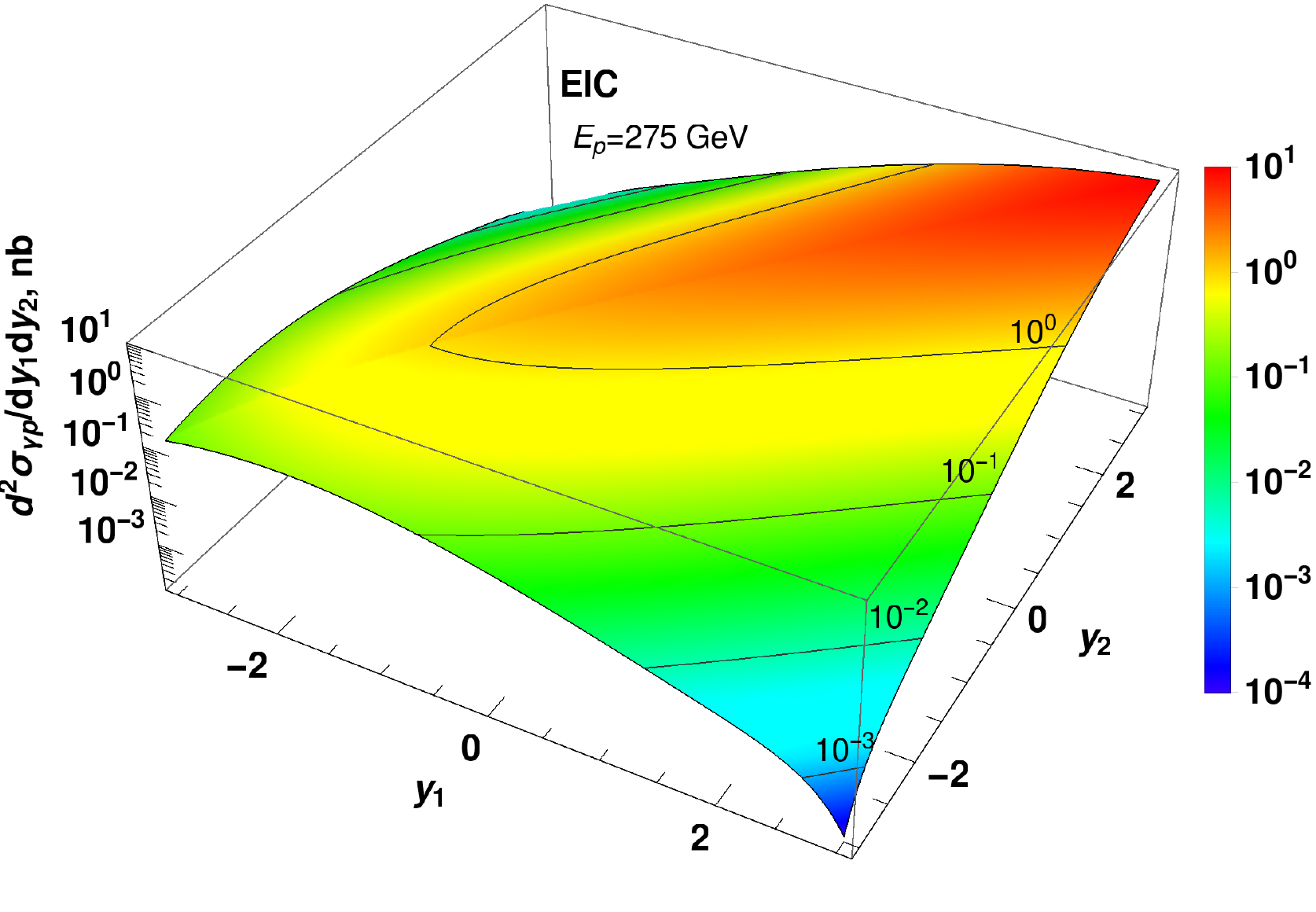}

\caption{\label{fig:RapiditiesAll-Integrated} (Color online) The dependence on rapidities
$y_{1},\,y_{2}$, of produced quarkonia, for the $p_{T}$-integrated
photoproduction cross-section $d\sigma_{\gamma p}/dy_{1}dy_{2}$.
The plot illustrates the fact that partons are produced with approximately
equal rapidities, $y_{1}\approx y_{2}$. For definiteness
we consider the proton with a typical energy of EIC kinematics ($E_{p}\sim275$~GeV
in the lab frame). For other proton energies the dependence has qualitatively
similar shape.}
\end{figure}

Finally, in Figures~\ref{fig:Rapidities-Integrated},~\ref{fig:dRapidities-Integrated},~\ref{fig:RapiditiesAll-Integrated}
we show the results for the cross-section $d\sigma_{\gamma p\to M_{1}M_{2}p}/dy_{1}dy_{2}$,
which is integrated over transverse momenta $\boldsymbol{p}_{i}^{\perp}$
of both quarkonia. This observable can be the most promising for
experimental studies, since it is easier to measure. We make the predictions
in the kinematics of the ultraperipheral $pA$ collisions at LHC,
as well as future electron-hadron colliders. Largely, the dependence
on $y_{1},\,y_{2}$ repeats similar dependence of the $p_{T}$-unintegrated
cross-sections. This happens because the $p_{T}$-integrated cross-sections
get its dominant contributions from the region of small $p_{T}\ll m_{Q}$,
where dependence on rapidity is mild. In Figures~\ref{fig:Rapidities-Integrated},~\ref{fig:dRapidities-Integrated}
we have also shown the cross-sections of ``master'' processes $ep\to eM_{1}M_{2}p$
and $Ap\to AM_{1}M_{2}p$. The expressions for these cross-sections differ from those
of $\gamma p\to M_{1}M_{2}p$ by a convolution with known kinematic
factors, which correspond to fluxes of equivalent photons generated
by the electron or heavy nucleus. These cross-sections have
completely different behavior on the rapidity $y_{1}=y_{2}$ of both quarkonia,
which can be understood from~(\ref{qPlus}-\ref{eq:LTSep}). Indeed,
mesons with higher lab-frame rapidities can be produced by photons
of higher energy $E_{\gamma}$, yet the flux of equivalent photons
created by a charged electron or ion is suppressed and vanishes when
the elasticity $y=E_{\gamma}/E_{e}$ approaches unity. Finally, 
Figure~\ref{fig:RapiditiesAll-Integrated} illustrates how the cross-section
behaves as a function of $y_{1},y_{2}$ in general, when $|y_{1}|\not=|y_{2}|$.
We can see that the cross-section has a typical ridge near $y_{1}\approx y_{2}$,
i.e. when quarkonia are produced with approximately the same rapidities.

\section{Conclusions}

\label{sec:Conclusions}In this paper we studied in detail the\emph{
}exclusive photoproduction of heavy charmonia pairs. This process
presents a lot of interest, both on its own, as a potential test of
quarkonia production mechanisms in small-$x$ kinematics, as well
as a background to exotic hadron production. We analyzed in detail
the leading order contributions and found that in this mechanism the
quarkonia pairs are produced with opposite $C$-parities, relatively
small opposite transverse momenta $p_{T}$, and small separation in
rapidity. This finding is explained by the fact that in the chosen kinematical region
the momentum transfer to the recoil proton is minimal. As expected, the
cross-section decreases rapidly as a function of $p_{T}$, and grows
as a function of photon-proton invariant energy ($\sim$quarkonia
rapidities), similar to single-photon production. However, the cross-section
decreases as a function of the rapidity difference between the quarkonia.
We estimated numerically the cross-section in the kinematics of ultraperipheral
$pA$ collisions at LHC, as well as in the kinematics of the future
electron-proton colliders, and found that the cross-section is sufficiently
large for experimental studies. Our evaluation is largely parameter-free
and relies only on the choice of the parametrization for the dipole
cross-section~(\ref{eq:CGCDipoleParametrization}) and wave functions
of quarkonia.

We need to mention that earlier studies of exclusive production focused
on production of quarkonia pairs with the same quantum numbers (e.g.
$J/\psi\,J/\psi$). In view of different quantum numbers, this process
predominantly proceed via exchange of \emph{two} photons at amplitude
level, like e.g. via photon-photon fusion $\gamma\gamma\to M_{1}M_{2}$~\cite{Goncalves:2015sfy,Goncalves:2019txs,Goncalves:2006hu,Baranov:2012vu,Yang:2020xkl}
or double photon scattering~\cite{Goncalves:2016ybl}. Due to extra
virtual photon in the amplitude, the cross-sections of such processes
are parametrically suppressed by $\sim\alpha_{{\rm em}}^{2}$ compared
to the cross-section of opposite $C$-parity quarkonia, and thus numerically
are significantly smaller. We hope that the process suggested in this
paper will be included in the program of the future EIC collider,
as well as ongoing studies at LHC in ultraperipheral kinematics.

Finally, we need to mention that it is quite straightforward to extend
the framework developed in this manuscript to the case of all-heavy
tetraquark production: for this it is only necessary that the product of final state quarkonia
wave functions in~(\ref{eq:Amp-1},~\ref{eq:Amp-2}) be replaced
with the wave function of the tetraquark state. Estimates of the cross-sections
for this case will be presented in a separate publication.

\section*{Acknowldgements}

We thank our colleagues at UTFSM university for encouraging discussions.
This research was partially supported by projects Proyecto ANID PIA/APOYO AFB180002
(Chile) and Fondecyt Regular (Chile) grants 1180232 and 1220242. The research of S. Andrade was partially supported by the Fellowship Program ANID BECAS/MAGÍSTER NACIONAL (Chile) 22200123.
Powered@NLHPC: This research was partially supported
by the supercomputing infrastructure of the NLHPC (ECM-02).
\appendix

\section{High energy scattering in the color dipole picture}

\label{subsec:Derivation} In this appendix for the sake of completeness
we briefly remind the general procedure which allows to express different
hard amplitudes in terms of the \emph{color singlet} forward dipole
scattering amplitude. While in the literature there are several equivalent
formulations~\cite{GLR,McLerran:1993ka,McLerran:1994vd,MUQI,MV,gbw01:1,Kopeliovich:2002yv,Kopeliovich:2001ee},
in what follows we will use the Iancu-Mueller approach~\cite{Iancu:2003uh}.

The natural hard scale, which controls the interaction of a heavy
quark with the gluonic field, is its mass $m_{Q}$. In the heavy quark
mass limit we may formally develop a systematic expansion over $\alpha_{s}\left(m_{Q}\right)\ll1$.
Furthermore, for small color singlet dipoles there is an additional
suppression by the dipole size, $r\sim1/m_{Q}$, so the interaction
of singlet dipoles with perturbative gluons is suppressed at least
as $\sim\alpha_{s}\left(m_{Q}\right)/m_{Q}$. However, the interaction
of gluons with each other, as well as with light quarks, remains strongly nonperturbative
in the deeply saturated regime,  so we expect that
the dynamics of the dipole amplitudes should satisfy the nonlinear
Balitsky-Kovchegov equation.

At very high energies the dynamics of partons can be described in the
eikonal approximation. The transverse coordinates of the high energy
partons remain essentially frozen during its propagation in the gluonic
field dipole of the target. Similarly, due to eikonal interactions we may 
disregard completely the change of the quark helicities. In
this picture the interaction of a dipole with the target is described
by the $S$-matrix element~\cite{Iancu:2003uh,Kovchegov:2012mbw}

\begin{equation}
S(y,\,\boldsymbol{x}_{Q},\,\boldsymbol{x}_{\bar{Q}})=\frac{1}{N_{c}}\left\langle {\rm tr}\left(V^{\dagger}(\boldsymbol{\boldsymbol{x}_{Q}})V(\boldsymbol{x}_{\bar{Q}})\right)\right\rangle \label{eq:S_matrix}
\end{equation}
where we use the notation $y=\ln(1/x)$ for the dipole rapidity, $\boldsymbol{x}_{Q}\,,\boldsymbol{x}_{\bar{Q}}$,
are the transverse coordinates of the partons (quark or antiquark),
and the factors $V^{\dagger}(\boldsymbol{x}_{Q})$ and $V(\boldsymbol{x}_{\bar{Q}})$
in~(\ref{eq:S_matrix}) are the Wilson lines, which describe the interaction
of the partons with the color field of a hadron. They can be expressed
as 
\begin{equation}
V\left(\boldsymbol{x}_{\perp}\right)=P\exp\left(ig\int dx^{-}A_{a}^{+}\left(x^{-},\,\boldsymbol{x}_{\perp}\right)t^{a}\right),\label{eq:Wilson}
\end{equation}
where $A_{\mu}^{a}$ is the gluonic field in a hadron. The impact
parameter dependent dipole amplitude $N(x,\,\boldsymbol{r},\,\boldsymbol{b})$
can be related to $S\left(y,\,\boldsymbol{x}_{Q},\,\boldsymbol{x}_{\bar{Q}}\right)$
as 
\begin{equation}
N\left(x,\,\boldsymbol{r},\,\boldsymbol{b}\right)=1-S\left(y,\,\boldsymbol{x}_{Q},\,\boldsymbol{x}_{\bar{Q}}\right),\label{eq:NS}
\end{equation}
where the variable $\boldsymbol{r}\equiv\boldsymbol{x}_{Q}-\boldsymbol{x}_{\bar{Q}}$
is the transverse size of the dipole, $\boldsymbol{b}\equiv z\,\boldsymbol{x}_{Q}+(1-z)\boldsymbol{x}_{\bar{Q}}$
is the transverse position of the dipole center of mass, and $z$
is the fraction of the light-cone momentum of a dipole which is carried
by the quark $Q$. In view of the weakness of the interaction between
heavy quarks and gluons, we can make an expansion of the exponent in~(\ref{eq:Wilson})
over $\alpha_{s}(m_{Q})$. In this approximation the effective interaction
of the quark or antiquark with the gluonic field of the proton can be
described by the factor $\pm i\,t^{a}\gamma_{a}\left(\boldsymbol{x}_{\perp}\right)$,
where $\boldsymbol{x}_{\perp}$ is the transverse coordinate of the
quark, 
\begin{equation}
\gamma_{a}(\boldsymbol{x})=g\int dx^{-}A_{a}^{+}(x^{-},\,\boldsymbol{x}),\label{eq:gamma}
\end{equation}
and $t_{a}$ are the ordinary color group generators of pQCD in the fundamental
representation. Inspired by the color structure of the interaction,
in what follows we will refer to these interactions as ``exchanges
of $t$-channel pomeron (gluons)'', tacitly assuming that it can
include cascades (showers) of particles. For the dipole scattering
amplitude~(\ref{eq:NS}), using (\ref{eq:S_matrix},~\ref{eq:gamma}),
we obtain

\begin{equation}
N\left(x,\,\boldsymbol{r},\,\boldsymbol{b}\right)\approx\frac{1}{2}\left[\gamma_{a}\left(\boldsymbol{x}_{Q}\right)-\gamma_{a}\left(\boldsymbol{x}_{\bar{Q}}\right)\right]^{2}.\label{eq:N_dip}
\end{equation}
For further evaluations it is more convenient to rewrite this result
in the form 
\begin{equation}
\gamma_{a}\left(\boldsymbol{x}_{1}\right)\gamma_{a}\left(\boldsymbol{x}_{2}\right)=-N\left(x,\,\boldsymbol{r}_{12},\,\boldsymbol{b}_{12}\right)+\frac{\rho\left(\boldsymbol{x}_{1}\right)+\rho\left(\boldsymbol{x}_{2}\right)}{2},\label{eq:N_dip2}
\end{equation}
where we defined a shorthand notation $\rho\left(\boldsymbol{x}_{a}\right)\equiv\left|\gamma_{a}(\boldsymbol{x})\right|^{2}$,
and $\boldsymbol{r}_{12}$, $\boldsymbol{b}_{12}$ are the distance
and center-of-mass of the quark-antiquark pair located at points $\boldsymbol{x}_{1},\,\boldsymbol{x}_{2}$.
For many processes the contributions $\sim\rho\left(\boldsymbol{x}_{i}\right)$
cancel, so the amplitude eventually can be represented as a linear
superposition of the dipole amplitudes $N\left(x,\,\boldsymbol{r},\,\boldsymbol{b}\right)$.
In what follows, we will see that the amplitude of the process considered
in this manuscript can be represented as a bilinear combination
of terms with structure $\sim\left[\gamma\left(\boldsymbol{x}_{i}\right)-\gamma\left(\boldsymbol{x}_{j}\right)\right]$.
For this special case the substitution of~(\ref{eq:N_dip2}) allows to
get a few important identities between bilinear expressions 
\begin{align}
\left[\gamma_{a}\left(\boldsymbol{x}_{1}\right)-\gamma\left(\boldsymbol{x}_{2}\right)\right] & \left[\gamma_{a}\left(\boldsymbol{x}_{3}\right)-\gamma_{a}\left(\boldsymbol{x}_{4}\right)\right]=N\left(x,\,\boldsymbol{r}_{23},\,\boldsymbol{b}_{23}\right)+N\left(x,\,\boldsymbol{r}_{14},\,\boldsymbol{b}_{14}\right)-N\left(x,\,\boldsymbol{r}_{13},\,\boldsymbol{b}_{13}\right)-N\left(x,\,\boldsymbol{r}_{24},\,\boldsymbol{b}_{24}\right),\label{eq:N_dip3}
\end{align}
\begin{align}
\left[\gamma_{a}\left(\boldsymbol{x}_{1}\right)-\gamma_{a}\left(\boldsymbol{x}_{2}\right)\right] & \left[\gamma_{a}\left(\boldsymbol{x}_{3}\right)+\gamma_{a}\left(\boldsymbol{x}_{4}\right)-2\gamma_{a}\left(\boldsymbol{x}_{5}\right)\right]=N\left(x,\,\boldsymbol{r}_{23},\,\boldsymbol{b}_{23}\right)+N\left(x,\,\boldsymbol{r}_{24},\,\boldsymbol{b}_{24}\right)-N\left(x,\,\boldsymbol{r}_{13},\,\boldsymbol{b}_{13}\right)-\\
\text{}-N\left(x,\,\boldsymbol{r}_{14},\,\boldsymbol{b}_{14}\right) & +2\left[N\left(x,\,\boldsymbol{r}_{15},\,\boldsymbol{b}_{15}\right)-N\left(x,\,\boldsymbol{r}_{25},\,\boldsymbol{b}_{25}\right)\right],\nonumber 
\end{align}
\begin{align}
 & \left[\gamma_{a}\left(\boldsymbol{x}_{1}\right)+\gamma_{a}\left(\boldsymbol{x}_{2}\right)-2\gamma_{a}\left(\boldsymbol{x}_{3}\right)\right]^{2}=2N\left(x,\,\boldsymbol{r}_{13},\,\boldsymbol{b}_{13}\right)+2N\left(x,\,\boldsymbol{r}_{23},\,\boldsymbol{b}_{23}\right)-N\left(x,\,\boldsymbol{r}_{12},\,\boldsymbol{b}_{12}\right),
\end{align}
where $\boldsymbol{r}_{ij}$ and $\boldsymbol{b}_{ij}$ are the relative
distance and center-of-mass of the quark-antiquark pair located at
points $\boldsymbol{x}_{i},\,\boldsymbol{x}_{j}$.

For the impact parameter independent ($\boldsymbol{b}$-integrated)
cross-section the results~(\ref{eq:N_dip}-\ref{eq:N_dip3}) can
be rewritten in a simpler form, 
\begin{equation}
N(x,\,\boldsymbol{r})=\frac{1}{2}\int d^{2}b\left|\gamma_{a}\left(x,\,\boldsymbol{b}-z\boldsymbol{r}\right)-\gamma_{a}\left(x,\,\boldsymbol{b}+\bar{z}\boldsymbol{r}\right)\right|^{2}.\label{eq:DipoleX}
\end{equation}
\begin{equation}
\int d^{2}\boldsymbol{b}\gamma_{a}(x,\,\boldsymbol{b})\gamma_{a}(x,\,\boldsymbol{b}+\boldsymbol{r})=-N(x,\,\boldsymbol{r})+\underbrace{\int d^{2}b\,\left|\gamma_{a}(x,\,\boldsymbol{b})\right|^{2}}_{={\rm const}}.\label{eq:SigmaDef}
\end{equation}
The value of the constant term in the right-hand side of~(\ref{eq:SigmaDef})
is related to the infrared behavior of the theory, and for the observables
which we consider in this paper, it cancels exactly. In what follows
we will apply this formalism to the evaluation of the exclusive dimeson
production amplitudes.

\section{Evaluation of the photon wave function}

For evaluation of the photon wave function we follow the standard rules of the
light--cone perturbaiton theory formulated in~\cite{Lepage:1980fj,Brodsky:1997de}.
The result for the $\bar{Q}Q$ component is well-known in the literature~\cite{Bjorken:1970ah,Dosch:1996ss},
yet below in Subsection~\ref{subsec:Basics} we will briefly repeat
its derivation in order to introduce notations. As we will see later
in Subsection~\ref{sec:WF}, the wave function of the $\bar{Q}Q\bar{Q}Q$-component
can be expressed in terms of the wave function of $\bar{Q}Q$-component.
In our evaluation we will focus on onshell transversely polarized
photons, which give the dominant contribution, unless some specific
cuts are imposed on its virtuality $Q^{2}$. The momentum of the photon~(\ref{eq:qPhoton})
introduced earlier simplifies in this case and has only light-cone
component in the plus-axis direction, 
\begin{equation}
q\approx\left(q^{+},\,0,\,\boldsymbol{0}_{\perp}\right).
\end{equation}
The polarization vector of the transversely polarized photon is given by 
\begin{align}
\varepsilon_{T}^{\mu}(q) & \equiv\left(0,\,\frac{\boldsymbol{q}_{\perp}\cdot\boldsymbol{\boldsymbol{\varepsilon}}_{\gamma}}{q^{+}},\,\boldsymbol{\boldsymbol{\varepsilon}}_{\gamma}\right)\approx\left(0,\,0,\,\boldsymbol{\boldsymbol{\varepsilon}}_{\gamma}\right),\label{eq:eDef}\\
\boldsymbol{\boldsymbol{\varepsilon}}_{\gamma} & =\frac{1}{\sqrt{2}}\left(\begin{array}{c}
1\\
\pm i
\end{array}\right),\quad\gamma=\pm1.
\end{align}
where in~(\ref{eq:eDef}) we took into account that $\boldsymbol{q}_{\perp}=0$.

Before the interaction with the target, the photon might fluctuate into 
virtual quark-antiquark pairs, as well as gluons. In what follows
we will use a convenient shorthand notation $\alpha_{i}=k_{i}/q^{+}$
for the fraction of light-cone momentum of the photon carried
by each parton, as well as $\boldsymbol{k}_{i\perp}$for the transverse
component of parton's momentum. In view of 4-momentum conservation
we expect that $\alpha_{i},\boldsymbol{k}_{i\perp}$ should satisfy
an identity 
\begin{align}
\sum_{i}\alpha_{i} & =1,\qquad\sum_{i}\boldsymbol{k}_{i\perp}=0,\label{eq:Conservation}
\end{align}
where summation is done over all partons. We may observe that the
vector $\boldsymbol{\boldsymbol{\varepsilon}}_{\gamma}$ satisfies
an identity 
\begin{equation}
\boldsymbol{\boldsymbol{\varepsilon}}_{\gamma}^{*}\equiv\boldsymbol{\boldsymbol{\varepsilon}}_{-\gamma},
\end{equation}
and its scalar product with any 2-vector $a$ yields 
\begin{equation}
\boldsymbol{\boldsymbol{\varepsilon}}_{\gamma}\cdot\boldsymbol{a}=\frac{a_{x}+i\gamma a_{y}}{\sqrt{2}}=\frac{\left|a\right|}{\sqrt{2}}e^{i\gamma\,{\rm arg}(a)},\,\,\,\,{\rm arg}(a)=\arctan\left(\frac{a_{y}}{a_{x}}\right).
\end{equation}

\subsection{$\bar{Q}Q$ component of the photon wave function}

\label{subsec:Basics} 
\begin{figure}
\includegraphics[scale=0.65]{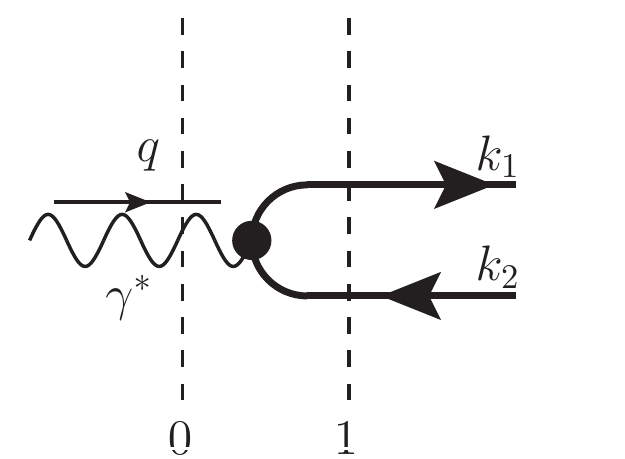}\includegraphics[scale=0.65]{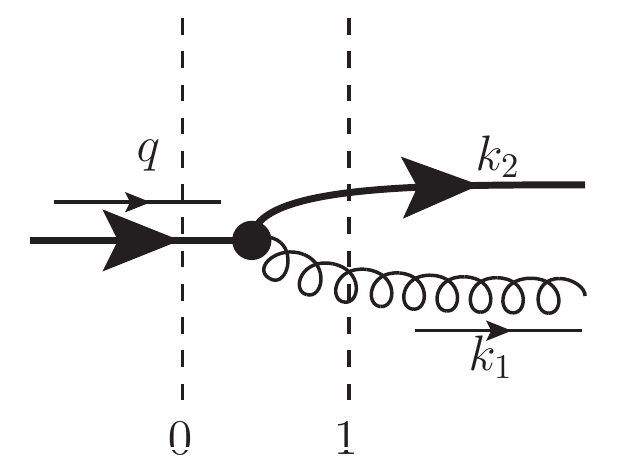}\caption{\label{fig:Photoproduction-QQ}Left plot: The leading order contributions
to the $\bar{Q}Q$-component of the photon wave function $\psi_{g\to\bar{Q}Q}$.
Right plot: the so-called gluon emission wave function, as defined
in~\cite{Lappi:2016oup}. The momenta $k_{i}$ shown in the right-hand
side are Fourier conjugates of the coordinates $x_{i}$.}
\end{figure}

In this section for the sake of completeness we would like to remind
the reader the main steps in the derivation of the $\bar{Q}Q$-component
photon wave function~\cite{Bjorken:1970ah,Dosch:1996ss} in the mixed
($\alpha,\,\boldsymbol{r}$) representation. In leading order
the subprocess $\gamma\to\bar{Q}Q$ gets contributions only from the
diagram shown in the left panel of the Figure~\ref{fig:Photoproduction-QQ}.
A bit later we will see that $\gamma\to\bar{Q}Q$, as well as the closely
related $g\to\bar{Q}Q$ subprocess, appears as constituent blocks
in the more complicated 4-quark wave function. For this reason, in order
to facilitate further discussion, temporarily in this section we will
assume that the photon momentum $q$ \emph{might} has a nonzero transverse
part $\boldsymbol{q}_{\perp}$, and will use notation $z=k_{1}^{+}/q^{+}$
for the fraction of light-cone momentum carried by the quark.
In momentum space the evaluation is straightforward, using the
rules from~~\cite{Lepage:1980fj,Brodsky:1997de,Lappi:2016oup} and
yields 
\begin{align}
\psi_{h,\bar{h}}^{\lambda}\left(z,\,k_{1},\boldsymbol{q}\right) & =-e_{q}\delta_{c\bar{c}}\frac{\bar{u}_{h}\left(k_{1}\right)\hat{\varepsilon}_{\lambda}(q)v_{\bar{h}}\left(q-k_{1}\right)}{\Delta_{01}^{-}\,\,\sqrt{k_{1}^{+}}\sqrt{q^{+}-k_{1}^{+}}}\label{eq:def}\\
\Delta_{01}^{-} & =-\frac{1}{2p^{+}}\,\frac{\boldsymbol{n}^{2}+m_{q}^{2}}{z\,(1-z)},
\end{align}
where $\lambda$ is the helicity of the incoming photon, $h,\,\bar{h}$,
are the helicities of the produced quark and antiquark, $c,\,\bar{c}$,
are the color indices of $Q$ and $\bar{Q}$, respectively, and $e_{q}$
is the electric charge corresponding to a given heavy flavor. The
momentum $\boldsymbol{n}$ is defined as $\boldsymbol{n}=\boldsymbol{k}_{1}-z\boldsymbol{q}_{\perp}=(1-z)\boldsymbol{k}_{1}-z\boldsymbol{k}_{2}$
and physically has the meaning of the transverse part of the relative
(internal) momentum of the $Q\bar{Q}$ pair. The numerator of~(\ref{eq:def})
can be written out explicitly using the rules from~\cite{Bjorken:1970ah,Dosch:1996ss},

\begin{align}
\bar{u}_{h}\left(k\right)\hat{\varepsilon}_{\lambda}(p)v_{\bar{h}}\left(p-k\right) & =\frac{2}{\sqrt{z(1-z)}}\left[\left(z\delta_{\lambda,h}-(1-z)\delta_{\lambda,-h}\right)\delta_{h,-\bar{h}}\boldsymbol{n}\cdot\boldsymbol{\varepsilon}_{\lambda}+\frac{1}{\sqrt{2}}m_{q}\,{\rm sign}(h)\delta_{\lambda,h}\delta_{h,\bar{h}}\right].
\end{align}
In configuration space the corresponding wave function can be
found making a Fourier transformation over the transverse momenta,
\begin{align}
 & \int\frac{d^{2}k_{1}}{(2\pi)^{2}}\frac{d^{2}k_{2}}{(2\pi)^{2}}e^{i\left(\boldsymbol{k}_{1}\cdot\boldsymbol{r}_{1}+\boldsymbol{k}_{2}\cdot\boldsymbol{r}_{2}\right)}\left(2\pi\right)^{2}\delta\left(\boldsymbol{k}_{1}+\boldsymbol{k}_{2}-\boldsymbol{q}\right)\psi_{h,\bar{h}}^{\lambda}\left(z,\,k_{1},\boldsymbol{q}\right)\label{eq:WF}\\
 & =e^{i\boldsymbol{q}\cdot\left(z\boldsymbol{r}_{1}+\bar{z}\boldsymbol{r}_{2}\right)}e_{q}\delta_{c\bar{c}}\Psi_{h\bar{h}}^{\lambda}\left(z,\,\boldsymbol{r}_{12},\,m_{q},\,m_{q}\right),\nonumber 
\end{align}
where the integral over $k_{2}$ was performed using the properties of
the $\delta$ function, and before the integration over $\boldsymbol{k}_{1}$
we shifted the integration variable as $\boldsymbol{k}_{1}\to\boldsymbol{n}+z\boldsymbol{q}$.
Explicitly, the integration over the variable $d^{2}\boldsymbol{n}$
yields 
\begin{equation}
\Psi_{h\bar{h}}^{\lambda}\left(z,\,\boldsymbol{r}_{12},\,m_{q},a\right)=-\frac{2}{(2\pi)}\left[\left(z\delta_{\lambda,h}-(1-z)\delta_{\lambda,-h}\right)\delta_{h,-\bar{h}}i\boldsymbol{\varepsilon}_{\lambda}\cdot\nabla-\frac{m_{q}}{\sqrt{2}}\,{\rm sign}(h)\delta_{\lambda,h}\delta_{h,\bar{h}}\right]K_{0}\left(a\,\boldsymbol{r}\right).\label{eq:PsiSplitting}
\end{equation}
The structure of the Eq\@.~(\ref{eq:WF}) clearly suggests that
in a mixed representation the variable $z\boldsymbol{r}_{1}+\bar{z}\boldsymbol{r}_{2}$
plays the role of the dipole center of mass, whereas $\boldsymbol{r}_{12}$
is its separation, in agreement with earlier findings from~\cite{Bartels:2003yj}.
For the incoming offshell photon with virtuality $-q^{2}=Q^{2}$, straightforward
integration yields in a similar fashion 
\begin{equation}
e^{i\boldsymbol{q}\cdot\left(z\boldsymbol{r}_{1}+\bar{z}\boldsymbol{r}_{2}\right)}e_{q}\delta_{c\bar{c}}\Psi_{h\bar{h}}^{\lambda}\left(z,\,\boldsymbol{r}_{12},\,m_{q},\,\sqrt{m_{q}^{2}-Q^{2}z(1-z)}\right)
\end{equation}
in the second line of~(\ref{eq:WF}). The extension of this result
for the production of a $Q\bar{Q}$ pair by a gluon is straightforward
and requires a simple replacement $e_{q}\delta_{c\bar{c}}\to g\,\left(t_{a}\right)_{c\bar{c}}$
.

Finally, we would like to discuss briefly the so-called parton-level
wave function of the gluon emission subprocess $q\to gq$, as introduced
in~\cite{Lappi:2016oup}. This object is useful for the analysis of different
amplitudes, as we will see in the next section. In the leading order
it gets contributions from the diagram shown in the right panel of
the Figure~\ref{fig:Photoproduction-QQ}. The evaluation of this
object are quite similar to the derivation of~(\ref{eq:def}-\ref{eq:PsiSplitting}).
In momentum space we obtain

\begin{align}
\tilde{\psi}_{c_{f}h_{f},\,c_{i}h_{i}}^{\lambda}\left(z,\,k_{1},\boldsymbol{q}\right) & =-gt_{c_{f}c_{i}}^{a}\frac{\bar{u}_{h_{f}}\left(q-k_{1}\right)\hat{\varepsilon}_{\lambda}\left(k_{1}\right)u_{h_{i}}\left(q\right)}{\Delta_{02}^{-}\,\sqrt{q^{+}}\sqrt{q^{+}-k_{1}^{+}}},\\
\Delta_{02}^{-} & =-\frac{1}{2p^{+}}\,\frac{\boldsymbol{n}^{2}+z^{2}m_{q}^{2}}{z\,(1-z)},\quad\boldsymbol{n}=\boldsymbol{k}_{1}-z\,\boldsymbol{q}
\end{align}
where $\lambda$ is the helicity of the outgoing gluon; $(h_{i},c_{i})$
and $(h_{f},c_{f})$ are the helicities and color indices of the incident
and final quark (before and after emission of a gluon); and similar
to the previous case we have introduced the momentum $\boldsymbol{n}=\boldsymbol{k}_{1}-z\boldsymbol{q}=(1-z)\boldsymbol{k}_{1}-z\boldsymbol{k}_{2}$
which corresponds to the relative motion of the quark and gluon after
emission of the latter. Using the rules from~\cite{Bjorken:1970ah,Dosch:1996ss},
we may rewrite the numerator as

\begin{align*}
\bar{u}_{h_{f}}\left(q-k_{1}\right)\hat{\varepsilon}_{\lambda}\left(k_{1}\right)u_{h_{i}}\left(q\right) & =\frac{2}{z\sqrt{1-z}}\left[\left(\delta_{\lambda,\,h_{i}}+(1-z)\delta_{\lambda,-h_{i}}\right)\delta_{h_{i},h_{f}}{\rm \boldsymbol{n}}\cdot\boldsymbol{\varepsilon}_{\lambda}-\frac{m_{q}}{\sqrt{2}}z^{2}\,{\rm sign}(h_{i})\delta_{\lambda,-h_{i}}\delta_{h_{i},-h_{f}}\right],
\end{align*}
In configuration space the corresponding wave function is given
by 
\begin{align}
 & \int\frac{d^{2}k_{1}}{(2\pi)^{2}}\frac{d^{2}k_{2}}{(2\pi)^{2}}e^{i\left(\boldsymbol{k}_{1}\cdot\boldsymbol{r}_{1}+\boldsymbol{k}_{2}\cdot\boldsymbol{r}_{2}\right)}\left(2\pi\right)^{2}\delta\left(\boldsymbol{k}_{1}+\boldsymbol{k}_{2}-\boldsymbol{q}\right)\psi_{c_{f}h_{f},\,c_{i}h_{i}}^{\lambda}\left(z,\,k_{1},\boldsymbol{q}\right)\label{eq:WF-1}\\
 & =e^{i\boldsymbol{q}\cdot\left(z\boldsymbol{r}_{1}+\bar{z}\boldsymbol{r}_{2}\right)}t_{c_{f}c_{i}}^{a}\Phi_{h_{f},\,h_{i}}^{\lambda}\left(z,\,\boldsymbol{r}_{12},\,m_{q},\,z\,m_{q}\right),\nonumber 
\end{align}
where the integral over $k_{2}$ was performed using the properties of
the wave function, and integration over the variable as $\boldsymbol{k}_{1}=\boldsymbol{n}+z\boldsymbol{q}$
yields 
\begin{equation}
\Phi_{h_{f},\,h_{i}}^{\lambda}\left(z,\,\boldsymbol{r}_{12},\,m_{q},a\right)=-\frac{2}{(2\pi)}\left[\left(\delta_{\lambda,h_{i}}+(1-z)\delta_{\lambda,-h_{i}}\right)\delta_{h_{i},h_{f}}i\boldsymbol{\varepsilon}_{\lambda}\cdot\nabla+\frac{m_{q}}{\sqrt{2}}z^{2}\,{\rm sign}(h_{i})\delta_{\lambda,-h_{i}}\delta_{h_{i},-h_{f}}\right]K_{0}\left(a\,\boldsymbol{r}\right).\label{eq:PhiEmission}
\end{equation}

For the case of incoming offshell quark with virtuality $Q^{2}$,
straightforward generalization show that the second line of~(\ref{eq:WF-1})
gets the form 
\begin{equation}
e^{i\boldsymbol{q}\cdot\left(z\boldsymbol{r}_{1}+\bar{z}\boldsymbol{r}_{2}\right)}\Phi_{h_{f},\,h_{i}}^{\lambda}\left(z,\,\boldsymbol{r}_{12},\,m_{q},\,\sqrt{m_{q}^{2}z-Q^{2}z(1-z)}\right)
\end{equation}

Similarly to the previous case, the structure of the Eq\@.~(\ref{eq:PhiEmission})
clearly suggests that in a mixed representation the variable $z\boldsymbol{r}_{1}+\bar{z}\boldsymbol{r}_{2}$
plays the role of the dipole center of mass, whereas $\boldsymbol{r}_{12}$
is its separation~\cite{Bartels:2003yj}.

\subsection{$\bar{Q}Q\bar{Q}Q$ component of the photon wave function}

\label{sec:WF}As was mentioned earlier in Section~\ref{subsec:Derivation},
in the eikonal approximation the amplitude of the subprocess $\gamma^{*}\to\bar{Q}Q\bar{Q}Q$
in configuration space can be represented as a convolution of the
wave function $\psi_{\bar{Q}Q\bar{Q}Q}^{(\gamma)}$ with linear combinations
of dipole amplitudes~(\ref{eq:N_dip3}). In leading order over
$\alpha_{s}$ the amplitude of the process is given by the two diagrams
shown in the Figure~(\ref{fig:Photoproduction-QQQQ}). It should
be understood that these diagrams should be supplemented by all possible
permutations of final state quarks. More precisely, for the production
of different heavy flavors (\emph{e.g}. $\bar{c}c\bar{b}b$) both
diagrams should be supplemented by contributions with permuted \emph{pairs}
of momenta $\left(k_{1},\,k_{2}\right)\leftrightarrow\left(k_{3},\,k_{4}\right)$.
For the same-flavor quarkonia pairs (\emph{e.g}. $\bar{c}c\bar{c}c$)
we should take into account contributions with independent permutations
of the quarks and antiquarks, $k_{1}\leftrightarrow k_{3}$ and $k_{2}\leftrightarrow k_{4}$.
The evaluation of the corresponding process follows the standard light--cone
rules formulated in~\cite{Lepage:1980fj,Brodsky:1997de}. We need
to mention that some blocks, which will be needed for the construction
of the amplitudem have already been evaluated in~\cite{Lappi:2016oup,Hanninen:2017ddy}
(although in the chiral limit only). In these section we extend those studies
and represent them in a form convenient for further analysis. According
to the general light-cone rules~\cite{Bjorken:1970ah,Dosch:1996ss},
in the evaluation of the diagrams in Figure~(\ref{fig:Photoproduction-QQQQ})
each propagator of the virtual (intermediate) parton has instantaneous
and non-instantaneous parts. For technical reasons it is convenient
to analyze separately the two types of contributions.

\subsubsection{Non-instantaneous contributions}

\begin{figure}
\includegraphics[scale=0.65]{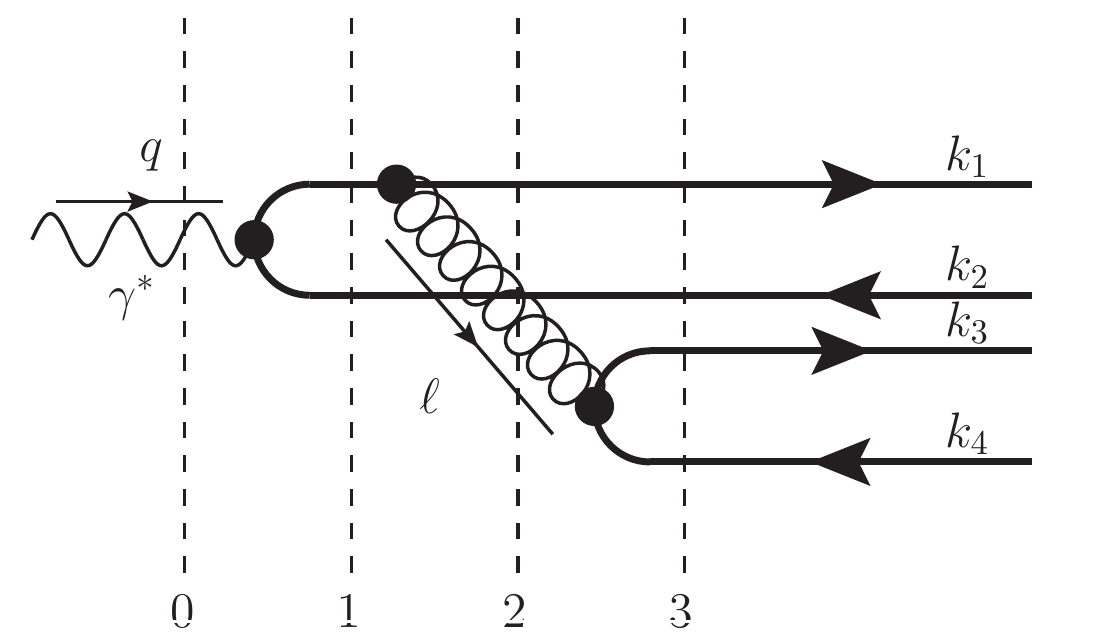}\includegraphics[scale=0.65]{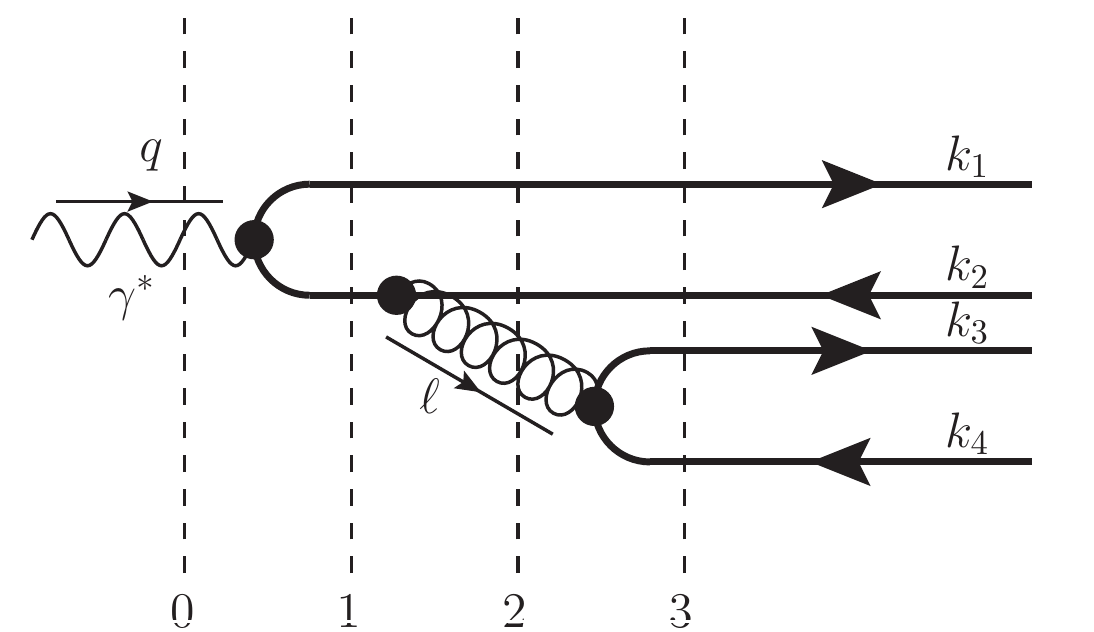}

\caption{\label{fig:Photoproduction-QQQQ}The leading order contribution to
the wave function $\psi_{\bar{Q}Q\bar{Q}Q}^{(\gamma)}$ defined in
the text. The momenta $k_{i}$ shown in the right-hand side are Fourier
conjugates of the coordinates $x_{i}$. It is implied that both diagrams
should be supplemented by all possible permutations of final state
quarks (see the text for more details).}
\end{figure}

In leading order over $\alpha_{s}$, the amplitude of the process
is given by the two diagrams shown in Figure~(\ref{fig:Photoproduction-QQQQ})
and depends on the momenta of the 4 quarks in the final state. In what
follows we will use the standard notation $\alpha_{i}=k_{i}/q^{+}$
for the fractions of photon momentum carried by each of these
fermions, as well as $\boldsymbol{k}_{i\perp}$ for the transverse
components of their momenta. We also will use a shorthand notation
$\ell=k_{3}+k_{3}$ for the momentum of the virtual gluon connecting
different quark lines. For the sake of generality we will assume that the
produced quark-antiquark pairs have different flavors, and will use the
notations $m_{1}$ for the current mass of the quark line connected
to a photon, and $m_{2}$ for the current masses of the quark-antiquark
pair produced from the virtual gluon.

Using the rules from~\cite{Bjorken:1970ah,Dosch:1996ss}, we may
obtain for the corresponding amplitude of the subprocess 
\begin{align}
\mathcal{A}_{c_{1}c_{2},c_{3}c_{2}}^{a_{1}a_{2},a_{3}a_{4}} & =-\frac{e_{q}g^{2}\,\left(t_{a}\right)_{c_{1}c_{2}}\otimes\left(t_{a}\right)_{c_{3}c_{4}}}{16\,\pi^{2}}\left(\frac{\bar{u}_{a_{1}}\left(k_{1}\right)\hat{\varepsilon}_{\lambda}^{*}(\ell)u_{b}\left(k_{1}+\ell\right)\,\,\bar{u}_{b}\left(q-k_{2}\right)\hat{\varepsilon}_{\gamma}(q)\,v_{a_{2}}\left(k_{2}\right)}{D_{11}D_{12}D_{13}\sqrt{k_{1}^{+}}\sqrt{k_{1}^{+}+\ell^{+}}\sqrt{q^{+}-k_{2}^{+}}\sqrt{k_{2}^{+}}}\right.+\label{eq:Amp_p}\\
 & -\left.\frac{\bar{u}_{a_{1}}\left(k_{1}\right)\hat{\varepsilon}_{\gamma}(q)v_{b}\left(q-k_{1}\right)\,\,\bar{v}_{b}\left(k_{2}+\ell\right)\hat{\varepsilon}_{\lambda}^{*}(\ell)\,v_{a_{2}}\left(k_{2}\right)}{D_{21}D_{22}D_{23}\sqrt{k_{1}^{+}}\sqrt{q^{+}-k_{1}^{+}}\sqrt{k_{2}^{+}+\ell^{+}}\sqrt{k_{2}^{+}}}\right)\frac{\bar{u}_{a_{3}}\left(k_{3}\right)\hat{\varepsilon}_{\lambda}\left(\ell\right)v_{a_{4}}\left(k_{4}\right)}{\left(k_{3}^{+}+k_{4}^{+}\right)\sqrt{k_{3}^{+}k_{4}^{+}}},\nonumber
\end{align}
where $a_{i}$ and $c_{i}$ are the helicity and color indices of
the final state quarks, and $D_{ij}$ are the conventional light-cone
denominators (with the first subscript index $i=1,2$ refers to the
first and the second diagram in Figure~\ref{fig:Photoproduction-QQQQ}
respectively, and the second index $j=1,2,3$ numerates proper cuts
shown with dashed vertical lines). Explicitly, these light-cone denominators are
given by 
\begin{align}
D_{11} & =-\frac{1}{2q^{+}}\left(\frac{\boldsymbol{k}_{2\perp}^{2}+m_{1}^{2}}{\alpha_{2}}+\frac{\left(\boldsymbol{k}_{1\perp}+\boldsymbol{\ell}\right)^{2}+m_{1}^{2}}{\alpha_{1}+z}\right)=-\frac{1}{2q^{+}}\frac{\boldsymbol{k}_{2\perp}^{2}+m_{1}^{2}}{\alpha_{2}\bar{\alpha}_{2}},\label{eq:D_11}
\end{align}
\begin{align}
D_{21} & =-\frac{1}{2q^{+}}\left(\frac{\boldsymbol{k}_{1\perp}^{2}+m_{1}^{2}}{\alpha_{1}}+\frac{\left(\boldsymbol{k}_{2\perp}+\boldsymbol{\ell}\right)^{2}+m_{1}^{2}}{\alpha_{2}+z}\right)=-\frac{1}{2q^{+}}\frac{\boldsymbol{k}_{1\perp}^{2}+m_{1}^{2}}{\alpha_{1}\bar{\alpha}_{1}},\label{eq:D_21}
\end{align}
\begin{align}
D_{12} & =D_{22}\equiv D_{2}=-\frac{1}{2q^{+}}\left(\frac{\boldsymbol{k}_{1\perp}^{2}+m_{1}^{2}}{\,\alpha_{1}}+\frac{\boldsymbol{k}_{2\perp}^{2}+m_{1}^{2}}{\,\alpha_{2}}+\frac{\boldsymbol{\ell}_{\perp}^{2}}{z}\right)=\label{eq:D_12}\\
 & =-\frac{1}{2q^{+}}\frac{\alpha_{2}\bar{\alpha}_{2}\left(\boldsymbol{k}_{1\perp}+\boldsymbol{k}_{2\perp}\frac{\alpha_{1}}{\bar{\alpha}_{2}}\right)^{2}+\frac{\alpha_{1}}{\bar{\alpha}_{2}}\left(1-\alpha_{1}-\alpha_{2}\right)\boldsymbol{k}_{2\perp}^{2}+m_{1}^{2}\left(\alpha_{1}+\alpha_{2}\right)\left(1-\alpha_{1}-\alpha_{2}\right)}{\alpha_{1}\alpha_{2}\left(1-\alpha_{1}-\alpha_{2}\right)}=\nonumber \\
 & =-\frac{1}{2q^{+}}\frac{\alpha_{1}\bar{\alpha}_{1}\left(\boldsymbol{k}_{2\perp}+\boldsymbol{k}_{1\perp}\frac{\alpha_{2}}{\bar{\alpha}_{1}}\right)^{2}+\frac{\alpha_{2}}{\bar{\alpha}_{1}}\left(1-\alpha_{1}-\alpha_{2}\right)\boldsymbol{k}_{1\perp}^{2}+m_{1}^{2}\left(\alpha_{1}+\alpha_{2}\right)\left(1-\alpha_{1}-\alpha_{2}\right)}{\alpha_{1}\alpha_{2}\left(1-\alpha_{1}-\alpha_{2}\right)}
\end{align}
\begin{align}
D_{13} & =D_{23}=D_{3}=-\frac{1}{2q^{+}}\left(\sum_{i=1}^{2}\frac{\boldsymbol{k}_{i\perp}^{2}+m_{1}^{2}}{2\,\alpha_{i}}+\sum_{i=3}^{4}\frac{\boldsymbol{k}_{i\perp}^{2}+m_{2}^{2}}{2\,\alpha_{i}}\right)=\label{eq:D_13}\\
 & =D_{12}-\frac{\left(\boldsymbol{k}_{3\perp}\alpha_{4}-\boldsymbol{k}_{4\perp}\alpha_{3}\right)^{2}+m_{2}^{2}\left(\alpha_{3}+\alpha_{4}\right)^{2}}{2q^{+}\alpha_{3}\alpha_{4}\left(\alpha_{3}+\alpha_{4}\right)}=D_{12}-\left(\frac{\alpha_{3}+\alpha_{4}}{\alpha_{3}\alpha_{4}}\right)\frac{\boldsymbol{q}_{34}^{2}+m_{2}^{2}}{2\,q^{+}},\nonumber \\
\boldsymbol{q}_{34} & =\frac{\boldsymbol{k}_{3\perp}\alpha_{4}-\boldsymbol{k}_{4\perp}\alpha_{3}}{\alpha_{3}+\alpha_{4}}.\label{eq:kRel}
\end{align}
To simplify the structure of the expressions~(\ref{eq:D_11}-\ref{eq:D_13}),
we introduced a shorthand notation $\bar{\alpha}_{i}\equiv1-\alpha_{i},\,\,i=1...4$.
The combination of momenta $\boldsymbol{q}_{34}$, defined in~(\ref{eq:kRel}),
represents the relative motion momenta of quarks 3 and 4 (Fourier
conjugate of a relative distance $\boldsymbol{r}_{3}-\boldsymbol{r}_{4}$).
Technically, the structure of the denominators, up to trivial redefinitions,
agrees with the findings of~\cite{Lappi:2016oup}. The expressions in the
numerator of~(\ref{eq:Amp_p}) can be written out explicitly using
the light-cone algebra from~\cite{Lepage:1980fj,Brodsky:1997de,Lappi:2016oup},
yielding for the amplitude

\begin{align}
\mathcal{A}_{c_{1}c_{2},c_{3}c_{4}}^{a_{1}a_{2},a_{3}a_{4}} & =\frac{1}{2\pi^{2}\,\left(q^{+}\right)^{2}}\frac{e_{q}\,g^{2}\,\left(t_{a}\right)_{c_{1}c_{2}}\otimes\left(t_{a}\right)_{c_{3}c_{4}}}{\sqrt{\alpha_{1}\alpha_{2}}\left(1-\alpha_{1}-\alpha_{2}\right)\left(\alpha_{3}+\alpha_{4}\right)\,D_{2}\left(\alpha_{1},\boldsymbol{k}_{1};\alpha_{2},\,\boldsymbol{k}_{2}\right)}\times\label{eq:Amp_p_explicit}\\
 & \times\left\{ \frac{1}{\boldsymbol{k}_{2\perp}^{2}+m_{1}^{2}}\sqrt{\frac{\alpha_{2}}{\alpha_{1}}}\left[\left(\alpha_{2}\delta_{\gamma,a_{2}}-\bar{\alpha}_{2}\delta_{\gamma,-a_{2}}\right)\delta_{b,-a_{2}}\boldsymbol{k}_{2}\cdot\boldsymbol{\varepsilon}_{\gamma}+\frac{m_{q}}{\sqrt{2}}\,{\rm sign}\left(a_{2}\right)\delta_{\gamma,a_{2}}\delta_{b,\,a_{2}}\right]\right.\times\nonumber \\
 & \times\left[\left(\bar{\alpha}_{2}\delta_{\lambda,\,a_{1}}+\alpha_{1}\delta_{\lambda,-a_{1}}\right)\delta_{a_{1},b}\boldsymbol{q}_{1}\cdot\boldsymbol{\varepsilon}_{\lambda}^{*}+\frac{m_{q}}{\sqrt{2}}\frac{\left(1-\alpha_{1}-\alpha_{2}\right)^{2}}{1-\alpha_{2}}\,{\rm sign}\left(-a_{1}\right)\delta_{\lambda,-a_{1}}\delta_{a_{1},-b}\right]\nonumber \\
 & -\frac{1}{\boldsymbol{k}_{1\perp}^{2}+m_{1}^{2}}\sqrt{\frac{\alpha_{1}}{\alpha_{2}}}\left[\left(\alpha_{1}\delta_{\gamma,a_{1}}-\bar{\alpha}_{1}\delta_{\gamma,-a_{1}}\right)\delta_{b,-a_{1}}\boldsymbol{k}_{1}\cdot\boldsymbol{\varepsilon}_{\gamma}+\frac{m_{q}}{\sqrt{2}}\,{\rm sign}\left(a_{1}\right)\delta_{\gamma,a_{1}}\delta_{a_{1},b}\right]\times\nonumber \\
 & \times\left.\left[\left(\bar{\alpha}_{1}\delta_{\lambda,\,a_{2}}+\alpha_{2}\delta_{\lambda,-a_{2}}\right)\delta_{a_{2},b}\boldsymbol{q}_{2}\cdot\boldsymbol{\varepsilon}_{\lambda}^{*}+\frac{m_{q}}{\sqrt{2}}\frac{\left(1-\alpha_{1}-\alpha_{2}\right)^{2}}{1-\alpha_{1}}\,{\rm sign}\left(-a_{2}\right)\delta_{\lambda,-a_{2}}\delta_{a_{2},-b}\right]\right\} \times\nonumber \\
 & \times\frac{\frac{2\left(\alpha_{3}+\alpha_{4}\right)}{\alpha_{3}\alpha_{4}}\left[\left(\frac{\alpha_{3}}{\alpha_{3}+\alpha_{4}}\delta_{\lambda,-a_{3}}-\frac{\alpha_{4}}{\alpha_{3}+\alpha_{4}}\delta_{\lambda,a_{3}}\right)\delta_{a_{3},-a_{4}}\boldsymbol{q}_{34}\cdot\boldsymbol{\varepsilon}_{-\lambda}+\frac{m_{q}}{\sqrt{2}}\,{\rm sign}(a_{3})\delta_{\lambda,a_{3}}\delta_{a_{3},a_{4}}\right]}{D_{2}\left(\alpha_{1},\boldsymbol{k}_{1};\alpha_{2},\,\boldsymbol{k}_{2}\right)-\frac{\boldsymbol{q}_{34}^{2}+m_{2}^{2}}{2q^{+}}\left(\frac{\alpha_{3}+\alpha_{4}}{\alpha_{3}\alpha_{4}}\right)}\nonumber 
\end{align}
where the momenta $\boldsymbol{q}_{i}$ are defined as 
\begin{align}
 & \boldsymbol{q}_{1}=-\left(\boldsymbol{k}_{1}+\frac{\alpha_{1}}{1-\alpha_{2}}\boldsymbol{k}_{2}\right),\qquad\boldsymbol{q}_{2}=-\left(\boldsymbol{k}_{2}+\frac{\alpha_{2}}{1-\alpha_{1}}\boldsymbol{k}_{1}\right).\label{eq:q_12}
\end{align}
We may observe that the amplitude (\ref{eq:Amp_p_explicit}) is antisymmetric
with respect to permutation of the momenta and helicities of the
first two quarks, $\left(\alpha_{1},\,\boldsymbol{k}_{1},\,a_{1}\right)\leftrightarrow\left(\alpha_{2},\,\boldsymbol{k}_{2},\,a_{2}\right)$,
and symmetric with respect to permutation of the the momenta and helicities
of the 3rd and 4th quarks, $\left(\alpha_{3},\,\boldsymbol{k}_{3},\,a_{3}\right)\leftrightarrow\left(\alpha_{4},\,\boldsymbol{k}_{4},\,a_{4}\right)$.
This symmetry simply reflects that the amplitude~(\ref{eq:Amp_p_explicit})
was evaluated as a sum of the left and the right diagrams in Figure~\ref{fig:Photoproduction-QQQQ},
which can be related by charge conjugation. This symmetry allows
to simplify some evaluations.

For evaluations in the dipole framework we need to rewrite the amplitude
in configuration space, making a Fourier transformation over the
transverse components, 
\begin{equation}
\psi_{\bar{Q}Q\bar{Q}Q}^{(\gamma)}\left(\left\{ \alpha_{i},\,\boldsymbol{x}_{i}\right\} \right)=\int\left(\prod_{i=1}^{4}\frac{d^{2}k_{i}}{\left(2\pi\right)^{2}}e^{i\boldsymbol{k}_{i}\cdot\boldsymbol{x}_{i}}\right)(2\pi)^{2}\delta^{2}\left(\sum\boldsymbol{k}_{i}\right)\mathcal{A}_{c_{1}c_{2},c_{3}c_{4}}^{a_{1}a_{2},a_{3}a_{4}}\left(\left\{ \alpha_{i},\,\boldsymbol{k}_{i}\right\} \right)\label{eq:PsiFourier}
\end{equation}

\begin{figure}
\includegraphics[scale=0.65]{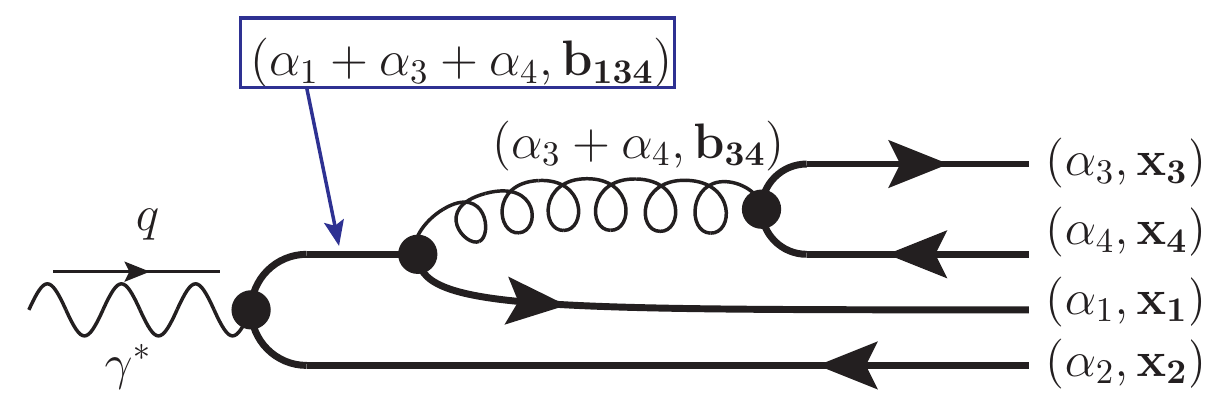}\includegraphics[scale=0.65]{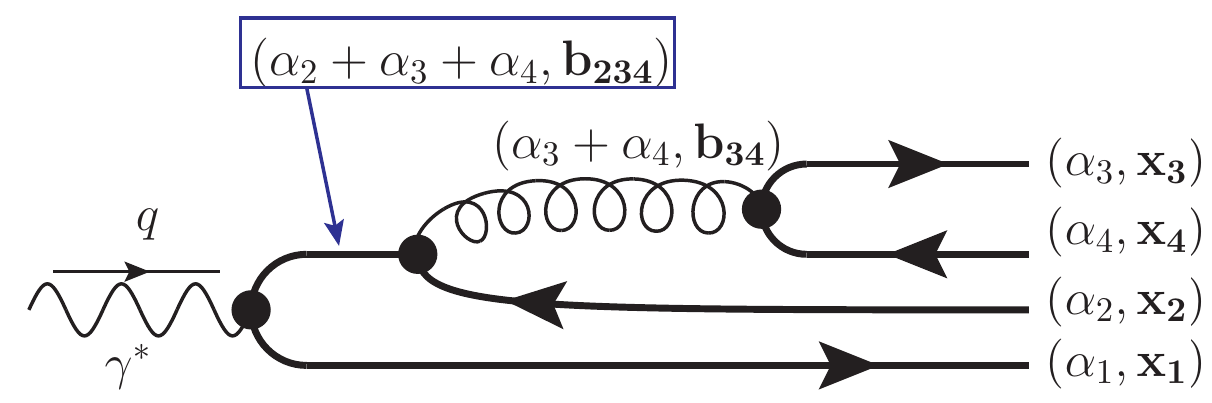}\caption{\label{fig:Photoproduction-QQQQ-1} Graphical illustration of the
transverse momentum dependence of the wave function $\psi_{\gamma\to\bar{Q}Q\bar{Q}Q}\left(\left\{ \alpha_{i},\,\boldsymbol{r}_{i}\right\} \right)$.
The letters $\boldsymbol{b}_{ij}$ and $\boldsymbol{b}_{ijk}$ stand
for the center of mass position of the partons $ij$ or $ijk$. See
the text for more details.}
\end{figure}

In view of momentum conservation~(\ref{eq:Conservation}), the wave
function $\psi_{\bar{Q}Q\bar{Q}Q}^{(\gamma)}$ will be invariant with
respect to global shifts 
\begin{equation}
\boldsymbol{x}_{i}\to\boldsymbol{x}_{i}+\boldsymbol{a}_{i},\quad\boldsymbol{a}_{i}={\rm const},
\end{equation}
i.e. should depend only on relative distances between quarks $\left|\boldsymbol{x}_{i}-\boldsymbol{x}_{j}\right|$.
After straightforward evaluation of the integrals and algebraic simplifications
it is possible to reduce~(\ref{eq:PsiFourier}) to the form 
\begin{equation}
\psi_{\bar{Q}Q\bar{Q}Q}^{(\gamma)}\left(\left\{ \alpha_{i},\,\boldsymbol{x}_{i}\right\} \right)=A\left(\left\{ \alpha_{i},\,\boldsymbol{x}_{i}\right\} \right)+B\left(\left\{ \alpha_{i},\,\boldsymbol{x}_{i}\right\} \right).\label{eq:PsiFull}
\end{equation}

where

\begin{align}
A\left(\left\{ \alpha_{i},\,\boldsymbol{r}_{i}\right\} \right) & =-\frac{2e_{q}\alpha_{s}\left(m_{Q}\right)\,\left(t_{a}\right)_{c_{1}c_{2}}\otimes\left(t_{a}\right)_{c_{3}c_{4}}}{\pi^{3}\left(1-\alpha_{1}-\alpha_{2}\right)^{2}\sqrt{\alpha_{1}\alpha_{2}}\,}\int\frac{q_{1}dq_{1}\,k_{2}dk_{2}}{\frac{\bar{\alpha}_{2}q_{1}^{2}}{\alpha_{1}\left(1-\alpha_{1}-\alpha_{2}\right)}+\frac{m_{1}^{2}\left(\alpha_{1}+\alpha_{2}\right)}{\alpha_{1}\alpha_{2}}+\frac{k_{2}^{2}}{\alpha_{2}\bar{\alpha}_{2}}}\times\label{eq:Amp_p_explicit-4-4}\\
 & \times\frac{1}{k_{2}^{2}+m_{1}^{2}}\sqrt{\frac{\alpha_{2}}{\alpha_{1}}}\left[\left(\alpha_{2}\delta_{\gamma,a_{2}}-\bar{\alpha}_{2}\delta_{\gamma,-a_{2}}\right)\left(\bar{\alpha}_{2}\delta_{\lambda,\,a_{1}}+\alpha_{1}\delta_{\lambda,-a_{1}}\right)\delta_{a_{1},-a_{2}}\times\right.\nonumber \\
 & \times\left(\boldsymbol{n}_{2,134}\cdot\boldsymbol{\varepsilon}_{\gamma}\right)\left(\boldsymbol{n}_{1,34}\cdot\boldsymbol{\varepsilon}_{\lambda}^{*}\right)k_{2}\,J_{1}\left(k_{2}\left|\boldsymbol{x}_{2}-\boldsymbol{b}_{134}\right|\right)q_{1}J_{1}\left(q_{1}\left|\boldsymbol{x}_{1}-\boldsymbol{b}_{34}\right|\right)+\nonumber \\
 & +\frac{m_{q}^{2}}{2}\,\delta_{\lambda,-a_{1}}\delta_{\gamma,a_{2}}\delta_{a_{1},-a_{2}}J_{0}\left(k_{2}\left|\boldsymbol{x}_{2}-\boldsymbol{b}_{134}\right|\right)J_{0}\left(q_{1}\left|\boldsymbol{x}_{1}-\boldsymbol{b}_{34}\right|\right)\frac{\left(1-\alpha_{1}-\alpha_{2}\right)^{2}}{1-\alpha_{2}}\nonumber \\
 & -\frac{im_{q}}{\sqrt{2}}\,{\rm sign}\left(a_{2}\right)\delta_{\gamma,a_{2}}\delta_{a_{1},a_{2}}\left(\bar{\alpha}_{2}\delta_{\lambda,\,a_{1}}+\alpha_{1}\delta_{\lambda,-a_{1}}\right)\times\nonumber \\
 & \times\boldsymbol{n}_{1,34}\cdot\boldsymbol{\varepsilon}_{\lambda}^{*}q_{1}J_{1}\left(q_{1}\left|\boldsymbol{x}_{1}-\boldsymbol{b}_{34}\right|\right)J_{0}\left(k_{2}\left|\boldsymbol{x}_{2}-\boldsymbol{b}_{134}\right|\right)\nonumber \\
 & -\frac{im_{q}}{\sqrt{2}}{\rm sign}\left(a_{1}\right)\delta_{\lambda,-a_{1}}\left(\alpha_{2}\delta_{\gamma,a_{2}}-\bar{\alpha}_{2}\delta_{\gamma,-a_{2}}\right)\delta_{a_{1},a_{2}}\frac{\left(1-\alpha_{1}-\alpha_{2}\right)^{2}}{1-\alpha_{2}}\times\nonumber \\
 & \times\left.\left(\boldsymbol{n}_{2,134}\cdot\boldsymbol{\varepsilon}_{\gamma}\right)k_{2}\,J_{1}\left(k_{2}\left|\boldsymbol{x}_{2}-\boldsymbol{b}_{134}\right|\right)J_{0}\left(q_{1}\left|\boldsymbol{x}_{1}-\boldsymbol{b}_{34}\right|\right)\right]\times\nonumber \\
 & \times\Psi_{a_{3},a_{4}}^{-\lambda}\left(\frac{\alpha_{3}}{\alpha_{3}+\alpha_{4}},\,\boldsymbol{r}_{34},\,m_{2},\,\sqrt{m_{2}^{2}+\frac{\alpha_{3}\alpha_{4}}{\alpha_{3}+\alpha_{4}}\left[\frac{\bar{\alpha}_{2}q_{1}^{2}}{\alpha_{1}\left(1-\alpha_{1}-\alpha_{2}\right)}+\frac{m_{1}^{2}\left(\alpha_{1}+\alpha_{2}\right)}{\alpha_{1}\alpha_{2}}+\frac{k_{2}^{2}}{\alpha_{2}\bar{\alpha}_{2}}\right]}\right)\nonumber 
\end{align}
and 
\[
B\left(\alpha_{1},\,\boldsymbol{x}_{1},\,\alpha_{2},\,\boldsymbol{x}_{2},\,\alpha_{3},\,\boldsymbol{x}_{3},\,\alpha_{4},\,\boldsymbol{x}_{4}\right)=-A\left(\alpha_{2},\,\boldsymbol{x}_{2},\,\alpha_{1},\,\boldsymbol{x}_{1},\,\alpha_{4},\,\boldsymbol{x}_{4},\,\alpha_{3},\,\boldsymbol{x}_{3}\right).
\]
The variable $\boldsymbol{b}_{j_{1}...j_{n}}$ corresponds to the position
of the center of mass of $n$ partons $j_{1},\,...j_{n}$ and was defined
earlier in~(\ref{eq:ImpFac}). The variables $\boldsymbol{n}_{i,j_{1}...j_{n}}=\left(\boldsymbol{x}_{i}-\boldsymbol{b}_{j_{1}...j_{n}}\right)/\left|\boldsymbol{x}_{i}-\boldsymbol{b}_{j_{1}...j_{n}}\right|$
are unit vectors pointing from quark $i$ towards the center-of-mass
of a system of quarks $j_{1}...j_{n}$. It is not possible to do
the remaining integrals over $q_{1},\,k_{2}$ analytically, nor present
the wave function~(\ref{eq:Amp_p_explicit-4-4}) as a convolution
of simpler ``elementary'' wave functions from Section~\ref{subsec:Basics}..
Technically, this happens because in the language of traditional Feynman
diagrams the intermediate (virtual) partons are offshell, and the
integration over $q_{1},k_{2}$ can be rewritten via integrals
over virtualities of intermediate particles. Nevertheless, the structure
of the coordinate dependence of $\psi_{\gamma\to\bar{Q}Q\bar{Q}Q}\left(\left\{ \alpha_{i},\,\boldsymbol{r}_{i}\right\} \right)$
can still be understood using the simple rules suggested in Section~\ref{subsec:Basics}.
Indeed, in the eikonal picture the transverse coordinates of all partons
are frozen. The tree-like structure of the leading order diagrams
1, 2, in Fig.~\ref{fig:Photoproduction-QQQQ} and the iterative evaluation
of the coordinate of the center of mass of two partons $\boldsymbol{b}_{ij}=\left(\alpha_{i}\boldsymbol{r}_{i}+\alpha_{j}\boldsymbol{r}_{j}\right)/\left(\alpha_{i}+\alpha_{j}\right)$
allows to reconstruct the transverse coordinates of all intermediate
partons, as shown in Figure~\ref{fig:Photoproduction-QQQQ-1}.
The variables $\boldsymbol{r}_{1}-\boldsymbol{b}_{34}$ and $\boldsymbol{r}_{2}-\boldsymbol{b}_{34}$
have the physical meaning of the relative distance between the recoil
quark or antiquark and the emitted gluon. Similarly, the variables
$\boldsymbol{r}_{1}-\boldsymbol{b}_{234}$ and $\boldsymbol{r}_{2}-\boldsymbol{b}_{134}$
can be interpreted as the size of the $\bar{Q}Q$ pair produced right
after splitting of the incident photon. These simple rules allow for the construction
of the heavy $\bar{Q}Q\bar{Q}Q$ production amplitude in the
gluonic field of the target.

The wave function $\psi_{\bar{Q}Q\bar{Q}Q}^{(\gamma)}\left(\left\{ \alpha_{i},\,\boldsymbol{r}_{i}\right\} \right)$
has a few singularities which require special attention in order to
guarantee that the amplitudes of the physical processes remain finite.
For the meson pair production, the choice of the quarkonia wave functions~(\ref{eq:WF_JPsi}-\ref{eq:WF_Etac}),
which vanish rapidly near the endpoints is sufficient in order to
guarantee finiteness of the amplitudes~(\ref{eq:Amp-1}-\ref{eq:Amp-2}).

\subsubsection{Instantaneous contributions}

\begin{figure}
\includegraphics[scale=0.65]{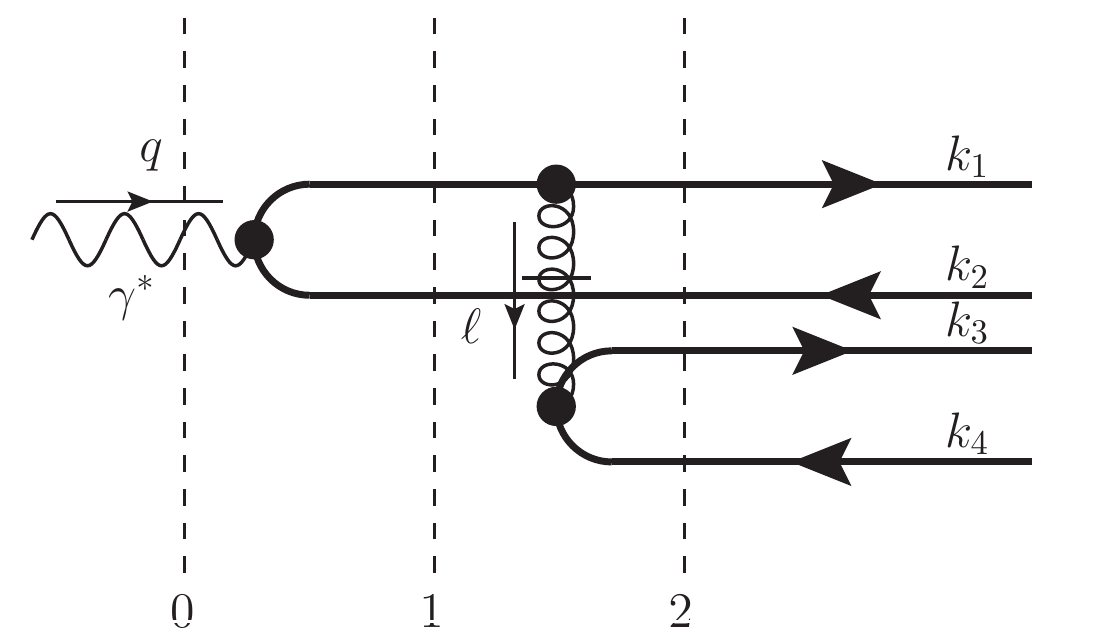}\includegraphics[scale=0.65]{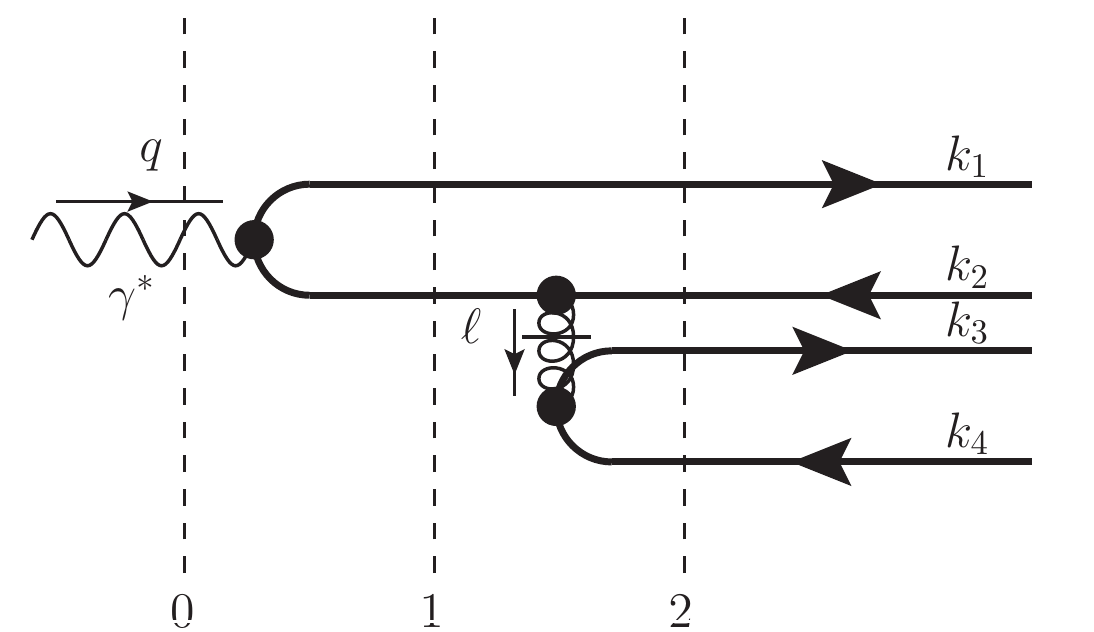}

\includegraphics[scale=0.65]{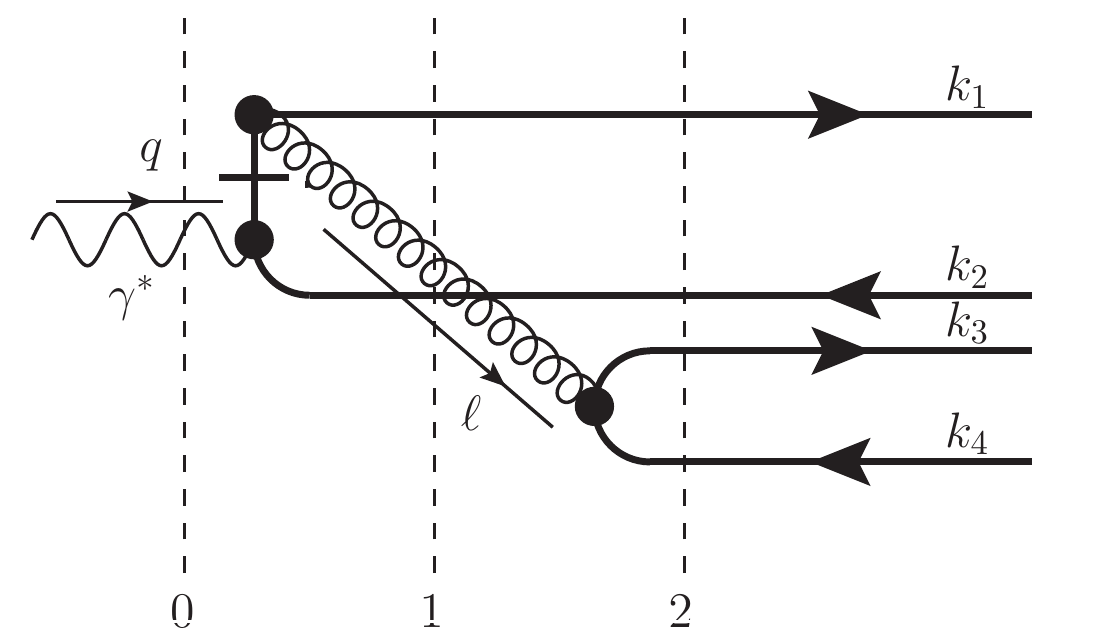}\includegraphics[scale=0.65]{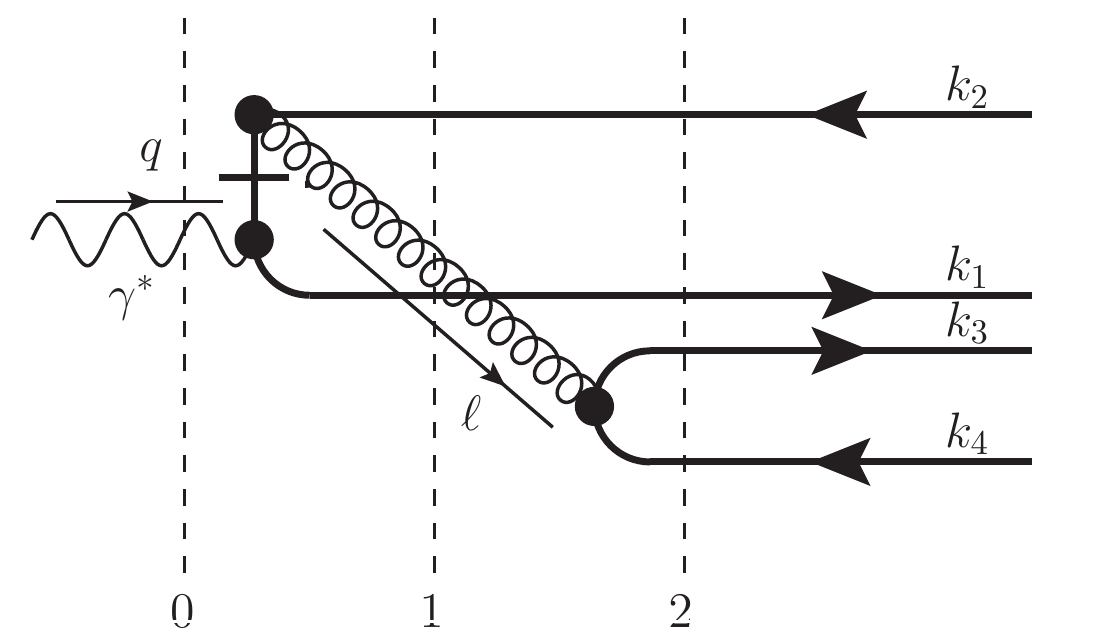}

\caption{\label{fig:Photoproduction-QQQQ-2}The instantaneous contributions
to the wave function $\psi_{\bar{Q}Q\bar{Q}Q}^{(\gamma)}$ defined
in the text. The upper and lower rows correspond to instantaneous
gluons and quarks respectively. The vertical dashed lines denote light-cone
denominators. The momenta $k_{i}$ shown in the right-hand side are
Fourier conjugates of the coordinates $x_{i}$. In what follows we
will refer to the diagrams in the first row as A1, B1, and the diagrams
in the second row as A2, B2, respectively.}
\end{figure}

According to canonical rules of the standard light--cone perturbation
theory~\cite{Lepage:1980fj,Brodsky:1997de}, the evaluations from
the previous section should be supplemented by the instantaneous contributions
of virtual partons. The propagators of the instantaneous offshell
quarks and gluons with momentum $k$ are given by 
\begin{equation}
S_{({\rm inst)}}(k)=\frac{\gamma^{+}}{2k^{+}}\equiv\frac{\left(n\cdot\gamma\right)}{2\left(k\cdot n\right)},\quad\Pi_{({\rm inst})}^{\mu\nu}=\frac{n_{\mu}n_{\nu}}{\left(k^{+}\right)^{2}}
\end{equation}
where $n^{\mu}$ is the light-cone vector in minus-direction. The
results for the instantaneous contributions of gluons are quite straightforward
to get, essentially repeating the evaluations from the previous subsection.
Since $\gamma_{+}\gamma_{+}=0$, there is no diagrams with two instantaneous
propagators (quark and gluon) connected to the same vertex. The numerators
of amplitudes with instantaneous propagators have simple structure
in view of identities~\cite{Lepage:1980fj,Brodsky:1997de,Lappi:2016oup}
$\bar{u}_{h_{f}}\left(p_{1}\right)\gamma_{+}u_{h_{i}}\left(p_{2}\right)=2\sqrt{p_{1}^{+}p_{2}^{+}}\delta_{h_{f},h_{i}}$and
$\bar{u}_{h}\left(p_{1}\right)\gamma_{+}v_{\bar{h}}\left(p_{2}\right)=2\sqrt{p_{1}^{+}p_{2}^{+}}\delta_{h,\,-\bar{h}}$.
The final result of the evaluation is 
\begin{align}
\psi_{\bar{Q}Q\bar{Q}Q}^{(\gamma)}\left(\left\{ \alpha_{i},\,\boldsymbol{r}_{i}\right\} \right) & =A_{g}\left(\left\{ \alpha_{i},\,\boldsymbol{r}_{i}\right\} \right)+B_{g}\left(\left\{ \alpha_{i},\,\boldsymbol{r}_{i}\right\} \right)+\label{eq:PsiFull-1}\\
 & +A_{q}\left(\left\{ \alpha_{i},\,\boldsymbol{r}_{i}\right\} \right)+B_{q}\left(\left\{ \alpha_{i},\,\boldsymbol{r}_{i}\right\} \right),\nonumber 
\end{align}
where the subscript indices $q,g$ in the right-hand side denote the
parton propagator, which should be taken instantaneous ($q$ for quark,
$g$ for gluon), and

\begin{align}
A_{g}\left(\left\{ \alpha_{i},\,\boldsymbol{r}_{i}\right\} \right) & =-\frac{e_{q}\alpha_{s}(m_{Q})\,\left(t_{a}\right)_{c_{1}c_{2}}\otimes\left(t_{a}\right)_{c_{3}c_{4}}}{\pi^{4}\left(1-\alpha_{1}-\alpha_{2}\right)^{3}\,}\int q_{1}dq_{1}\,k_{2}dk_{2}J_{0}\left(q_{1}\left|\boldsymbol{r}_{1}-\boldsymbol{b}_{34}\right|\right)\times\\
 & \times\frac{1}{\boldsymbol{k}_{2\perp}^{2}+m_{1}^{2}}\left[\left(\alpha_{2}\delta_{\gamma,a_{1}}-\bar{\alpha}_{2}\delta_{a_{1},-\gamma}\right)\delta_{a_{1},-a_{2}}i\boldsymbol{n}_{2,134}\cdot\boldsymbol{\varepsilon}_{\gamma}k_{2}J_{1}\left(k_{2}\left|\boldsymbol{r}_{2}-\boldsymbol{b}_{134}\right|\right)\right.\nonumber \\
 & \left.+\frac{m_{q}}{\sqrt{2}}\,{\rm sign}\left(a_{1}\right)\delta_{\gamma,a_{1}}\delta_{a_{1},a_{2}}J_{0}\left(k_{2}\left|\boldsymbol{r}_{2}-\boldsymbol{b}_{134}\right|\right)\right]\alpha_{3}\alpha_{4}\delta_{a_{3},-a_{4}}K_{0}\left(a_{34}\,\boldsymbol{r}_{34}\right).\nonumber 
\end{align}
\begin{align}
A_{q}\left(\left\{ \alpha_{i},\,\boldsymbol{r}_{i}\right\} \right) & =-\frac{e_{q}\alpha_{s}\left(m_{q}\right)\,\left(t_{a}\right)_{c_{1}c_{2}}\otimes\left(t_{a}\right)_{c_{3}c_{4}}}{2\pi^{4}\left(1-\alpha_{1}-\alpha_{2}\right)^{2}\bar{\alpha}_{2}}\delta_{a_{1},-a_{2}}\delta_{\gamma,-a_{1}}\int q_{1}dq_{1}\,k_{2}dk_{2}\frac{J_{0}\left(q_{1}\left|\boldsymbol{r}_{1}-\boldsymbol{b}_{34}\right|\right)J_{0}\left(k_{2}\left|\boldsymbol{r}_{2}-\boldsymbol{b}_{134}\right|\right)}{D_{2}\left(\alpha_{1},\boldsymbol{k}_{1};\alpha_{2},\,\boldsymbol{k}_{2}\right)}\times\\
 & \times\left[-\left(\alpha_{3}\delta_{-\gamma,a_{3}}-\alpha_{4}\delta_{\gamma,a_{3}}\right)\delta_{a_{3},-a_{4}}i\boldsymbol{\varepsilon}_{\gamma}\cdot\boldsymbol{n}_{34}a_{34}K_{1}\left(a_{34}\,\boldsymbol{r}_{34}\right)-\frac{m_{q}(\alpha_{3}+\alpha_{4})}{\sqrt{2}}\,{\rm sign}(a_{3})\delta_{\gamma,-a_{3}}\delta_{a_{3},a_{4}}K_{0}\left(a_{34}\,\boldsymbol{r}_{34}\right)\right]\nonumber \\
 & a_{34}\left(q_{1},\,k_{2}\right)\equiv\sqrt{m_{2}^{2}+\frac{\alpha_{3}\alpha_{4}}{\alpha_{3}+\alpha_{4}}\left[\frac{\bar{\alpha}_{2}q_{1}^{2}}{\alpha_{1}\left(1-\alpha_{1}-\alpha_{2}\right)}+\frac{m_{1}^{2}\left(\alpha_{1}+\alpha_{2}\right)}{\alpha_{1}\alpha_{2}}+\frac{k_{2}^{2}}{\alpha_{2}\bar{\alpha}_{2}}\right]}
\end{align}
and the functions $B_{q},\,B_{g}$ can be obtained from $A_{q},A_{g}$
using 
\begin{equation}
B_{i}\left(\alpha_{1},\,\boldsymbol{x}_{1},\,\alpha_{2},\,\boldsymbol{x}_{2},\,\alpha_{3},\,\boldsymbol{x}_{3},\,\alpha_{4},\,\boldsymbol{x}_{4}\right)=-A_{i}\left(\alpha_{2},\,\boldsymbol{x}_{2},\,\alpha_{1},\,\boldsymbol{x}_{1},\,\alpha_{4},\,\boldsymbol{x}_{4},\,\alpha_{3},\,\boldsymbol{x}_{3}\right),\quad i=q,g.
\end{equation}

\section{Scattering amplitudes in the eikonal approximation}

\label{sec:DipAmp}As was discussed in Section~\ref{subsec:Derivation},
in configuration space the interaction of the target with heavy quarks
reduces to a mere multiplication by the factor $\pm\gamma\left(\boldsymbol{x}_{\perp}\right)$.
For evaluation of the scattering amplitude it is very instructive
to use the light-cone evolution picture of the process, as shown in
Figure~\ref{fig:Photoproduction-QQQQ}, tacitly assuming that
the cuts (vertical dashed lines) in that figure separate different
successive stages of the scattering process.

\begin{figure}
\includegraphics[scale=0.45]{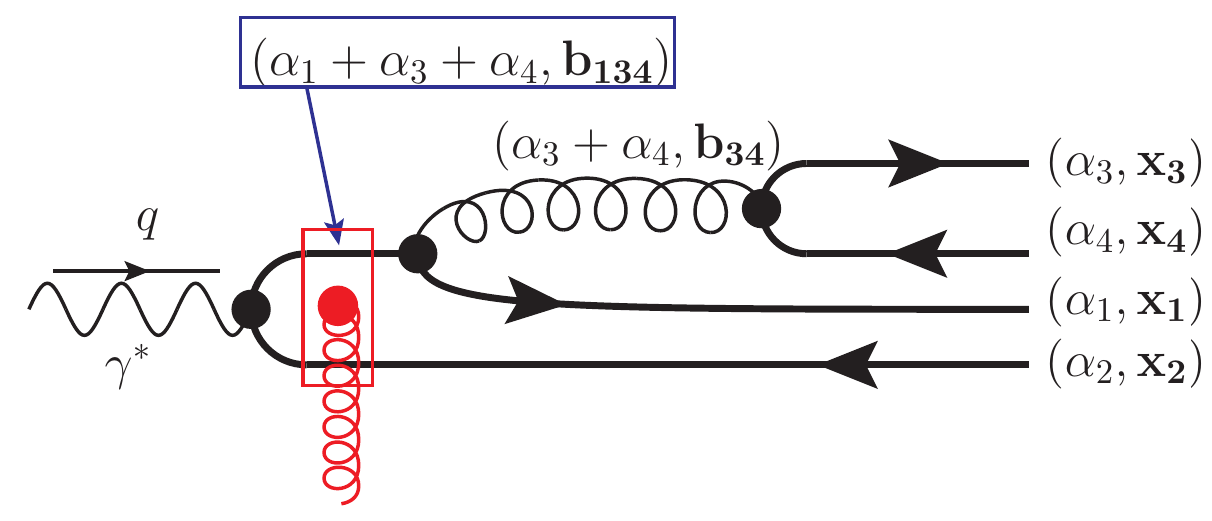}\includegraphics[scale=0.45]{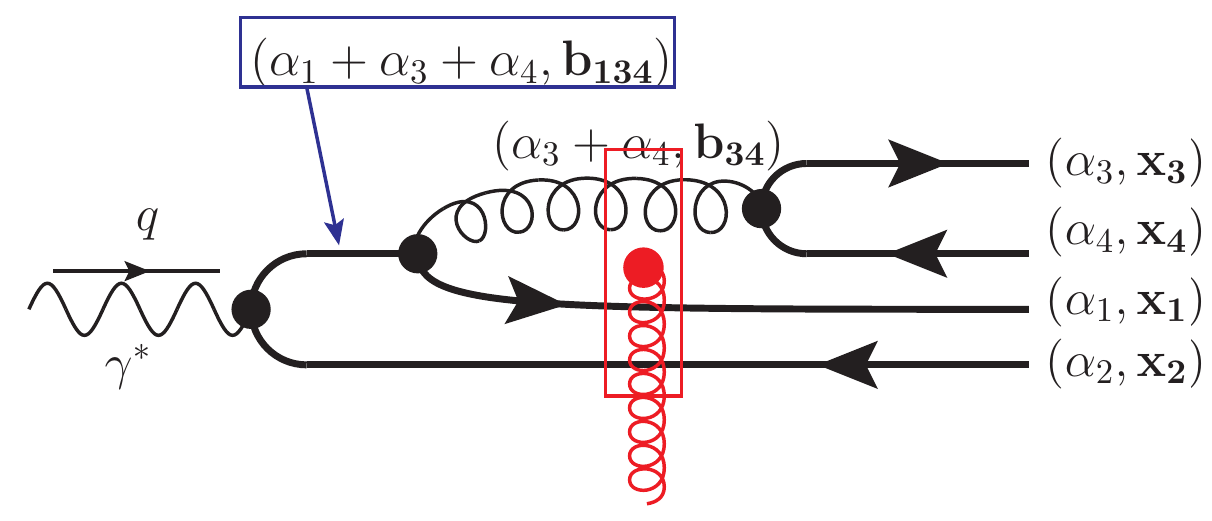}\includegraphics[scale=0.45]{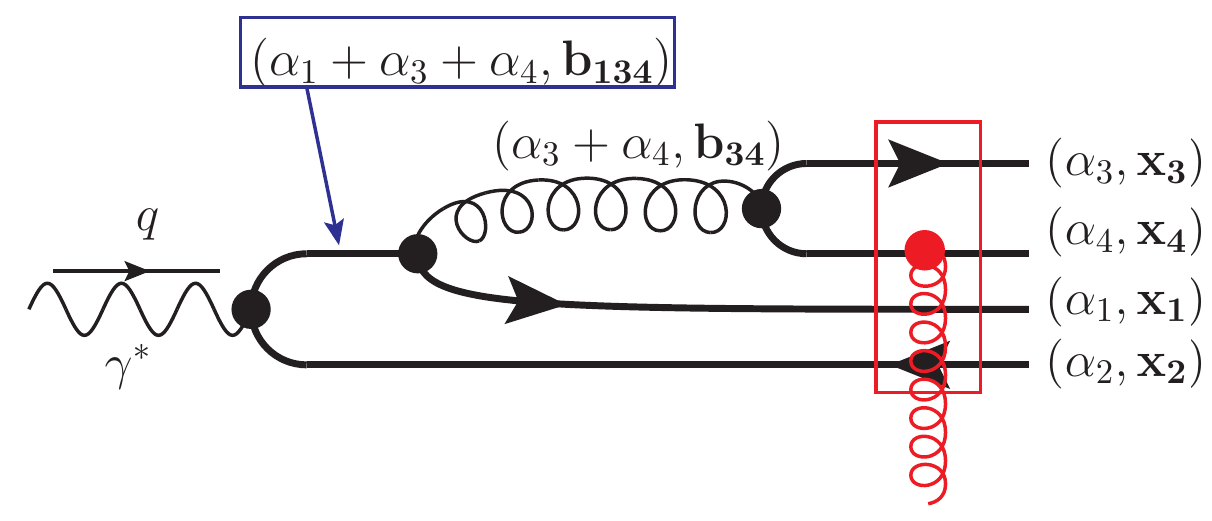}\caption{\label{fig:Photoproduction-QQQQ-1-1}Schematic illustration of the
diagrams which contribute to a $\gamma\to\bar{Q}Q\bar{Q}Q$ subprocess,
via single-gluon exchange in $t$-channel. For the sake of simplicity
we omitted a proton blob in the lower part. The square box with gluon
connected in the middle stands for a coupling of a dipole (sum of
the couplings to all partons which pass through the block, $\sim\sum(\pm)\gamma(\boldsymbol{x}_{i})t_{a}$).
The center-of-mass $\boldsymbol{b}_{i_{1}...i_{n}}$ of a system of
partons $i_{1}...i_{n}$ is defined in~(\ref{eq:ImpFac}). In all
plots it is implied the inclusion of diagrams which can be obtained
by inversion of heavy quark lines (\textquotedblleft charge conjugation\textquotedblright ).}
\end{figure}

We will start assuming first a single-gluon interaction
with high-energy partons. The colorless photon creates a pair of quark
and antiquarks with transverse coordinates $(\boldsymbol{b}_{134}$,$\boldsymbol{x}_{2})$
or $(\boldsymbol{x}_{1},\,\boldsymbol{b}_{234}$$)$ respectively,
as shown in Figure\ref{fig:Photoproduction-QQQQ-1-1}. The eikonal
interaction can occur at any of the three stages of the process,
so the Born amplitude of such process includes a sum of contributions
due to interactions at all stages, 
\begin{equation}
\mathcal{A}=\mathcal{A}_{1}+\mathcal{A}_{2}+\mathcal{A}_{3},
\end{equation}
where the corresponding contributions $\mathcal{A}_{1,2,3}$ are given
by 
\begin{align}
\mathcal{A}_{1} & =\psi_{\bar{Q}Q\bar{Q}Q}^{(\gamma)}\left(\alpha_{1},\boldsymbol{x}_{1};\,\alpha_{2},\,\boldsymbol{x}_{2};\,\alpha_{3},\,\boldsymbol{x}_{3};\,\alpha_{4},\,\boldsymbol{x}_{4};\,q\right)\times\\
 & \times\sum_{acd}\left[\gamma_{c}\left(\boldsymbol{b}_{134}\right)-\gamma_{c}\left(\boldsymbol{x}_{2}\right)\right]\left(\frac{if_{acd}+d_{acd}}{2}\right)\left(t_{d}\right)_{c_{1}c_{2}}\left(t_{a}\right)_{c_{3}c_{4}}\nonumber \\
 & -\left(1\leftrightarrow2,\,\,3\leftrightarrow4\right),\nonumber \\
\mathcal{A}_{2} & =\psi_{\bar{Q}Q\bar{Q}Q}^{(\gamma)}\left(\alpha_{1},\boldsymbol{x}_{1};\,\alpha_{2},\,\boldsymbol{x}_{2};\,\alpha_{3},\,\boldsymbol{x}_{3};\,\alpha_{4},\,\boldsymbol{x}_{4};\,q\right)\times\\
 & \times\left\{ \sum_{acd}\left[\gamma_{c}\left(\boldsymbol{x}_{1}\right)+\gamma_{c}\left(\boldsymbol{x}_{2}\right)-2\gamma_{c}\left(\boldsymbol{x}_{34}\right)\right]\left(\frac{if_{acd}}{2}\right)\left(t_{d}\right)_{c_{1}c_{2}}\left(t_{a}\right)_{c_{3}c_{4}}\right.+\nonumber \\
 & +\sum_{acd}\left[\gamma_{c}\left(\boldsymbol{x}_{1}\right)-\gamma_{c}\left(\boldsymbol{x}_{2}\right)\right]\left(\frac{d_{acd}}{2}\right)\left(t_{d}\right)_{c_{1}c_{2}}\left(t_{a}\right)_{c_{3}c_{4}}+\nonumber \\
 & +\left.\sum_{acd}\left[\gamma_{c}\left(\boldsymbol{x}_{1}\right)-\gamma_{c}\left(\boldsymbol{x}_{2}\right)\right]\delta_{c_{1}c_{2}}\left(t_{c}\right)_{c_{3}c_{4}}\right\} \nonumber \\
 & -\left(1\leftrightarrow2,\,\,3\leftrightarrow4\right),\nonumber \\
\mathcal{A}_{3} & =\psi_{\bar{Q}Q\bar{Q}Q}^{(\gamma)}\left(\alpha_{1},\boldsymbol{x}_{1};\,\alpha_{2},\,\boldsymbol{x}_{2};\,\alpha_{3},\,\boldsymbol{x}_{3};\,\alpha_{4},\,\boldsymbol{x}_{4};\,q\right)\times\\
 & \times\left\{ \sum_{acd}\left[\gamma_{c}\left(\boldsymbol{x}_{1}\right)+\gamma_{c}\left(\boldsymbol{x}_{2}\right)-\gamma_{c}\left(\boldsymbol{x}_{3}\right)-\gamma_{c}\left(\boldsymbol{x}_{4}\right)\right]\left(\frac{if_{cad}}{2}\right)\left(t_{d}\right)_{c_{1}c_{2}}\left(t_{a}\right)_{c_{3}c_{4}}\right.+\nonumber \\
 & +\left.\sum_{acd}\left[\gamma_{c}\left(\boldsymbol{x}_{1}\right)-\gamma_{c}\left(\boldsymbol{x}_{2}\right)+\gamma_{c}\left(\boldsymbol{x}_{3}\right)-\gamma_{c}\left(\boldsymbol{x}_{4}\right)\right]\left(\frac{d_{acd}}{2}\right)\left(t_{d}\right)_{c_{1}c_{2}}\left(t_{a}\right)_{c_{3}c_{4}}\right\} \nonumber \\
 & -\left(1\leftrightarrow2,\,\,3\leftrightarrow4\right),\nonumber 
\end{align}
We may observe that all factors $\gamma_{c}(\boldsymbol{x}_{i})$
always appear in combination $\gamma_{c}(\boldsymbol{x}_{i})-\gamma_{c}(\boldsymbol{x}_{j})$,
which guarantees that in the heavy quark mass limit, when the distances
between the quarks are small, the corresponding amplitude is suppressed
at least as $\sim1/m_{Q}$. The three-gluon coupling $\sim\gamma\left(\boldsymbol{x}_{34}\right)$
always appears in combination $\left[\gamma_{c}\left(\boldsymbol{x}_{1}\right)+\gamma_{c}\left(\boldsymbol{x}_{2}\right)-2\gamma_{c}\left(\boldsymbol{x}_{34}\right)\right],$
in agreement with the earlier findings of~\cite{Kopeliovich:2002yv}.
\begin{figure}
\includegraphics[scale=0.25]{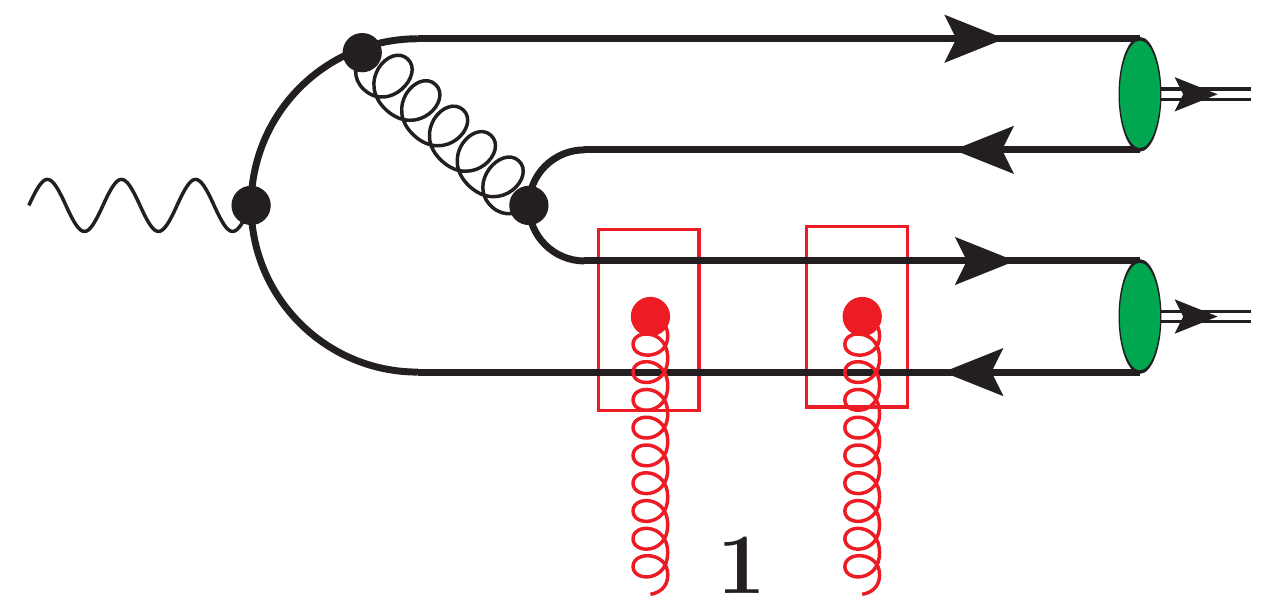}\includegraphics[scale=0.25]{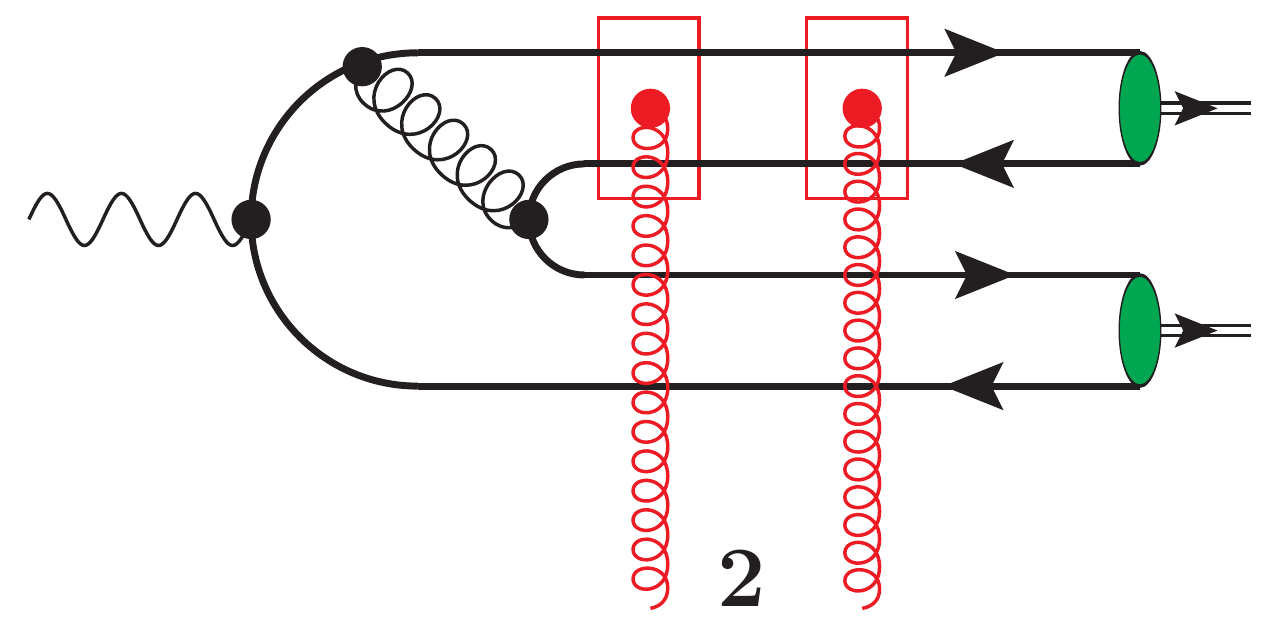}\includegraphics[scale=0.25]{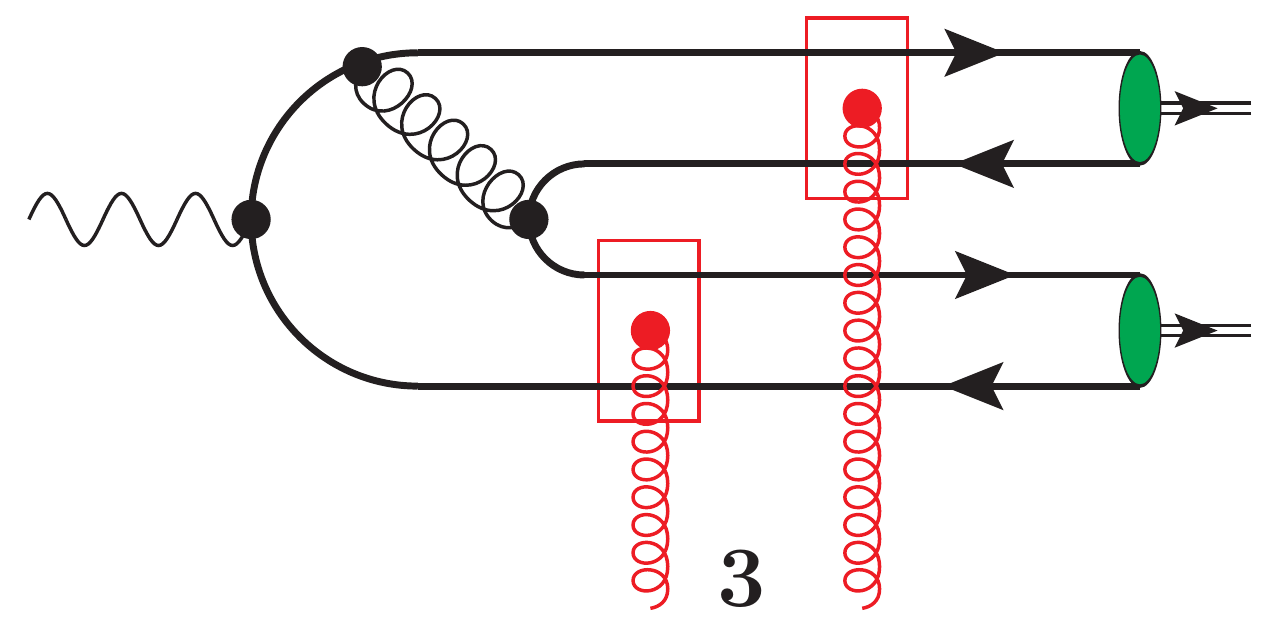}\includegraphics[scale=0.25]{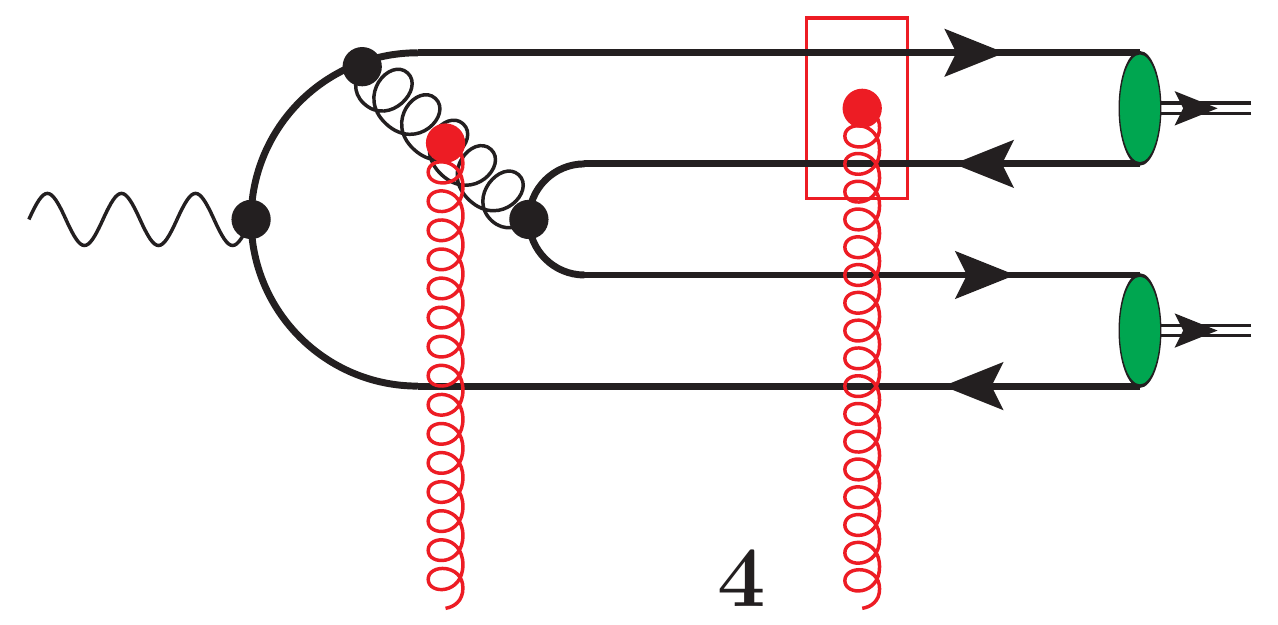}\includegraphics[scale=0.25]{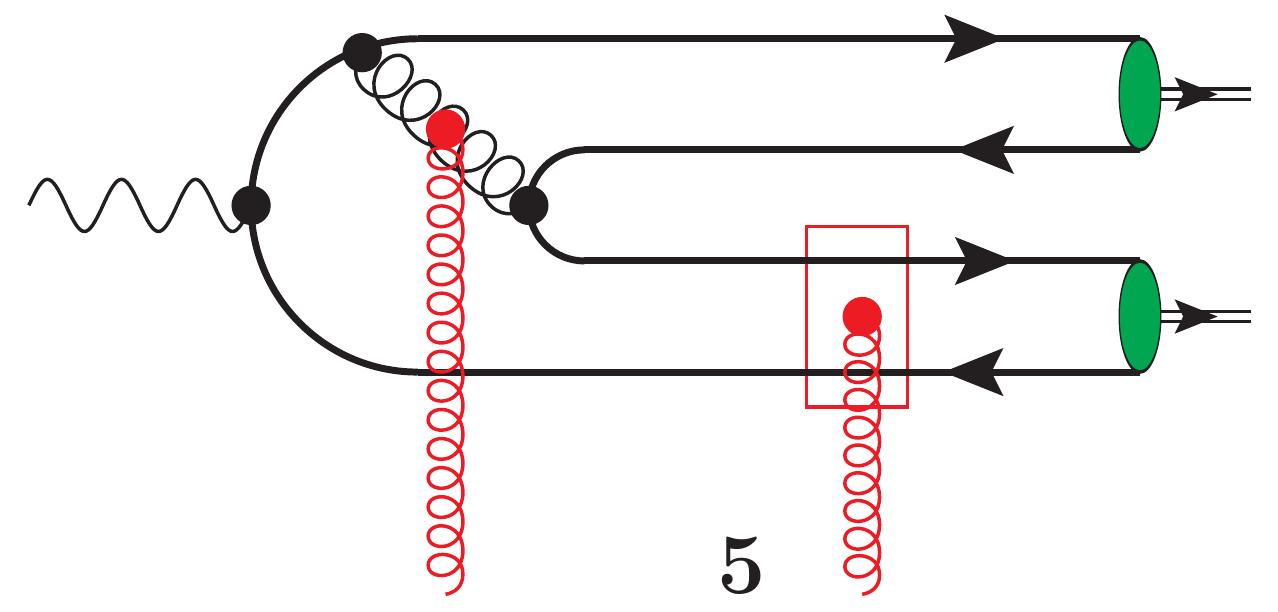}

\includegraphics[scale=0.25]{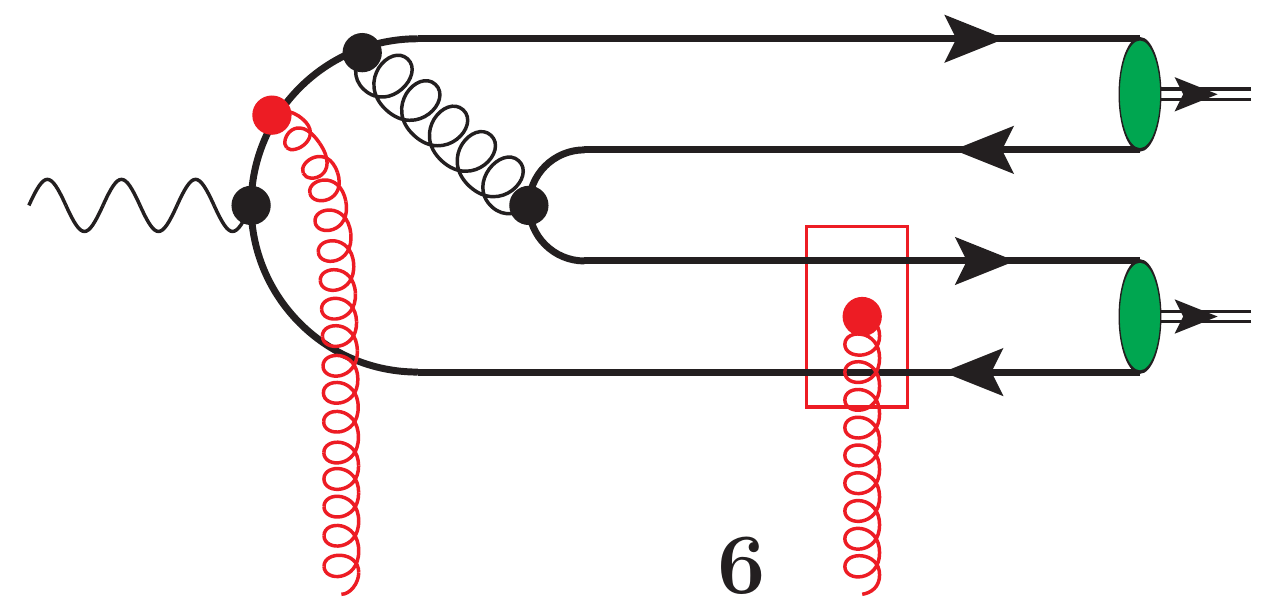}\includegraphics[scale=0.25]{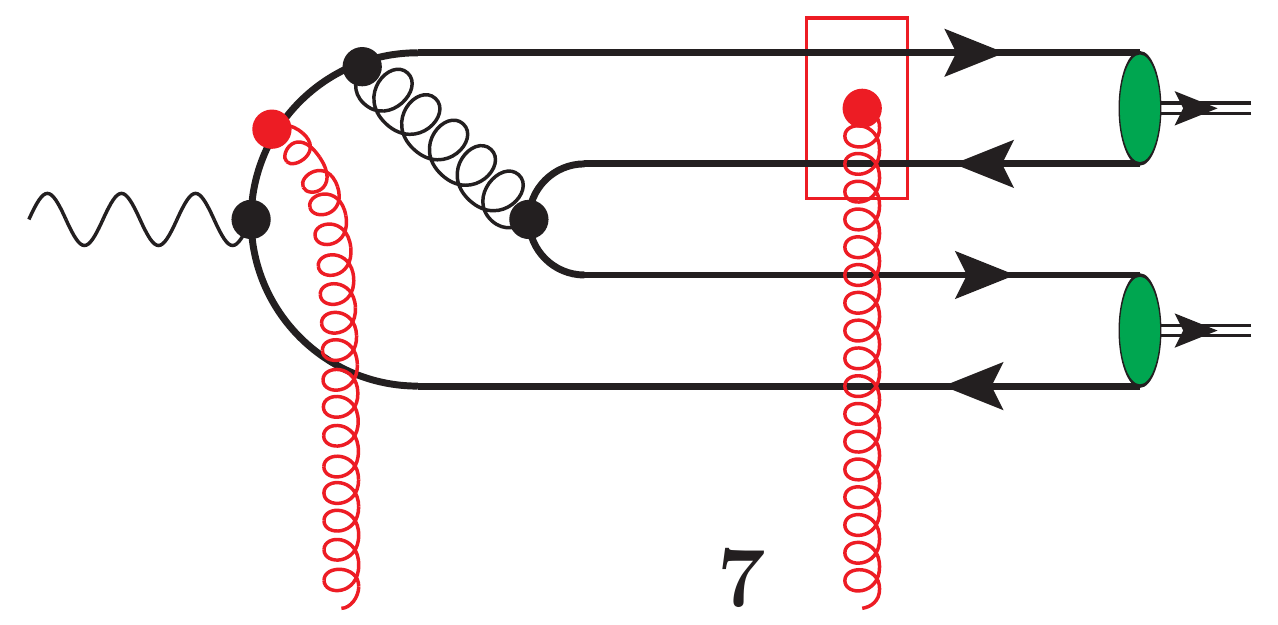}\includegraphics[scale=0.25]{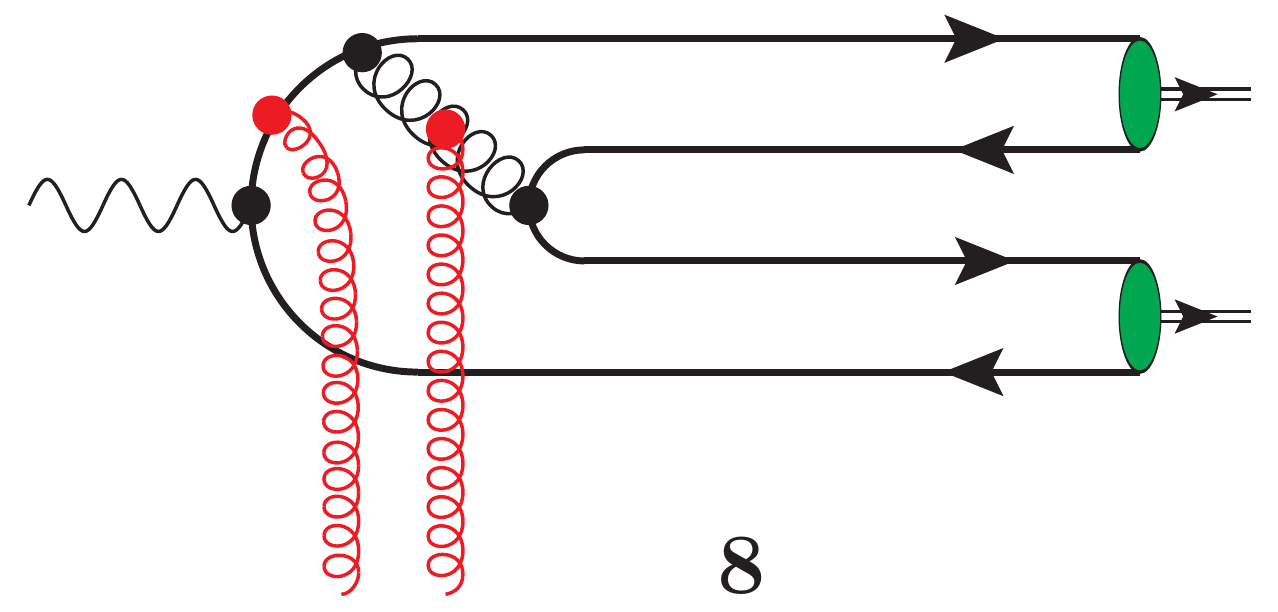}\includegraphics[scale=0.25]{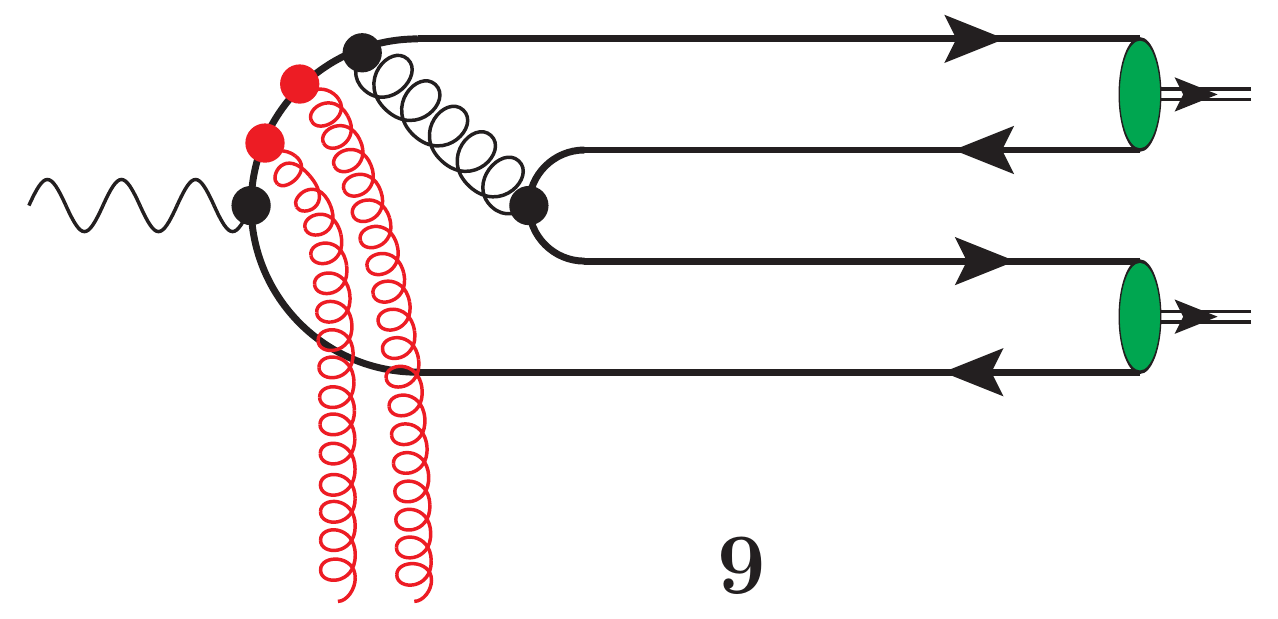}\includegraphics[scale=0.25]{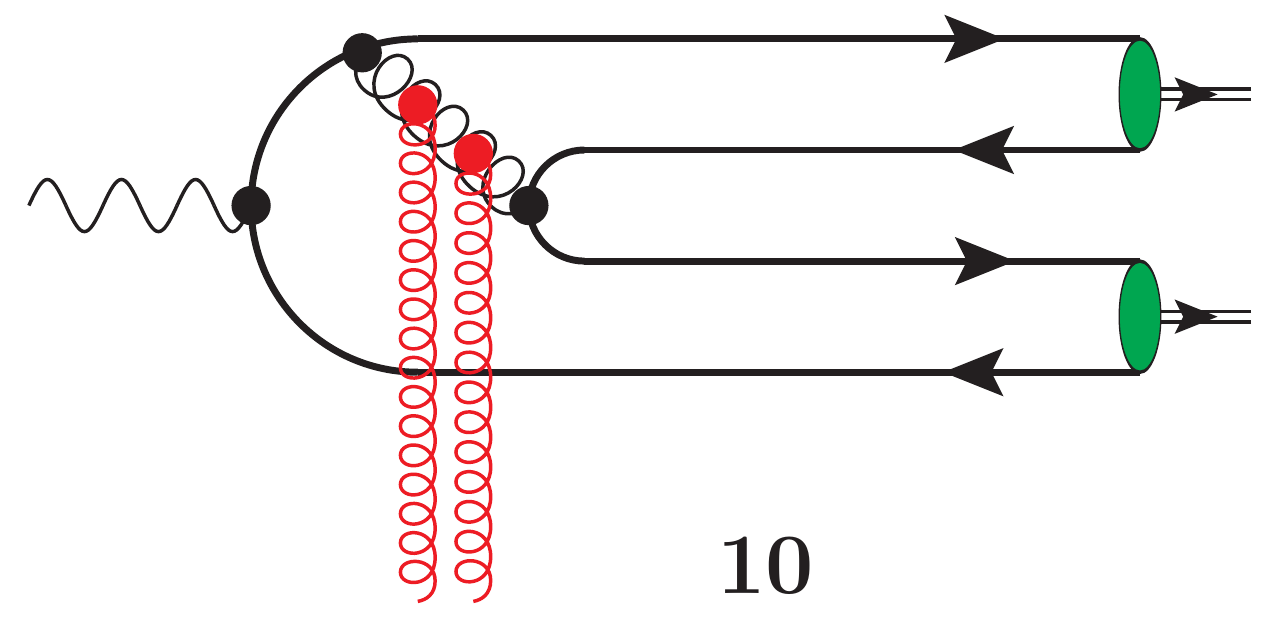}\caption{\label{fig:Photoproduction-1}Schematic illustration of the diagrams
which contribute to meson pair production. For the sake of simplicity
we approximated a pomeron with a pair of $t$-channel gluons, and
omitted all possible gluon exchanges between them (as well as a proton
blob in the lower part). The square box with gluon connected in the
middle stands for a dipole coupling (sum of the couplings of
a quark and antiquark which pass through the block, $\sim\left(\gamma(\boldsymbol{x}_{Q})-\gamma(\boldsymbol{x}_{\bar{Q}})\right)t_{a}$).
In all plots it is implied the inclusion of diagrams which can be obtained
by inversion of the heavy quark lines (\textquotedblleft charge conjugation\textquotedblright ).}
\end{figure}

For the case of two gluon exchanges, we may repeat the same evaluations,
taking into account the set of diagrams shown in Figure~\ref{fig:Photoproduction-1}.
The final result of this evaluation is 
\begin{align}
\mathcal{A} & =\psi_{\bar{Q}Q\bar{Q}Q}^{(\gamma)}\left(\alpha_{1},\boldsymbol{x}_{1};\,\alpha_{2},\,\boldsymbol{x}_{2};\,\alpha_{3},\,\boldsymbol{x}_{3};\,\alpha_{4},\,\boldsymbol{x}_{4};\,q\right)\times\label{eq:Amp_gg}\\
 & \times\left\{ \mathcal{C}_{1}\left[\left(\gamma\left(\boldsymbol{x}_{1}\right)-\gamma\left(\boldsymbol{x}_{4}\right)\right)^{2}+\left(\gamma\left(\boldsymbol{x}_{3}\right)-\gamma\left(\boldsymbol{x}_{2}\right)\right)^{2}\right]+\right.\\
 & +2\mathcal{C}_{2}\left(\gamma\left(\boldsymbol{x}_{1}\right)-\gamma\left(\boldsymbol{x}_{4}\right)\right)\left(\gamma\left(\boldsymbol{x}_{3}\right)-\gamma\left(\boldsymbol{x}_{2}\right)\right)-\nonumber \\
 & +\mathcal{C}_{3}\left(\gamma\left(\boldsymbol{x}_{1}\right)+\gamma\left(\boldsymbol{x}_{2}\right)-2\gamma\left(\boldsymbol{x}_{34}\right)\right)\left(\gamma\left(\boldsymbol{x}_{1}\right)+\gamma\left(\boldsymbol{x}_{2}\right)-\gamma\left(\boldsymbol{x}_{3}\right)-\gamma\left(\boldsymbol{x}_{4}\right)\right)\nonumber \\
 & +\mathcal{C}_{1}\left(\gamma\left(\boldsymbol{b}_{134}\right)-\gamma\left(\boldsymbol{x}_{2}\right)\right)\left(\gamma\left(\boldsymbol{x}_{3}\right)-\gamma\left(\boldsymbol{x}_{2}\right)\right)+\nonumber \\
 & +\mathcal{C}_{2}\left(\gamma\left(\boldsymbol{b}_{134}\right)-\gamma\left(\boldsymbol{x}_{2}\right)\right)\left(\gamma\left(\boldsymbol{x}_{1}\right)-\gamma\left(\boldsymbol{x}_{4}\right)\right)+\nonumber \\
 & -\mathcal{C}_{3}\left(\gamma\left(\boldsymbol{b}_{134}\right)-\gamma\left(\boldsymbol{x}_{2}\right)\right)\left(\gamma\left(\boldsymbol{x}_{1}\right)+\gamma\left(\boldsymbol{x}_{2}\right)-2\gamma\left(\boldsymbol{x}_{34}\right)\right)+\nonumber \\
 & +\left.\mathcal{C}_{1}\left(\gamma\left(\boldsymbol{b}_{134}\right)-\gamma\left(\boldsymbol{x}_{2}\right)\right)^{2}+\mathcal{C}_{4}\left(\gamma\left(\boldsymbol{x}_{1}\right)+\gamma\left(\boldsymbol{x}_{2}\right)-2\gamma\left(\boldsymbol{x}_{34}\right)\right)^{2}\right\} \nonumber 
\end{align}
where the color factors $\mathcal{C}_{1,2,3}$ were defined earlier
in Section~\ref{subsec:Evaluation}, in the text under Eq.~(\ref{eq:Amp-2}).
With the help of (\ref{eq:N_dip},~\ref{eq:N_dip3}) it is possible
to rewrite the amplitude~(\ref{eq:Amp_gg}) as 
\begin{align}
\mathcal{A} & =\psi_{\bar{Q}Q\bar{Q}Q}^{(\gamma)}\left(\alpha_{1},\boldsymbol{x}_{1};\,\alpha_{2},\,\boldsymbol{x}_{2};\,\alpha_{3},\,\boldsymbol{x}_{3};\,\alpha_{4},\,\boldsymbol{x}_{4};\,q\right)\times\label{eq:Amp_gg-1}\\
 & \left\{ -2\mathcal{C}_{1}\left[N\left(x,\,\boldsymbol{r}_{14},\,\boldsymbol{b}_{14}\right)+N\left(x,\,\boldsymbol{r}_{23},\,\boldsymbol{b}_{23}\right)\right]+\right.\nonumber \\
 & +2\mathcal{C}_{2}\left[N\left(x,\,\boldsymbol{r}_{34},\,\boldsymbol{b}_{34}\right)+N\left(x,\,\boldsymbol{r}_{12},\,\boldsymbol{b}_{12}\right)\right.\left.-N\left(x,\,\boldsymbol{r}_{13},\,\boldsymbol{b}_{13}\right)-N\left(x,\,\boldsymbol{r}_{24},\,\boldsymbol{b}_{24}\right)\right]\nonumber \\
 & +\mathcal{C}_{3}\left[-2N\left(x,\boldsymbol{r}_{12},\boldsymbol{b}_{12}\right)+N\left(x,\boldsymbol{r}_{13},\boldsymbol{b}_{13}\right)+N\left(x,\boldsymbol{r}_{14},\boldsymbol{b}_{14}\right)+N\left(x,\boldsymbol{r}_{23},\boldsymbol{b}_{23}\right)+N\left(x,\boldsymbol{r}_{24},\boldsymbol{b}_{24}\right)\right.\nonumber \\
 & \left.+2N\left(x,\boldsymbol{\boldsymbol{r}}_{1,34},\boldsymbol{b}_{134}\right)+2N\left(x,\boldsymbol{\boldsymbol{r}}_{2,34},\boldsymbol{b}_{234}\right)-2N\left(x,\boldsymbol{\boldsymbol{r}}_{3,34},\boldsymbol{b}_{334}\right)-2N\left(x,\boldsymbol{\boldsymbol{r}}_{4,34},\boldsymbol{b}_{344}\right)\right]\nonumber \\
 & +\mathcal{C}_{1}\left[N\left(x,\,\boldsymbol{r}_{2,134},\boldsymbol{b}_{1234}\right)+N\left(x,\,\boldsymbol{r}_{23},\,\boldsymbol{b}_{23}\right)-N\left(x,\,\boldsymbol{r}_{3,134},\,\boldsymbol{b}_{1334}\right)\right]\nonumber \\
 & +\mathcal{C}_{2}\left[N\left(x,\,\boldsymbol{r}_{4,134},\,\boldsymbol{b}_{1344}\right)+N\left(x,\,\boldsymbol{r}_{12},\,\boldsymbol{b}_{12}\right)-N\left(x,\,\boldsymbol{r}_{1,134},\,\boldsymbol{b}_{1,134}\right)-N\left(x,\,\boldsymbol{r}_{24},\,\boldsymbol{b}_{24}\right)\right]\nonumber \\
 & -\mathcal{C}_{3}\left[-N\left(x,\boldsymbol{r}_{1,134},\boldsymbol{b}_{1134}\right)-N\left(x,\boldsymbol{r}_{2,134},\boldsymbol{b}_{1234}\right)+N\left(x,\boldsymbol{r}_{12},\boldsymbol{b}_{12}\right)\right.\nonumber \\
 & \,\,\,\,\,\left.+2N\left(x,\boldsymbol{r}_{34,134},\boldsymbol{b}_{34,134}\right)-2N\left(x,\boldsymbol{r}_{2,34},\boldsymbol{b}_{234}\right)\right]+\mathcal{C}_{1}N\left(x,\,\boldsymbol{r}_{2,134},\boldsymbol{b}_{1234}\right)\nonumber \\
 & +\left.\mathcal{C}_{4}\left[2N\left(x,\,\boldsymbol{r}_{1,34},\,\boldsymbol{b}_{134}\right)+2N\left(x,\,\boldsymbol{r}_{2,34},\,\boldsymbol{b}_{234}\right)-N\left(x,\,\boldsymbol{r}_{12},\,\boldsymbol{b}_{12}\right)\right]\right\} \nonumber 
\end{align}
where 
\begin{align}
\boldsymbol{r}_{1,34} & =\boldsymbol{r}_{1}-\frac{\alpha_{3}\boldsymbol{r}_{3}+\alpha_{4}\boldsymbol{r}_{4}}{\alpha_{3}+\alpha_{4}}=\frac{\alpha_{3}\boldsymbol{r}_{13}+\alpha_{4}\boldsymbol{r}_{14}}{\alpha_{3}+\alpha_{4}},\\
\boldsymbol{r}_{2,34} & =\frac{\alpha_{3}\boldsymbol{r}_{23}+\alpha_{4}\boldsymbol{r}_{24}}{\alpha_{3}+\alpha_{4}},\\
\boldsymbol{r}_{3,34} & =\frac{\alpha_{4}\boldsymbol{r}_{34}}{\alpha_{3}+\alpha_{4}},\\
\boldsymbol{r}_{4,34} & =-\frac{\alpha_{3}\boldsymbol{r}_{34}}{\alpha_{3}+\alpha_{4}}=-\frac{\alpha_{3}}{\alpha_{4}}\boldsymbol{r}_{3,34},\\
\boldsymbol{r}_{34,134} & =\frac{\alpha_{3}\boldsymbol{r}_{3}+\alpha_{4}\boldsymbol{r}_{4}}{\alpha_{3}+\alpha_{4}}-\frac{\alpha_{1}\boldsymbol{r}_{1}+\alpha_{3}\boldsymbol{r}_{3}+\alpha_{4}\boldsymbol{r}_{4}}{\alpha_{1}+\alpha_{3}+\alpha_{4}}=\\
 & =-\frac{\alpha_{1}\left(\alpha_{3}\boldsymbol{r}_{13}+\alpha_{4}\boldsymbol{r}_{14}\right)}{\left(\alpha_{3}+\alpha_{4}\right)\left(\alpha_{1}+\alpha_{3}+\alpha_{4}\right)}\nonumber 
\end{align}
\begin{align}
\boldsymbol{r}_{1,134} & =\boldsymbol{r}_{1}-\frac{\alpha_{1}\boldsymbol{r}_{1}+\alpha_{3}\boldsymbol{r}_{3}+\alpha_{4}\boldsymbol{r}_{4}}{1-\alpha_{2}}=\frac{\left(1-\alpha_{2}-\alpha_{3}\right)\boldsymbol{r}_{1}-\alpha_{3}\boldsymbol{r}_{3}-\alpha_{4}\boldsymbol{r}_{4}}{1-\alpha_{2}}\\
 & =\frac{\alpha_{3}\boldsymbol{r}_{13}+\alpha_{4}\boldsymbol{r}_{14}}{1-\alpha_{2}},\nonumber\\
\boldsymbol{r}_{2,134} & =\boldsymbol{r}_{2}-\frac{\alpha_{1}\boldsymbol{r}_{1}+\alpha_{3}\boldsymbol{r}_{3}+\alpha_{4}\boldsymbol{r}_{4}}{1-\alpha_{2}}=\frac{\left(1-\alpha_{2}\right)\boldsymbol{r}_{2}-\alpha_{3}\boldsymbol{r}_{3}-\alpha_{1}\boldsymbol{r}_{1}-\alpha_{4}\boldsymbol{r}_{4}}{1-\alpha_{2}}\\
 & =\frac{\alpha_{1}\boldsymbol{r}_{21}+\alpha_{3}\boldsymbol{r}_{23}+\alpha_{4}\boldsymbol{r}_{24}}{1-\alpha_{2}}\nonumber
  \end{align}
\begin{align}
\boldsymbol{r}_{3,134} & =\boldsymbol{r}_{3}-\frac{\alpha_{1}\boldsymbol{r}_{1}+\alpha_{3}\boldsymbol{r}_{3}+\alpha_{4}\boldsymbol{r}_{4}}{1-\alpha_{2}}=\frac{\left(1-\alpha_{2}-\alpha_{3}\right)\boldsymbol{r}_{3}-\alpha_{1}\boldsymbol{r}_{1}-\alpha_{4}\boldsymbol{r}_{4}}{1-\alpha_{2}}\\
 & =\frac{\alpha_{1}\boldsymbol{r}_{31}+\alpha_{4}\boldsymbol{r}_{34}}{1-\alpha_{2}}\nonumber \\
\boldsymbol{r}_{4,134} & =\boldsymbol{r}_{4}-\frac{\alpha_{1}\boldsymbol{r}_{1}+\alpha_{3}\boldsymbol{r}_{3}+\alpha_{4}\boldsymbol{r}_{4}}{1-\alpha_{2}}=\frac{\left(1-\alpha_{2}-\alpha_{4}\right)\boldsymbol{r}_{4}-\alpha_{1}\boldsymbol{r}_{1}-\alpha_{3}\boldsymbol{r}_{3}}{1-\alpha_{2}}\\
 & =\frac{\alpha_{1}\boldsymbol{r}_{41}+\alpha_{3}\boldsymbol{r}_{43}}{1-\alpha_{2}}=-\frac{\alpha_{1}\boldsymbol{r}_{14}+\alpha_{3}\boldsymbol{r}_{34}}{1-\alpha_{2}},\nonumber 
\end{align}
If we introduce a variable $\boldsymbol{R}=\sum\alpha_{i}\boldsymbol{r}_{i}$
then we may rewrite the above-given expressions as 
\begin{align}
r_{1,134} & =\boldsymbol{r}_{1}-\frac{\boldsymbol{R}-\alpha_{2}\boldsymbol{r}_{2}}{1-\alpha_{2}}=\frac{\bar{\alpha}_{2}\boldsymbol{r}_{1}+\alpha_{2}\boldsymbol{r}_{2}-\boldsymbol{R}}{1-\alpha_{2}},\\
r_{2,134} & =\boldsymbol{r}_{2}-\frac{\boldsymbol{R}-\alpha_{2}\boldsymbol{r}_{2}}{1-\alpha_{2}}=\frac{\boldsymbol{r}_{2}-\boldsymbol{R}}{1-\alpha_{2}},\\
r_{3,134} & =\boldsymbol{r}_{3}-\frac{\boldsymbol{R}-\alpha_{2}\boldsymbol{r}_{2}}{1-\alpha_{2}}=\frac{\bar{\alpha}_{2}\boldsymbol{r}_{3}+\alpha_{2}\boldsymbol{r}_{2}-\boldsymbol{R}}{1-\alpha_{2}},\\
r_{4,134} & =\boldsymbol{r}_{4}-\frac{\boldsymbol{R}-\alpha_{2}\boldsymbol{r}_{2}}{1-\alpha_{2}}=\frac{\bar{\alpha}_{2}\boldsymbol{r}_{4}+\alpha_{2}\boldsymbol{r}_{2}-\boldsymbol{R}}{1-\alpha_{2}}.
\end{align}
Using the values of color factors $\mathcal{C}_{1}=\left(N_{c}^{2}-1\right)/4N_{c}=\mathcal{C}_{2}+\mathcal{C}_{3}$,
$\mathcal{C}_{2}=-1/4N_{c}$, $\mathcal{C}_{3}=N_{c}/4$, $\mathcal{C}_{4}\equiv N_{c}/2=2\mathcal{C}_{3}$,
and identifying the coefficient in front of $\psi_{\bar{Q}Q\bar{Q}Q}^{(\gamma)}$
in (\ref{eq:Amp_gg-1}) with $\tilde{\sum}_{\ell n}\sigma_{\ell}\sigma_{n}\,c_{\ell n}\gamma\left(\boldsymbol{b}_{\ell}\right)\gamma\left(\boldsymbol{b}_{n}\right)$
in~(\ref{eq:Amp-1}), we get the final result~(\ref{eq:N_dip-1}).
The evaluation of the amplitude (\ref{eq:Amp-2}) follows the same
algorithm; technically it is significantly simpler, because the production
of two colorless $\bar{Q}Q$ requires in this topology that each of
the $t$-channel gluons should be attached to different quark loops,
thus significantly reducing the number of possible diagrams. After
straightforward algebraic simplifications, we can get
the final result for this case~(\ref{eq:N_dip-2}). 

 \end{document}